\documentclass[aip,
jcp,
reprint,
superscriptaddress,
amsmath,amssymb,floatfix,pra,
longbibliography
]{revtex4-2}
\usepackage{printlen}
\usepackage{lipsum}
\usepackage{graphicx}
\usepackage{dcolumn}
\usepackage{comment}
\usepackage{bm}
\usepackage{color}
\usepackage{graphicx}
\usepackage{dcolumn}
\usepackage{comment}
\usepackage{bm}
\usepackage{xcolor}
\definecolor{darkblue}{rgb}{0,0,0.6}
\usepackage{hyperref}
\hypersetup{colorlinks,linkcolor=darkblue,citecolor=darkblue,urlcolor=darkblue}

\usepackage[normalem]{ulem}

\definecolor{lightgray}{gray}{0.8}

\begin{document}

\title{Hyperuniformity and conservation laws in non-equilibrium systems}

\author{Rapha\"el Maire}
\thanks{\href{mailto:raphael.maire@universite-paris-saclay.fr}{raphael.maire@universite-paris-saclay.fr}}
\affiliation{Universit\'e Paris-Saclay, CNRS, Laboratoire de Physique des Solides, 91405 Orsay, France}

\author{Ludivine Chaix}

\affiliation{Institut Curie, Universit\'e PSL, Sorbonne Université, CNRS UMR168, Physique des Cellules et Cancer, 75005 Paris, France}

\date{\today}

\begin{abstract}

We demonstrate that hyperuniformity, the suppression of density fluctuations at large length scales, emerges generically from the interplay between conservation laws and non-equilibrium driving. The underlying mechanism for this emergence is analogous to self-organized criticality. Based on this understanding, we introduce four non-equilibrium models that consistently demonstrate hyperuniformity. Furthermore, we show that systems with an arbitrary number of conserved mass multipole moments exhibit an arbitrary strong tunable hyperuniform scaling, with the structure factor following $S(k) \sim k^m$, where $m$ is set by the number of conserved multipoles. Finally, we find that hyperuniformity arising from a combination of conserved noise and partially conserved average motion is not robust against non-linear perturbations. Notably, non-linear damping destroys hyperuniformity in hyperuniform fluids. These results highlight the central role of conservation laws in stabilizing hyperuniformity and reveal a unifying mechanism for its emergence in non-equilibrium systems.

\end{abstract}

\maketitle

\section{Introduction}

Complex systems, from flocks of birds~\cite{cavagna2017dynamic} and financial markets~\cite{mandelbrot2007misbehavior}, to neuronal networks in the brain~\cite{friedman2012universal} often exhibit long-range correlations that challenge the intuition of equilibrium physics. At equilibrium, such large-scale order is a hallmark of criticality, a special state that typically requires fine-tuning of system parameters like temperature~\cite{chaikin1995principles}. Yet, in many natural and artificial systems operating far from equilibrium, highly correlated states appear robustly, without any obvious fine-tuning. This observation suggests the existence of deep organizing principles governing non-equilibrium matter. Self-organized criticality emerged as the classic paradigm to explain this phenomenon, proposing that dissipative systems can spontaneously evolve toward a critical-like state without any external tuning~\cite{bak1988self}. One striking manifestation of such emergent organization is given by non-equilibrium hyperuniform fluids~\cite{lei2024non}. These systems exhibit very strong long-range correlations, characterized by an anomalous suppression of density fluctuations on large length scales~\cite{torquato2018hyperuniform, salvalaglio2024persistent}. Disregarding non-equilibrium fluids for a moment, a trivial example of a hyperuniform system is given by perfect crystals. However, this extreme order is fragile: in equilibrium systems with short-range interactions, any thermal fluctuation readily destroys hyperuniformity~\cite{kim2018effect}, and its persistence at finite temperature typically requires long-range interactions~\cite{torquato2018hyperuniform,hansen2013theory}. Therefore, a typical emergence of hyperuniformity requires driving the system out of equilibrium.
\\
Indeed, a wide variety of driven systems such as center of mass conserving random organization models~\cite{hexner2017noise,PhysRevE.99.022115}, related lattice models~\cite{bertrand2019nonlinear, mukherjee2024anomalous,hazra2025hyperuniformity, dandekar2020exact} and active or granular matter with reciprocal interactions~\cite{lei2024non, kuroda2023microscopic,kuroda2025singulardensitycorrelationschiral, keta2024long, lei2019hydrodynamics, lei2019nonequilibrium,lei2023random, maire2025dynamical,li2025fluidization,gao2025liquidgascriticalityhyperuniformfluids, liu2023local, liu2025hyperuniform} have been shown to display hyperuniformity. Strikingly, these systems robustly \textit{self-organize} into a hyperuniform state without requiring fine-tuning or long-range interactions. This emergent behavior consistently arises from an interplay between conservation laws and non-equilibrium driving~\cite{lei2019hydrodynamics}, a mechanism we detail below. Beyond these examples, hyperuniformity also arises in the late-time, self-similar regime of phase separation~\cite{de2024hyperuniformity, padhan2025suppression, padhan2025hyperuniformity, wilken2023spatial, zheng2024universal, salvalaglio2020hyperuniform} and in systems with long-range or hydrodynamic interaction~\cite{oppenheimer2022hyperuniformity, huang2021circular,levesque2000charge, ganguly2020ground,yashunsky2024topological, backofen2024nonequilibrium, nizam2021dynamic, thambi2025clustering}. In addition, hyperuniform states can emerge at critical points of absorbing phase transitions~\cite{hexner2015hyperuniformity,mitra2021hyperuniformity,anand2025emergent,chen2024emergent,wilken2021random,wilken2020hyperuniform,weijs2015emergent} or in other fine-tuned settings~\cite{boltz2024hyperuniformity,Castillo_2019, hexner2017enhanced}. Beyond fluids, hyperuniformity also emerges in systems at jamming~\cite{shang2025jamming, atkinson2016critical, wang2025hyperuniform, wilken2023dynamical, wilken2021random, hexner2018two, ozawa2017exploring, ikeda2015thermal, maher2023hyperuniformity, dam2025hyperuniformity} and in a diverse range of athermal systems in Nature~\cite{zhu2023hyperuniformity, philcox2023disordered, ezoe2025weighted}, often manifesting as the solution to an underlying optimization problem~\cite{dong2023hyperuniform, PhysRevE.89.022721, Ge_2023, PhysRevLett.133.028401, tang2024tunable, mayer2015well, emery2024complex}.

The significance of hyperuniformity extends well beyond structural order. The long-range correlations it induces profoundly reshape classical problems in statistical mechanics. They can invalidate the Mermin–Wagner theorem~\cite{PhysRevLett.131.047101,enhancing2024Maire, kuroda2024long,ikeda2024harmonic, keta2024long}, accelerate crystallization~\cite{mkhonta2024liquid} alter nucleation processes~\cite{Lei_Ni_2023, di2016self}, suppress capillary waves~\cite{maire2025hyperuniforminterfacesnonequilibriumphase} and reduce the upper critical dimension of $O(N)$ models~\cite{gao2025liquidgascriticalityhyperuniformfluids,ikeda2023correlated}. On the applied side, hyperuniform materials exhibit remarkable optical and acoustic properties~\cite{man2013isotropic,florescu2009designer,diego2025hypersonic, milovsevic2019hyperuniform,barsukova2025stealthy,aubry2020experimental,gkantzounis2017hyperuniform, hong2024topological, dale2022hyperuniform}, enhanced transport properties when used as disordered media~\cite{zhang2016transport, shi2025three, Liang_Wang_Song_2024}, and unconventional mechanical characteristics~\cite{xu2017microstructure, torquato2022extraordinary}. They have further been investigated as platforms for selective Raman sensors~\cite{khairunnisa2025hyperuniform}, efficient catalysts~\cite{holm2020nanoscale}, and high-efficiency photovoltaic devices~\cite{tavakoli2022over}.

Despite these theoretical and technological advances, a fundamental question remains: under what conditions can hyperuniformity emerge spontaneously, without fine-tuning or imposed long-range interactions? In this article, we address this question with two complementary aims. First, we clarify the deep connections between hyperuniformity, conservation laws, self-organized criticality, and non-equilibrium dynamics. Second, we exploit these insights to construct and characterize new classes of systems that naturally evolve toward hyperuniform states. 

To this end, we describe in Sec.~\ref{sec: 1} how conservation laws together with non-equilibrium driving give rise to the majority of hyperuniform states observed in driven and active matter; this discussion recalls the mechanism described in Ref.~\onlinecite{lei2019hydrodynamics} and emphasizes its link to self-organized criticality while being written for a non-specialist audience. Building on this framework, we introduce four novel systems exhibiting hyperuniformity. In Sec.~\ref{sec: 2} we highlight the role of non-equilibrium dynamics in center of mass conserving systems and use this insight to explore models with multiple conserved quantities such as quadrupole or octupole moments. This enables the dynamical realization of arbitrarily strong hyperuniform scaling in fluids. Finally, Sec.~\ref{sec: 3} demonstrates that hyperuniform states in systems lacking conservation laws are often fragile to the introduction of non-linearities, such as non-linear damping, underscoring the crucial role of conservation laws in stabilizing hyperuniformity.

\section{Disordered hyperuniformity in non-equilibrium systems}\label{sec: 1}
\subsection{Definitions}

Consider the empirical density field $\tilde\rho(\bm r)$ of a $d$-dimensional system of $N$ particles:
\begin{equation}
    \tilde \rho(\bm r)=\sum_{i=1}^N\delta(\bm r- \bm r_i),
\end{equation}
where $\bm r_i$ denotes the position of particle $i$. The spectrum of this density field is called the structure factor $S(\bm k)$:
\begin{equation}
    \label{eq: structure factor}
    S(\bm k)=\dfrac{1}{N}\left|\sum_{i=1}^N e^{i\bm k\cdot \bm r}\right|^2,
\end{equation}
which quantifies the intensity of density fluctuations at wavevector $\bm k$. A system is said to be hyperuniform if $S(\bm k)\to 0$ as $\bm k \to 0$, indicating the suppression of density fluctuations on large length scales~\cite{torquato2018hyperuniform}. Since we will focus on isotropic systems, we shall typically consider the angular average of $S(\bm k)$, denoted by a slight abuse of notation $S(k)$.

In equilibrium fluids with short-ranged interactions, the long-wavelength structure factor follows the Ornstein-Zernike form~\cite{hansen2013theory}:
\begin{equation}
    \dfrac{S( k)}{\rho_0 k_B T} = \dfrac{\kappa_T}{1 + (\xi k)^2},
\end{equation}
where $\rho_0$ is the mean density, $k_{\rm B}$ the Boltzmann constant, $T$ the temperature, $\kappa_T$ the isothermal compressibility, and $\xi$ a correlation length. This relation implies that hyperuniformity can arise only in the limit of zero compressibility at finite temperature. Otherwise, $S(k)$ remains finite and approaches a constant value for $k \ll \xi$, proportional to the variance of particle number within a given large region~\cite{mackay2024countoscope}.

These constraints imposed by equilibrium render the emergence of hyperuniform states exceedingly restrictive; we therefore turn to non-equilibrium systems, which can self-organize into hyperuniform states, as we now show.

\subsection{Typical non-equilibrium mechanism leading to the \texorpdfstring{$\bm S\bm(\bm k\bm)\bm\sim \bm k^{\bm 2}$}{S(k) ~ k\^2} scaling}\label{sec: recap}

We recall the model introduced by Lei and Ni in Ref.~\onlinecite{lei2019hydrodynamics} to illustrate a very general mechanism leading to hyperuniformity. Consider $N$ hard-disks of diameter $\sigma$ and mass $m$ confined in a periodic box of volume $L^d$. Upon collision, two particles $i$ and $j$ undergo an energy-injecting collision:
\begin{equation}
    \dfrac{1}{2}m \bm v_i'^2 + \dfrac{1}{2}m \bm v_j'^2 = \dfrac{1}{2}m \bm v_i^2 + \dfrac{1}{2}m \bm v_{j}^2+\Delta E,
\end{equation}
where $\Delta E$ is the constant energy gained per collision. Together with momentum conservation, this uniquely determines the post-collisional velocities $\bm v'$ from the pre-collisional ones $\bm v$. Between these active collisions, each particle undergoes dissipative free flight:
\begin{equation}
    \label{eq: dissipative friction}
    \bm v_i(t)=\bm v_i(t=0) e^{-\gamma t},
\end{equation}
with $\gamma$ a viscous drag. Thus, particles gain energy upon colliding but gradually lose it during free flight.

This dynamics can be described by a Boltzmann equation and successive integration of its centered velocity moments, combined with a Chapman–Enskog expansion, yields closed evolution equations for the mesoscopic density field $\rho(\bm r)$, velocity field $\bm u(\bm r)$, and temperature field  $T(\bm r)$ defined as the kinetic energy per degree of freedom~\cite{garzo2018enskog}. A simpler phenomenological approach, however, is to modify the equilibrium fluctuating Navier–Stokes equations to incorporate energy injection at collisions and dissipation during free flight. This leads to~\cite{maire2025dynamical}:
\begin{equation}
    \label{eq: hydro field}
    \begin{split}
        \partial_t\rho + \bm u \cdot \bm \nabla \rho =& -\rho \bm \nabla \cdot \bm u,\\
        \partial_t\bm u + \bm u \cdot \bm \nabla \bm u =& -\gamma \bm u + \rho^{-1} \bm \nabla \cdot \left(\bm \Pi+\bm \Pi^{\rm{rand}}\right),\\
        \partial_t T + \bm u \cdot \bm \nabla T =& -m\rho^{-1}\left[\bm \nabla \cdot  \bm J- \bm \Pi : \bm \nabla \bm u\right] +\\& \Delta \dot{T} + \Delta \dot{T}^{\rm rand}.
    \end{split}
\end{equation}
The first equation expresses advection and dilatation of the density by the velocity field. The second describes the velocity dynamics, damped by $\gamma$ and driven by a stress tensor $\bm\Pi=\bm \Pi^{\rm{diss}}+\bm \Pi^{\rm{rev}}$. The dissipative part includes the viscosities $\bm \Pi^{\rm diss}=\mathcal{O}(\bm \nabla \bm u)$, while the reversible part $\bm \Pi^{\rm rev}=-p\bm 1+\mathcal{O}(\bm \nabla^2\rho)$ includes the thermodynamic pressure $p$. Due to the discrete nature of collisions, the mesoscopic velocity field also experiences a Gaussian random stress $\bm \Pi^{\rm rand}$, with zero mean and correlations~\cite{landau2013statistical}:
\begin{equation}
    \begin{split}
    \langle\Pi^{\rm rand}_{ij}(\bm r, t)\Pi^{\rm rand}_{kl}(\bm r', t')\rangle&=2T\left[\eta\left(\delta_{ik}\delta_{jl}+\delta_{il}\delta_{jk}\right)\right.+\\&\left.\mu\delta_{ij}\delta_{kl}\right]\delta(\bm r - \bm r')\delta(t-t').
    \end{split}
\end{equation}
At equilibrium, the coefficients $\eta$ and $\mu$ would be \textit{related} to shear and bulk viscosities via the fluctuation–dissipation theorem. Out of equilibrium, such a relation is not expected, though the tensorial structure remains unchanged, since $\bm\Pi$ must be symmetric in typical fluids~\cite{fruchart2023odd}. Importantly, because this noise is a stress, it conserves momentum locally, as dictated by its collisional origin. However, the presence of the global damping term $-\gamma \bm u$ prevents conservation of the total momentum field. This feature will prove to be central to the emergence of hyperuniformity. The final equation in Eqs.~\eqref{eq: hydro field} governs the temperature field, which is advected by $\bm u$, driven by a heat current $\bm J$, the stress tensor and by the non-equilibrium term $\Delta \dot T$ accounting for temperature change at each point in space due to damping and active collisions. Unlike the velocity field, the noise $\Delta \dot T^{\rm rand}$ does not need to appear as a current, as collisions do not conserve energy.

Since neither velocity nor temperature are conserved, they can be adiabatically eliminated when focusing on the long-wavelength behavior~\cite{brito2013hydrodynamic}. We fix the temperature at its steady value $T_0$, defined by $\Delta \dot T(\rho_0, T_0)\equiv0$. For clarity and reasons that will become clear later, we first retain the velocity field but assume its gradients remain small due to damping. With these approximations, we obtain the equations introduced in Ref.~\onlinecite{lei2019hydrodynamics}:
\begin{equation}
    \label{eq: hydro field 2}
    \begin{split}
        \partial_t\rho =& - \bm \nabla \cdot ( \rho \bm u),\\
        \partial_t\bm u=& -\gamma \bm u - \rho^{-1} \bm \nabla p  +\rho^{-1}\bm \nabla \cdot \bm \Pi^{\rm{rand}}.\\
    \end{split}
\end{equation}
The system is now modeled as being exactly at $T=T_0$. In this simplified framework, the stochastic noise becomes the sole source of energy injection, as all other terms are conservative or dissipative. This contrasts with the full description provided by Eqs.~\eqref{eq: hydro field}, where energy is also deterministically supplied by the term $\Delta \dot T$. It is important to understand that the noise $\bm \Pi^{\rm rand}$ arises from the discrete nature of collisions and not from their non-equilibrium or non-conservative character. It \textit{persists} at equilibrium, although adiabatic elimination is then not feasible and detailed balance ensures that, on average, it neither injects nor removes energy from the system~\cite{sekimoto2010stochastic}.

Eqs.~\eqref{eq: hydro field 2} highlight an important scale separation: energy is injected locally by the noise at the collisional scale but dissipated uniformly across all scales. Indeed, the noise term $\bm \nabla \cdot \bm \Pi^{\rm rand}$ vanishes at large scales, as its Fourier correlations scale as $\bm k^2$, while damping acts equally at all $\bm k$. This produces a depletion of energy at large wavelengths. Indeed, linearizing around $\rho=\rho_0$ and $\bm u=0$ yields:
\begin{equation}
   \dfrac{1}{2}\rho_0 \langle u_\parallel(\bm k) u_\parallel(-\bm k)\rangle=\dfrac{1}{2}\dfrac{\left.\partial_\rho p\right|_{\rho_0, T_0}}{\rho_0}S(\bm k)=\dfrac{T_0(2\eta + \mu)\bm k^2}{2\rho_0\gamma},
\end{equation}
where $u_\parallel$ denotes the component of $\bm u$ parallel to $\bm k$, and the structure factor is computed as $S(\bm k)=\langle \rho(\bm k)\rho(-\bm k)\rangle$. As expected, the energy spectrum vanishes at small $\bm k$ with: $\rho_0 \langle u_\parallel(\bm k) u_\parallel(-\bm k)\rangle/2\sim\bm k^2$. Since density fluctuations are excited only through this velocity field, long-wavelength modes are suppressed, yielding a hyperuniform state with $S(k)\sim k^2$, provided the average density remains homogeneous. This is the result of Ref.~\onlinecite{lei2019hydrodynamics}.

Before showing the generality of this mechanism leading to hyperuniformity, we make a few remarks.

The argument above holds only away from criticality, where $\left.\partial_\rho p\right|_{\rho_0, T_0}=0$. At the critical point in $d\leq 2$, critical density fluctuations scale as $k^{-2}$, which exactly cancels the hyperuniform $k^2$ suppression, yielding instead a flat structure factor \cite{gao2025liquidgascriticalityhyperuniformfluids}.

If collisions did not conserve momentum, the noise in the velocity field could not be written as a stress but would instead contain a noise which does not conserve the velocity field:
\begin{equation}
    \label{eq: hydro field non reciprocal}
    \begin{split}
        \partial_t\rho =& - \bm \nabla \cdot ( \rho \bm u),\\
        \partial_t\bm u=& -\gamma \bm u - \rho^{-1} \bm \nabla p  +\rho^{-1}\bm \nabla \cdot \bm \Pi^{\rm{rand}}+\bm\zeta^{\rm rand}.\\
    \end{split}
\end{equation}
At large scales, $\bm \zeta^{\rm rand}$ acts as white noise, $\langle \zeta_i^{\rm rand}(\bm r, t)\zeta_j^{\rm rand}(\bm r, t)\rangle=2D\delta_{ij}\delta(\bm r - \bm r')\delta(t-t')$ which immediately destroys hyperuniformity, leading to $S(k)\sim D + \alpha k^2$. Thus, although the velocity field itself is not conserved, it is crucial that the noise exciting it conserve momentum. 
Generically, any type of noise that does not conserve the momentum will destroy hyperuniformity~\cite{enhancing2024Maire, lei2019hydrodynamics,hexner2017noise,kuroda2024long}, making this mechanism difficult to realize experimentally~\cite{enhancing2024Maire}.

Since $\bm u$ is not a conserved field, it can be adiabatically eliminated. Setting $\partial_t \bm u=0$ in Eq.~\eqref{eq: hydro field 2} yields a single density equation~\cite{lei2019hydrodynamics}:
\begin{equation}
    \label{eq: linearized laplacian noise}
    \gamma\partial_t\rho = \bm\nabla^2p +\bm\nabla^2\zeta,
\end{equation}
with $\zeta$ a Gaussian noise of zero mean and variance $\langle \zeta(\bm r,t)\zeta(\bm r',t') \rangle = 2T_0(2\eta + \mu)\delta(\bm r-\bm r')\delta(t-t')$. This is a fluctuating diffusion equation, distinguished from its equilibrium counterpart by the appearance of a Laplacian noise rather than a divergence noise~\cite{hexner2017noise}. We note that Eq.~\eqref{eq: linearized laplacian noise} strictly conserves the center of mass of the system~\cite{hexner2017noise}, unlike the underdamped hydrodynamics of Eq.~\eqref{eq: hydro field 2}, where it is only bounded, with its velocity exponentially decaying. Compared to the underdamped hydrodynamic equation, the overdamped description no longer allows one to interpret hyperuniformity as the depletion of kinetic energy at large length scales. Nevertheless, hyperuniformity can still be understood as the result of weak driving on large scales. Indeed, taking the Fourier transform of Eq.~\eqref{eq: linearized laplacian noise} and linearizing gives:
\begin{equation}
    \label{eq: k dependent temperature}
    \gamma\partial_t\rho = -\left.\partial_\rho p\right|_{\rho_0, T_0}\bm k^2 +\sqrt{2(2\eta + \mu)T_0\bm k^4}\xi.
\end{equation}
This equation can formally be viewed as satisfying a fluctuation-dissipation relation if the thermal bath is assigned an effective $\bm k-$dependent temperature~\cite{enhancing2024Maire,gao2025liquidgascriticalityhyperuniformfluids} proportional to $T_0 \bm{k}^2$ and vanishing on large length scales. Since the density field is effectively driven at small $\bm k$ by a bath at $T=0$, hyperuniformity emerges.

If we retain the full reversible stress and preserve non-linear terms, we arrive at a generalized equation for the hyperuniform state:
\begin{equation}
    \label{eq: model B HU}
    \gamma\partial_t\rho = -\bm\nabla\cdot\left(\rho \bm \nabla\dfrac{\delta F}{\delta \rho}-\bm \nabla \cdot \bm\Pi^{\rm neq}\right) +\bm\nabla^2\zeta,
\end{equation}
where $F$ is a variational Landau free energy and $\bm{\Pi}^{\rm neq}$ is a non-equilibrium stress that cannot be derived from $F$, as in active model B+~\cite{cates2024active}. Eq.~\eqref{eq: model B HU} can therefore be regarded as \textit{hyperuniform model B}, a natural starting point for the theoretical analysis of hyperuniform fluids arising from the aforementioned mechanism~\cite{maire2025hyperuniforminterfacesnonequilibriumphase,gao2025liquidgascriticalityhyperuniformfluids}.

In this context, hyperuniformity originates from the breakdown of the usual fluctuation–dissipation theorem~\cite{marconi2008fluctuation}. In the hydrodynamic description, fluctuations originate at the collisional scale while dissipation acts uniformly at all scales through global damping. The same mismatch is present in hyperuniform model B via the effective $\bm k$-dependent temperature, due to the Laplacian noise. At equilibrium, such a separation of scale between the noise and the damping is forbidden. Either $\gamma = 0$, in which case both dissipation (viscosities) and noise correlations scale as $k^2$ and are restricted to short wavelengths; or $\gamma \neq 0$, in which case consistency requires the presence of an additional white noise acting uniformly across all scales, as in colloidal suspensions. In both scenarios, the steady state shows no large-scale depletion of energy.\\
The emergence of hyperuniformity from the violation of the fluctuation-dissipation theorem driven by the interplay between conservation laws and non-equilibrium dynamics, mirrors in a reversed manner, the mechanism by which long-range correlations arise in certain models of self-organized criticality~\cite{bak1988self}. In many such systems, the stochastic driving terms typically violate the conservation laws obeyed by the deterministic dynamics~\cite{bonachela2009self, grinstein1990conservation, garrido1990long, van1999randomly, plati2021long, simha2002hydrodynamic, kundu2016long, spohn1983long, dorfman1994generic, grinstein1991generic, bak1992self, giusfredi2024localization}. A canonical example is provided by sandpile models, where the order parameter is conserved under the deterministic rules but not by stochastic grain addition~\cite{grinstein1990conservation}. Such mismatches generically induce strong fluctuations, producing a divergence of the structure factor scaling as $|\bm k|^{-2}$. In contrast, the mechanism of hyperuniformity described here can be interpreted as a modified form of self-organized criticality, one that suppresses (rather than amplifies) large-scale fluctuations. This suppression arises from the action of a momentum conserving noise on a non-conserved average term, leading to a depletion of fluctuations at long wavelengths.
\\
The mechanism described above is not specific to the model we used for illustration. The hydrodynamic approach demonstrates that these findings apply broadly to any system with a similar separation of scales between noise and dissipation. In the next section, we show how this understanding allows one to readily construct hyperuniform models.

\subsection{Numerical simulations of new non-equilibrium hyperuniform fluids}

\begin{figure*}
    \centering
    \includegraphics[width=0.999\linewidth]{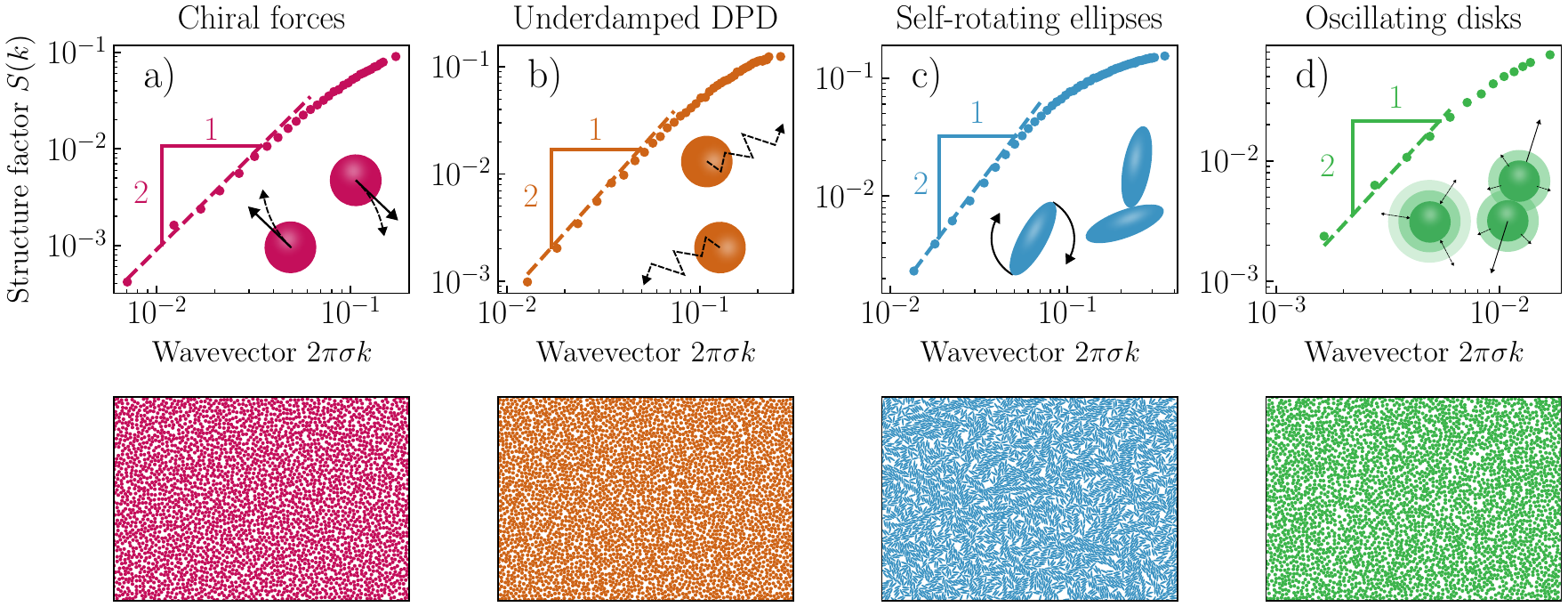}
    \caption{Simulations of four models in 2D periodic square boxes displaying hyperuniformity. Top panels: radially averaged structure factor $S(k)$. Bottom panels: corresponding snapshots of the systems. Details of the models and simulation parameters are given in Appendix~\ref{app: chiral active}, \ref{app: dpd}, \ref{app: rods} and \ref{app: oscillating radius} for a), b), c) and d) respectively.}
    \label{fig: 4systems}
\end{figure*}

Based on the hydrodynamic description of Sec.~\ref{sec: recap}, three ingredients are required to obtain hyperuniformity: (i) global damping, (ii) the effective noise of the mesoscopic description must arise from momentum conserving or center of mass conserving interactions, (iii) the system must be able to sustain an active state in the presence of (i) without violating (ii). The first condition is easily satisfied, either through overdamped dynamics (where momentum conservation translates into center of mass conservation) or underdamped dynamics with viscous drag. The second and third conditions imply that activity and \textit{diffusion} must arise from particle pairs, as unbounded self-propelled particles or external momentum non-conserving noise cannot yield hyperuniformity. This restriction is the most stringent one. Based on these principles, we propose four new systems, relevant to active matter or theoretical exploration, that display hyperuniformity. For each, we outline only the essential features; detailed simulation methods and models used are given in Appendix~\ref{app: details}.

\emph{Noiseless chiral active particles --} Particles of diameter $\sigma$ interact through chiral, non-conservative forces acting tangentially but reciprocally between neighbors: $\bm F^\perp_{ij}=-\bm F^\perp_{ji}=|\bm F^\perp_{ij}|\hat{\bm r}_{ij}\times \bm z$, with $\hat{\bm r}_{ij}$ the unit vector along $\bm r_{ij}$ and $\bm z$ the unit vector perpendicular to the $xy$ plane. These provide the momentum-conserving energy injection, while viscous damping $-\gamma \bm v$ dissipates energy. Together, these suffice to produce hyperuniformity as shown in Fig.~\ref{fig: 4systems}a) where the structure factor scales as $\bm k^2$ on large length scales. This model, motivated by living active matter~\cite{tan2022odd} and colloidal spinners~\cite{massana2021arrested}, is directly relevant to active systems and has recently been explored in the contexts of chirality-induced phase separation~\cite{caprini2025Bubble} and edge currents~\cite{caporusso2024phase}. We note that the introduction of an external white Gaussian noise destroys hyperuniformity beyond a given length scale, as in Eq.~\ref{eq: hydro field non reciprocal}.

\emph{Underdamped dissipative particle dynamics --} Noise does not always destroy hyperuniformity. When it respects momentum (or center-of-mass) conservation, it can in fact generate hyperuniform states. A simple computational example is provided by underdamped dissipative particle dynamics:
\begin{align}
m\dot{\bm v}_i=&-\gamma m\bm v_i- \sum_{j\neq i}\Big[\partial_{\bm r_i}U(\bm r_{ij})+\Gamma w(r_{ij}) \left( \hat{\bm{r}}_{ij} \cdot \bm{v}_{ij} \right) \hat{\bm{r}}_{ij}\nonumber\\&\qquad + \sqrt{2\Gamma T w( r_{ij})}\theta_{ij} \hat{\bm r}_{ij}\Big],\label{eq: simple enough}
\end{align}
where $U$ is a short-ranged repulsive potential, $w(r)$ is a short-ranged weight function, $\Gamma$ is a viscosity-like damping, $\bm v_{ij}=\bm v_i-\bm v_j$, and $\theta_{ij}$ is a Gaussian noise with:
\begin{equation}
    \langle \theta_{ij}(t) \rangle = 0, \quad \langle \theta_{ij}(t)\theta_{kl}(t') \rangle = (\delta_{ik}\delta_{jl} + \delta_{il}\delta_{jk})\delta(t-t').
\end{equation}
The noise and its associated damping act pairwise and oppositely, thereby conserving locally the momentum. In the absence of global damping $\gamma$, they are commonly used to model coarse-grained systems while preserving their hydrodynamic modes~\cite{groot1997dissipative}. With global damping however, this dynamics leads to hyperuniformity with $S(k)\sim k^2$, as given in Fig.~\ref{fig: 4systems}b). In the overdamped limit $m\to0$, Eq.~\eqref{eq: simple enough} represents perhaps the simplest continuous-time realization of hyperuniformity, directly connected to random organization models~\cite{anand2025emergent}. Its explicit noise structure also makes it amenable to a Dean’s derivation of the fluctuating hydrodynamic equations of these hyperuniform fluids (Eq.~\eqref{eq: hydro field 2}), as done for a similar model in Ref.~\onlinecite{anand2025emergent}. This derivation and a discussion concerning the limitation of Dean's method are given in Appendix~\ref{app: dean}. By incorporating attractive forces into Eq.~\eqref{eq: simple enough}, one can readily study phase separation, criticality, interfaces, and nucleation in hyperuniform fluids.

\emph{Self-rotating active ellipses --} Our third system is similar to the one of Ref.~\onlinecite{liu2023local, liu2025hyperuniform} and consists of overdamped ellipses driven by a constant self-rotation $\omega_0$, causing every particle to rotate independently. Interparticle interactions are purely Hamiltonian and steric, thereby conserving center of mass. Energy is injected through the active torque, converted into translational motion via collisions between ellipses, and dissipated through the overdamped dynamics of the system. This system also exhibits hyperuniformity (Fig.~\ref{fig: 4systems}c), provided no external noise is introduced. The model demonstrates that overdamped motion is fully compatible with hyperuniformity, and that interparticle interactions need not be intrinsically out of equilibrium: it suffices that activity is sustained through local momentum-conserving collisions in combination with global damping for hyperuniformity to emerge.  At high densities, such a system forms an active nematic, where defects are continuously generated by the interplay between active torque and steric interactions. This state of matter is generally accompanied by the inverse of hyperuniformity: giant number fluctuations~\cite{sousa2025selfpropulsiveactivenematics, PhysRevLett.113.038302, ramaswamy2003active}.

\emph{Oscillating-radius particles --} Finally, we consider underdamped hard-disks whose radius oscillates in time, providing the energy injection. This model, closely related to one presented in Ref.~\onlinecite{ikeda2024harmonic}, is conceptually similar to the collisional energy-injection model discussed earlier, and likewise yields hyperuniformity (Fig.~\ref{fig: 4systems}d). It falls into the broader class of pulsating active matter, which has attracted increasing attention recently~\cite{zhang2023pulsating, banerjee2024hydrodynamics,tjhung2017discontinuous}.

Many previously studied non-equilibrium systems displaying hyperuniformity can be understood through the same mechanism~\cite{hexner2017noise,PhysRevE.99.022115, mukherjee2024anomalous,hazra2025hyperuniformity, dandekar2020exact, lei2024non, kuroda2023microscopic,kuroda2025singulardensitycorrelationschiral, keta2024long, lei2019hydrodynamics, lei2019nonequilibrium,lei2023random, maire2025dynamical,li2025fluidization,gao2025liquidgascriticalityhyperuniformfluids}. For instance, Keta and Henkes showed that biological cells on a substrate with damping and pairwise reciprocal interactions naturally arrange into hyperuniform patterns~\cite{keta2024long}. Similarly, chiral particles generically form hyperuniform states in the absence of external noise, owing to pairwise momentum-conserving active forces and the fact that external driving is not required to sustain activity~\cite{gao2025liquidgascriticalityhyperuniformfluids}. Even chiral self-propelled particles can display hyperuniformity: a particle running on a deterministic circular trajectory of radius $R$ can be mapped to a fictitious particle of radius $R$ whose center undergoes momentum-conserving collisions~\cite{lei2019nonequilibrium, kuroda2023microscopic}. Since circular self-propulsion confines the center of mass within a bounded region, while diffusion originates from momentum-conserving interparticle collisions, this type of self-propulsion does not hinder hyperuniformity. This stands in contrast to more classical self-propulsion mechanisms, or to the limit $R\to\infty$~\cite{kuroda2023anomalous}.

\section{Generic emergence of arbitrary strong hyperuniformity in non-equilibrium multipole conserving systems}\label{sec: 2}
\subsection{Non-equilibrium center of mass conserving systems}\label{sec: levine}

We have seen that, on large length scales, hyperuniformity can emerge from the interplay between non-equilibrium dynamics and a center of mass conserving evolution equation, exemplified by the hyperuniform model B (Eq.~\eqref{eq: model B HU}). In the linear regime, this reduces to a diffusion equation with an unconventional noise term:
\begin{equation}
    \label{eq: model B HU linear}
    \partial_t\rho = \kappa\bm \nabla^2\rho +\sqrt{2D}\bm \nabla^2\zeta,
\end{equation}
where $\zeta$ is a Gaussian white noise of zero mean and unit variance. This equation was originally derived for a lattice process in which two particles occupying the same site hop to opposite neighboring sites, thereby conserving the center of mass~\cite{hexner2017noise}. 

Eq.~\eqref{eq: model B HU linear} can also be justified on the basis of universality. It is the simplest equation for a single scalar that conserves the center of mass $\bm{\mathcal Q}_1(t)=\int \bm r\rho(\bm r, t)d\bm r$. Indeed, we obtain~\cite{hexner2017noise}:
\begin{equation}
    \label{eq: proof conservation}
    \begin{split}
        \dot{\bm{\mathcal Q}}_1(t)=&\int \bm r \left(\kappa\bm \nabla^2\rho +\sqrt{2D}\bm \nabla^2\zeta\right)d\bm r=0+\mathcal{B.T},
    \end{split}
\end{equation}
where $\mathcal{B.T}$ denotes boundary terms arising from integration by parts, which vanish in periodic or infinite domains. In finite systems with walls, however, the center of mass is no longer conserved, just as momentum is not conserved in molecular dynamics simulations of systems confined within a reflective box. A divergence noise $\bm\nabla\cdot\bm\eta$ would fail to conserve the center of mass in the bulk, and is thus excluded from Eq.~\eqref{eq: model B HU linear}.

This derivation, however, must tacitly assume a non-equilibrium setting. An equilibrium system with center of mass conservation is not hyperuniform, because its statistical properties are governed by the Hamiltonian and Gibbs measure, which are insensitive to dynamical constraints~\footnote{One could argue that these systems might be described by a generalized Gibbs ensemble or by a generalized microcanonical ensemble. For instance, if $\mathcal{\bm Q}_n$ is conserved, the microcanonical entropy would take the form~\cite{landau2013statistical}: $S(E, N, V, \mathcal{\bm Q}_n)=\log\left(\Omega(E, N, V, \mathcal{\bm Q}_n)\right) = \log\left(\sum_{\rm state} \delta(E - E^{\rm state}) \delta(\mathcal{\bm Q}_n - \mathcal{\bm Q}_n^{\rm state})\right)$. However, the additional constraints are likely subextensive and thus vanish in the thermodynamic limit~\cite{niiyama2009effect}. For example, in standard molecular dynamics simulations of systems with short-range interactions, the center-of-mass velocity, linear momentum, and angular momentum are naturally conserved. However, even when these quantities are not explicitly included in the microcanonical entropy, accurate predictions of statistical observables can still be obtained~\cite{calvo2002sampling}. Similarly, the conservation of these quantities depend on the boundary conditions, which are known to often be irrelevant in the thermodynamics limit. The question is more delicate for systems with long-range interactions~\cite{laliena1999effect}.}.

\begin{figure}
    \centering
    \includegraphics[width=0.999\linewidth]{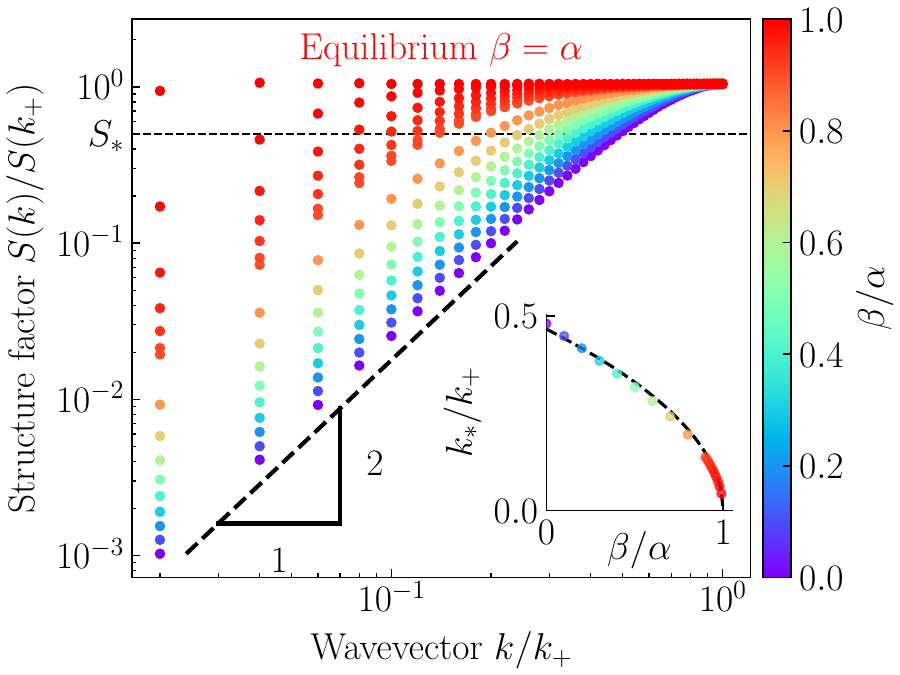}
    \caption{Relation between detailed balance breaking and hyperuniformity. The main panel shows the structure factor $S(k)$ for a 1D lattice model with center of mass conservation, for different values of the jump-rate ratio $\beta/\alpha$. The case $\beta/\alpha=1$ corresponds to equilibrium and yields a flat, non-hyperuniform spectrum. Inset: largest wavevector $k=k_*$ such that $S(k_*)/S(k_+)<S_*=0.5$, with $k_+$ the maximal lattice wavevector, as a function of the non-equilibrium control parameter $\beta/\alpha$. The dashed curve is the prediction $k_*\propto \sqrt{1-\beta/\alpha}$}
    \label{fig: DB broken}
\end{figure}

Assuming ergodicity in equilibrium center of mass conserving systems (which is far from guaranteed~\cite{babbar2025classical}), standard statistical-mechanical constraints must apply, including the fluctuation–dissipation theorem. Since Laplacian noise is the simplest choice compatible with center of mass conservation, the fluctuation-dissipation theorem requires that the deterministic linear term be related to the square of the noise correlation. This yields a bi-Laplacian dynamics:
\begin{equation}
    \label{eq: model bilaplacian linear}
    \partial_t\rho = -\kappa'\bm \nabla^4\rho +\sqrt{2D}\bm \nabla^2\zeta,
\end{equation}
which produces a flat structure factor, as expected in equilibrium.

From this perspective, it is not the Laplacian noise in hyperuniform model B (Eq.~\eqref{eq: model B HU}) that signals the lack of equilibrium, but rather the diffusive deterministic term itself. In other words, diffusion in a center of mass conserving system is intrinsically non-equilibrium.

Interestingly, the linearized dynamics not only conserve the center of mass but also, on average, the third multipole moment $\bm{\mathcal Q}_3=\int \bm r\otimes\bm r\otimes\bm r\rho(\bm r)d\bm r$, although this symmetry is broken by the noise. This constitutes an emergent, or ``weak'' symmetry of the system, valid only on average, in contrast to ``strong'' symmetries that are exact along every trajectory~\cite{huang2023generalized}. This weak symmetry is rooted in microscopic time-reversal invariance and detailed balance~\cite{guo2022fracton}.

To illustrate this mechanism, one can consider ordinary diffusion. On a 1D lattice, let particles hop to the left with rate $\alpha$, and to the right with rate $\beta$. When $\alpha\neq\beta$, the system is driven out of equilibrium and exhibits a macroscopic mass current. When $\alpha=\beta$, detailed balance holds: the macroscopic current vanishes and the large-scale dynamics reduce to diffusion, which conserves the center of mass on average. At continuum scales, this dynamics is described by:
\begin{equation}
\partial_t\rho = -\partial_x \left[A(\beta-\alpha)\rho - B\partial_x\rho + \sqrt{2D}\eta\right],
\end{equation}
with $A$ and $B$ constants, $D$ dependent on $\alpha$ and $\beta$ and $\eta$ a white Gaussian noise. Detailed balance ($\alpha=\beta$) enforces the weak symmetry associated with diffusive behavior as the advective term vanishes.

The same mechanism applies to center of mass conserving systems. Consider again the 1D lattice process introduced above: when two or more particles occupy a site $x$, a pair may hop to $x-1$ and $x+1$ with rate $\alpha n_x(n_{x}-1)$, where $n_x$ is the occupation number. The reverse process, in which a particle from $x-1$ and a particle from $x+1$ simultaneously hop to site $x$, occurs at a rate $\beta n_{x-1}n_{x+1}$. Detailed balance is recovered~\cite{han2024scaling} when $\alpha=\beta$. We simulate this model at high density via Gillespie algorithm~\cite{Gillespie2007Stochastic}, with $\beta\leq\alpha$ to avoid clustering, and measure the structure factor for various values of $\beta/\alpha$. Results are presented in Fig.~\ref{fig: DB broken}. As expected, the equilibrium case yields a flat spectrum, while any violation of detailed balance produces hyperuniformity. In the high-density and linear regime, we can show (see Appendix~\ref{app: detail lattice}) that the coarse-grained density satisfies:
\begin{equation}
    \label{eq: equation for dipole conservation}
    \partial_t\rho = \partial_x^2\left((1-\beta/\alpha)(A\rho - B\partial_x\rho) - C\partial_x^2\rho  + \sqrt{2D}\eta\right),
\end{equation}
with constants $A, B, C$ and $D$ a function of the rates. The corresponding structure factor is: 
\begin{equation}
    S(k)=\dfrac{Dk^2}{A(1-\beta/\alpha)+Ck^2}.
\end{equation}
Hyperuniformity arises as long as detailed balance is broken.

Defining $k_*$ as the crossover wavevector below which $S(k_*)<S_*$ for some small threshold $S_*$ which would indicate the beginning of the hyperuniform scaling, we find $k_*\propto \sqrt{1-\beta/\alpha}$, which vanishes as $\beta/\alpha\to 1$, consistent with simulations (Fig.~\ref{fig: DB broken}, inset).

Thus, hyperuniformity emerges generically in center of mass conserving systems that break detailed balance. This naturally raises the question: Do systems conserving higher-order multipole moments display analogous behavior? In the following, we demonstrate that this is indeed the case.

\subsection{Non-equilibrium multipole conserving systems}

We now consider conservation of higher-order multipole moments. The $M$-th multipole moment of the density field is defined as:
\begin{equation}
    \label{eq: definition multipole}
    \bm{\mathcal{Q}}_M=\int \underbrace{\bm r\otimes \dots \otimes \bm r}_{M\rm ~times}\rho(\bm r)d\bm r.
\end{equation}
To explore the consequences of such constraints, we construct 1D periodic lattice models whose dynamics locally conserve all multipoles of order \textit{below or equal} to $M$. Each lattice site $x$ hosts $n_x$ particles, and an event corresponds to a local redistribution of particles specified by a kernel $\bm{\mathcal L}\equiv [\delta_{-m'}, \delta_{-m'+1}, \dots, \delta_m]$ where $\delta_j$ denotes the particle number change at a given site. We set $m=m'$ for events spanning an odd number of sites, and $m=m'+1$ for those involving an even number. We define the operator $\mathcal L_x$ which acts locally around site $x$ as:
\begin{equation}
    \begin{split}
    \mathcal L_x\bm n =&\mathcal L_x[n_0,\dots,n_x,\dots, n_N]\\
    =&[n_0, \dots, n_{x-m'}+\delta_{-m'}, \dots,\\&n_x+\delta_0,\dots, n_{x+m}+\delta_{m}, \dots, n_N].
    \end{split}
\end{equation}
The lattice dynamics of Sec.~\ref{sec: levine} can be described by $\bm{\mathcal L}=[1,-2,1]$ which corresponds to two particles leaving a site $x$ and hopping to $x\pm1$.

An event associated to $\mathcal L_x$ occurs at a rate $\mathcal R_{\mathcal L_x }^{\{\alpha\}}$ given by:
\begin{equation}
    \label{eq: rate}
    \mathcal R_{\mathcal L_x }^{\{\alpha\}}= \alpha \prod_{j=-m'}^{m} n_{x+j}^{\underline{\delta_j}},
\end{equation}
where:
\begin{equation}
    n^{\underline{\delta}} \equiv 
    \begin{cases}
        n(n-1)\cdots(n - |\delta| + 1), & \text{if } \delta < 0, \\
        1, & \text{otherwise}.
    \end{cases}
\end{equation}
Thus, the more particles available at departure sites, the higher the event rate.

The reverse event $\bm{\mathcal L}^{-1}=-\bm{\mathcal L}$ occurs with rate $\mathcal R_{\mathcal L_x^{-1}}^{\{\beta\}}$. When $\alpha=\beta$, detailed balance holds and the system is at equilibrium; otherwise, it is driven out of equilibrium.

Conservation of the $M$-th multipole requires:
\begin{equation}
    \label{eq: condition multipole}
    \mathcal Q_M(\mathcal L_x \bm n) = \mathcal Q_M(\bm n)\Rightarrow \sum_{j=-m'}^{m}j^M\delta_j=0.
\end{equation}
This condition is independent of the choice of origin, provided all lower-order multipoles are also conserved. Using Eq.~\eqref{eq: condition multipole}, one can systematically construct dynamics conserving all moments up to order $M$. Some minimal examples are (see Appendix~\ref{app: proof}):
\begin{center}
\begin{tabular}{c|c}
$M$ & $\bm{\mathcal L}$ \\
\hline
0 & $[+1, -1]$ \\
1 & $[+1, -2, +1]$ \\
2 & $[-1, +3, -3, +1]$ \\
3 & $[-1, +4, -6, +4, -1]$ \\
7 & $[-1, +8, -28, +56, -70, +56, -28, +8, -1]$ \\
$K$ & \small $(-1)^{\lfloor  K/2 \rfloor}\Big[\binom{K+1}{0}, \dots,(-1)^{m}\binom{K+1}{m}, \dots, (-1)^{K+1}\binom{K+1}{K+1}\Big]$
\end{tabular}
\end{center}

As $M$ increases, the number of constraints grows linearly, requiring less local and more particle-involving kernels to satisfy conservation. 

\begin{figure*}
    \centering
    \includegraphics[width=0.999\linewidth]{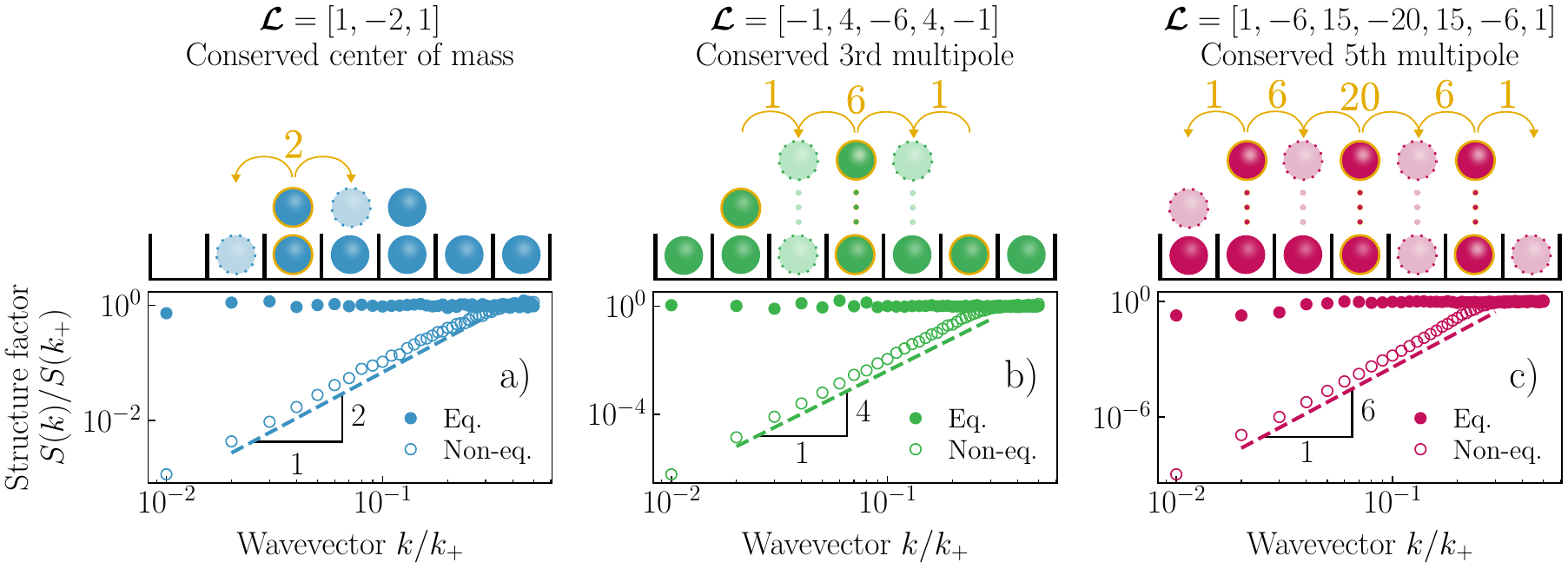}
    \caption{Steady-state structure factor $S(k)$ for one-dimensional lattice models with conservation laws up to different multipole orders. (a) Dipole (center of mass) conservation; (b) conservation up to the third multipole; (c) conservation up to the fifth multipole. For each case we simulate an equilibrium system ($\alpha=\beta$) and a non-equilibrium system ($\beta=0$). The cartoons above each panel depict the underlying non-equilibrium hopping process.}
\label{fig: fractons}
\end{figure*}

We now simulate these lattice models, comparing equilibrium dynamics where $\alpha=\beta$ with strongly non-equilibrium dynamics $\beta=0$. However, all our conclusions will also hold as long as $\alpha > \beta$. We note that the cases $\beta>\alpha$ leads to a diverging accumulations of particles on sites with our choice of kernel. The results of the simulations are presented in Fig.~\ref{fig: fractons}.  In Fig.~\ref{fig: fractons}a), we simulate again the event $\bm{\mathcal L}=[1, -2, 1]$, as a comparative baseline against other events conserving higher order multipole. This specific event only conserves the center of mass and, as we previously found, the non-equilibrium case ($\beta=0$) leads to a structure factor in $S(k)\sim k^2$ while the equilibrium ($\alpha=\beta$) structure factor is flat. For $M=3$, the dynamics conserve particle number, center of mass, quadrupole, and octupole moments. As shown in Fig.~\ref{fig: fractons}b), equilibrium again yields a flat spectrum, whereas the non-equilibrium case produces stronger hyperuniformity with $S(k)\sim k^4$. For $M=5$, additional conservation of the hexadecapole and dotriacontapole moments leads to an even stronger suppression of fluctuations in the non-equilibrium case with $S(k)\sim k^6$ found in simulation as given in Fig.~\ref{fig: fractons}c). In Appendix~\ref{app: even}, we also show that lattice dynamics conserving up to the 2nd and 4th moment, leads respectively to a hyperuniform scaling of $S(k)\sim k^2$ and $S(k)\sim k^4$. 

Therefore, we hypothesize that homogeneous non-equilibrium hyperuniform fluids conserving up to the $M$ multipole are generically hyperuniform with: 
\begin{equation}
    S(k)\sim
    \begin{cases}
        k^{M+1} &\text{ when } M \text{ is odd},\\
        k^M  &\text{ when } M \text{ is even},  
    \end{cases}
\end{equation}
with $M$ the order of the highest multipole conserved. Thus, out of equilibrium, arbitrarily many conservation laws can produce arbitrarily high hyperuniform exponents.

To understand these results, we can use an argument similar to the one presented in Eq.~\eqref{eq: proof conservation} and obtain the most general evolution equation that conserves all moments up to order $M$. In any dimension, we find:
\begin{equation}
    \label{eq: general conservation equation}
    \dot{\bm{\mathcal{Q}}}_{n\leq M}=0+\mathcal{B.T}\Rightarrow\partial_t\rho= \partial_{i_0}\partial_{i_1}\dots\partial_{i_M} \mathcal{J}_{i_0, i_1, \dots, i_M},
\end{equation}
where repeated indices are summed over. $\partial_{i_n}$ is a spatial derivative and $\mathcal J$ is a tensor current related to the mass current $\bm J$ by: $J_i=-\partial_{i_1}\dots\partial_{i_M} \mathcal{J}_{i, i_1, \dots, i_M}$ such that Eq.~\eqref{eq: general conservation equation} can be written as $\partial_t\rho = -\bm \nabla\cdot \bm J$. This class of models known as fractons and typically arising via quasiparticles in quantum systems~\cite{pretko2020fracton} is characterized by stringent dynamical constraints: particles can move only collectively in order to satisfy the conservation laws. In the linear regime, this results in subdiffusion~\cite{gromov2020fracton}. While we focus on 1D examples for clarity, the results we present generalize to higher dimensions under rotational symmetry. However, the fluctuating hydrodynamics of these systems can easily break down in lower dimensions due to an arbitrarily large upper critical dimension, which increases with $M$, the order of the highest conserved multipole moment~\cite{glorioso2022breakdown}.

Decomposing $\mathcal J$ into an average and fluctuating contributions yields:
\begin{equation}
    \partial_t\rho(x, t) = \partial_x^{M+1}\mathcal J^{\rm av}(x, t) + \partial_x^{M+1}\mathcal J^{\rm rand}(x, t),
\end{equation}
the latter having Gaussian white correlations: 
\begin{equation}
    \label{eq: current random}
    \langle \mathcal J^{\rm rand}(x, t)\mathcal J^{\rm rand}(x', t')\rangle=2D\delta(t-t')\delta(x-x'),
\end{equation}
At equilibrium, within the linear regime, the fluctuation-dissipation theorem requires $J^{\rm av}(x, t)\propto \partial_x^{M+1}\rho$. This yields:
\begin{equation}
     \label{eq: equilibrium evolution}
     \partial_t\rho(x, t) = (-1)^M\kappa\partial_x^{2(M+1)}\rho(x, t) + \partial_x^{M+1}\mathcal J^{\rm rand}(x, t),
\end{equation}
with $\kappa$ a constant. This equilibrium system is of course predicted to have a flat structure factor.

Out of equilibrium, owing to the absence of a fluctuation-dissipation theorem, lower-order terms are allowed:
\begin{equation}
    \label{eq: evolution equation fracton}
    \begin{split}
    \partial_t\rho(x, t) =& \partial_x^{M+1}(A\rho +B\partial_x\rho+\dots)+\partial_x^{M+1}\mathcal J^{\rm rand}(x, t),
    \end{split}
\end{equation}
with $A$ and $B$ having well-chosen sign to ensure linear stability. The resulting structure factor is:
\begin{equation}
    \label{eq: main message structure factor}
    \begin{split}
        S(k)&=\dfrac{D k^{2(M+1)}}{\textrm{Re}\left[(ik)^{M+1}\left(A+ikB+\dots\right)\right]}\\
        &=
        \begin{cases}
            \dfrac{D}{|A|} k^{M+1}&\text{ when } M \text{ is odd},\\
            \dfrac{D}{|B|}k^{M}  &\text{ when } M \text{ is even}.  
        \end{cases}
    \end{split}
\end{equation}
This matches our lattice simulations and confirms that hyperuniform scaling exponents are dictated by the highest conserved multipole. A more rigorous approach than simply guessing the equations is to coarse-grain the lattice dynamics. By doing so in Appendix~\ref{app: detail lattice}, we can demonstrate that the resulting continuum description is indeed given by Eq. \eqref{eq: evolution equation fracton}. This method shows that, as expected, all terms of order below $\mathcal{O}(\partial^{M+1})$ vanish identically due to the \textit{microscopic} conservation laws. It also reveals that all non-equilibrium constants which violate detailed balance in the hydrodynamic equations (such as $A$, $B$, and so on), are directly proportional to the difference between the parameters $\alpha$ and $\beta$. Consequently, these non-equilibrium terms vanish in the equilibrium limit $\alpha=\beta$, leaving only a flat structure factor. We note that the interpretation of hyperuniformity generation as an effective thermalization with a bath at a $\bm k$-dependent temperature $\sim D k^{M+1}$ still applies.

As shown above, the emergence of hyperuniformity in non-equilibrium systems with multipole conservation is a generic consequence of the structure of the hydrodynamic equations. For instance, because moment conservation naturally arises in quantum matter, one may ask whether our conclusions remain valid in a quantum setting. From a hydrodynamic perspective, there is no reason to expect a different outcome. Indeed, replacing the classical hopping particles on the lattice by quantum particles (thereby introducing effects such as Bose enhancement or Pauli blocking~\cite{le2000thermal}) does not alter the central conclusion: a breakdown of detailed balance generically produces non-equilibrium terms in the hydrodynamic equations that permit hyperuniform states (see also Appendix.~\ref{app: quantum}).

Another realization of these hydrodynamic equations in continuous time and space can be envisioned using a system with overdamped dynamics. One direct way to enforce the conservation of an arbitrary multipole is to project a conventional overdamped dynamics, at each instant, onto the manifold of configurations with the desired multipoles fixed. Such a projection would need to act locally. This approach is however computationally arduous and conceptually inelegant. A more natural route is to impose the conservation laws via symmetries and Noether’s theorem~\cite{jose1998classical}. Consider a Hamiltonian dynamics for $N$ particles:
\begin{equation}
    \dot{\bm r}_i=\partial_{\bm{p}_i}H,\quad \dot{\bm p}_i=-\partial_{\bm{r}_i}H,
\end{equation}
where $H$ is the Hamiltonian and $\bm p$ is the canonical momentum. Noether’s theorem then implies that conservation of all multipoles of order $n\leq M$:
\begin{align}
    \dot{\bm{\mathcal{Q}}}_{n\leq M}&= \dfrac{d}{dt}\left(\sum_i  \underbrace{\bm r_i\otimes\dots\otimes\bm r_i}_{n \text{ times}}\right) = 0,
\end{align}
arises from the invariance of the Hamiltonian under momentum shifts of the form~\cite{prakash2024classical}:
\begin{align}
    H\left(\bm{r}_i,\bm p_i\right) &= H\!\left(\bm{r}_i ,  \bm p_i  + \sum_{k=0}^{M-1}\bm{\varepsilon}^{(k)}_{\nu_1\cdots\nu_k}  r_i^{\nu_1}\cdots r_i^{\nu_k}\right),
\end{align}
where each $\bm{\varepsilon}^{(k)}$ is a symmetric vectorial tensor of rank $k$. For example, exact conservation of the center of mass corresponds to the constant shift: $\bm p_i\mapsto\bm p_i+\bm\varepsilon$. A representative Hamiltonian invariant under these transformations is:
\begin{equation}
    \label{eq: hamiltonian fracton}
    H = \dfrac{1}{2}\sum_{i\neq j} m(\bm p_i-\bm p_j)^2K(|\bm r_i-\bm r_j|) + V(|\bm r_i-\bm r_j|),
\end{equation}
where $K(r)$ decays rapidly for $r\neq 0$ ensuring that the momentum sector remains local in real space. Such a system, provided it is ergodic, exhibits equilibrium-like static properties. Moreover, as $\bm p$ is conserved, the generalized momentum -- which is no longer simply proportional to velocity $\dot {\bm r}$ -- constitutes an additional slow field alongside the density on hydrodynamical scales~\cite{osborne2022infinite}. To realize a genuinely non-equilibrium continuum model of the type in Eq.~\eqref{eq: evolution equation fracton} one may damp the generalized momentum in a manner that preserves the imposed conservation laws (although we may note that, in Sec.~\ref{sec: recap}, a non-conserved but bounded center of mass already sufficed). Simultaneously, the system must be driven, for example by a non-conservative chiral or non-reciprocal force, to sustain an active steady state. We do not pursue this construction here and leave this promising direction for future work.

Beyond these specific realizations of hyperuniformity, an analysis of the multipole moments provide a broader framework to understand hyperuniformity in other systems. Following Ref.~\onlinecite{gabrielli2008tilings}, we assume the Taylor expansion of the structure factor converges, and obtain:
\begin{equation}
    \begin{split}
    S(\bm k)=&S_0+\left.\dfrac{1}{2}k_ik_j\dfrac{\partial^2S(\bm k)}{\partial{ k_i}\partial{k_j}}\right|_{\bm k = 0}+\\&\left.\dfrac{1}{24}k_ik_jk_lk_k\dfrac{\partial^4S(\bm k)}{\partial{k_i}\partial{k_j}\partial{ k_l}\partial{k_k}}\right|_{\bm k = 0}+\dots
    \end{split}
\end{equation}
The dipole moment field, $\bm{ q}_1(\bm r) \equiv \bm r \rho(\bm r)$, has a power spectrum given by: $\bm S_1(\bm k)= N^{-1}\int e^{-i \bm{k} \cdot \bm{r}} e^{i \bm{k} \cdot \bm{r}'} \bm r\otimes \bm r' \langle \rho(\bm r)\rho(\bm r') \rangle = -(\partial_{\bm k}\otimes\partial_{\bm k}) S(\bm k)$, which directly relates to the coefficient of the $|\bm k|^2$ term in the expansion of $S(\bm k)$. The same reasoning applies to higher-order multipole fields of order $m$, with power spectrum $\bm S_m$. Consequently, any system whose structure factor admits such an expansion and scales as $|\bm k|^{2m}$ is not only hyperuniform but also exhibits suppressed multipole fluctuations for all orders $n<m$, i.e., $\bm{S}_{n<m}(\bm{k}\to 0)=0$. Building on this idea, Ref.~\onlinecite{gabrielli2008tilings} constructed static distribution of points with arbitrarily high hyperuniform scaling via a constrained tiling of the plane, providing a geometric and static counterpart to our analysis, which focuses on the dynamical realization of such fluctuation suppression.  More generally, this analysis could be helpful to understand systems displaying strong hyperuniformity~\cite{tang2024tunable, wang2025hyperuniform}.

In a recent work, Mukherjee \textit{et al.} quantified the \textit{mass} current autocorrelation of systems with a non-equilibrium center of mass conserving dynamics~\cite{mukherjee2024anomalous}. They found that it differs from the one obtained in equilibrium diffusive system. We generalize their analysis to dynamics that conserve higher-order multipoles. We define the autocorrelation $C_J=\langle J(x,t)J(x, 0) \rangle$ in 1D with $J$ the mass current obtained from $\partial_t \rho = -\partial_x J$, and compute it using Eqs.~\eqref{eq: current random}, \eqref{eq: equilibrium evolution} and \eqref{eq: evolution equation fracton}. We find:
\begin{center}
\begin{tabular}{c|c|c}
$M$ & Equilibrium $C_J(t)$ &~ Hyperuniform $C_J(t)$ \\
\hline
$M$ even $\vphantom{\dfrac 2 3}$& $\propto t^{-\frac{4M+3}{2(M+1)}}$ & $\propto t^{-\frac{3M+1}{M+2}}$\\
$M$ odd & $\propto t^{-\frac{4M+3}{2(M+1)}}$ & $\propto t^{-\frac{3M+2}{M+1}}$ \\
\hline\hline
$M=0$ $\vphantom{1^{\frac{1}{2^2}}}$& $\propto t^{-3/2}$  & $\propto t^{-1/2}$  \\
$M=1$ & $\propto t^{-7/4}$  & $\propto t^{-5/2}$  \\
$M=2$ & $\propto t^{-11/6}$  & $\propto t^{-7/4}$  \\
$M\to \infty$& $\propto t^{-2}$ & $\propto t^{-3}$  
\end{tabular}
\end{center}
We recover $C_J(t)\sim t^{-3/2}$ for equilibrium systems with diffusion~\cite{dorfman2021contemporary} or peculiar non-equilibrium systems without center of mass conservation but with a diffusive deterministic term such as the Manna model~\cite{mukherjee2023dynamic}. We also recover the result $C_J(t)\sim t^{-5/2}$ of Ref.~\onlinecite{mukherjee2024anomalous} for a non-equilibrium dynamics conserving the center of mass but with diffusive deterministic evolution. Beyond reproducing known results, our analysis unveils a general trend: current fluctuations decorrelate more rapidly as the order of the highest conserved multipole $M$ increases. In the asymptotic limit $M\to\infty$, the decay exponent saturates to $-2$ for equilibrium systems and $-3$ for their non-equilibrium hyperuniform counterparts. Although derived here for one-dimensional systems, these scaling relations can be readily generalized to higher dimensions.

Finally, we comment on absorbing phase transitions in these systems. For concreteness, consider again the lattice model with $\beta=0$ and $\bm{\mathcal L}=[1, -2, 1]$. At low densities, the dynamics may halt, since no site contains two or more particles, leading to a permanently frozen absorbing state. Increasing the density beyond a critical threshold $\rho_c$ restores activity. Near $\rho_c$, the long-wavelength behavior is described by a coarse-grained field theory involving both the conserved particle density $\rho$ and the density of active particles $\rho_a$, defined as those with nonzero hopping probability. A minimal phenomenological description is~\cite{henkel2008non}:
\begin{equation}
    \label{eq: cdp}
    \begin{split}
        \partial_t\rho_a &= (a+b\delta \rho)\rho_a - c\rho_a^2 + \kappa\bm \nabla^2\rho_a + \sqrt{2D\rho_a}\eta,\\
        \partial_t\delta\rho &= \kappa'\bm\nabla^2\rho_a,
    \end{split}
\end{equation}
with $\delta\rho(\bm r, t) = \rho(\bm r, t) - \rho_0$. Here $a(\rho_0)+b(\rho_0)\delta\rho$ controls the growth rate of activity, and $c$ saturates this growth. Particle diffusion occurs solely via the active population, which is subject to multiplicative noise with correlations proportional to the local activity. $\eta$ is a white Gaussian noise with unit variance.

Eqs.~\eqref{eq: cdp} describe the conserved directed percolation universality class and can be derived from simple lattice rules (e.g. $\bm{\mathcal L}=[1, -1, -1, 1]$ with $\beta=0$)~\cite{wiese2016coherent}. Is the universality class modified if the microscopic dynamics conserve the $M$-th multipole? The active population can still diffuse, as local conservation laws do not constrain this non-conserved subset, but the coupling between density and activity must respect multipole conservation. In particular, the diffusive coupling of the conserved density is replaced by a higher-order spatial derivative, giving:
\begin{equation}
    \label{eq: cdp mth multipole}
    \begin{split}
        \partial_t\rho_a &= (a+b\rho)\rho_a - c\rho_a^2 + \kappa\bm \nabla^{2}\rho_a + \sqrt{2D\rho_a}\eta,\\
        \partial_t\delta\rho &= (-1)^{\frac{M-1}{2}}\kappa'\bm\nabla^{M+1}\rho_a,
    \end{split}
\end{equation}
where $M$ is assumed odd so that the linear operator produces dissipative relaxation. Power counting (See Appendix~\ref{app: power counting}) shows that $\rho_a^2$ is irrelevant near the Gaussian fixed point for $d>4$, as in ordinary conserved directed percolation. However, the non-linear coupling between $\delta\rho$ and $\rho_a$ becomes irrelevant above $d=5-M$. Hence, for $d>5-M$, the critical behavior may revert to the non-conserved directed percolation universality class. Below this dimension, the system may flow to a new universality class, distinct from conserved directed percolation when $M\neq 1$, potentially exhibiting hyperuniform criticality. Verifying these scenarios requires extensive numerical work; moreover, the renormalization group analysis of conserved directed percolation is notoriously challenging~\cite{wiese2024hyperuniformity, ma2025hyperuniformity}, so we leave this avenue for future studies.

\section{Fragility of hyperuniformity without conservation laws}\label{sec: 3}

We have seen that a physically motivated route to hyperuniformity is the interplay between conservation laws and non-equilibrium driving. For center of mass conserving systems, this mechanism yields Eq.~\eqref{eq: model B HU linear}, which in Fourier space reads:
\begin{equation}
    \partial_t\rho(\bm k, t)=-\kappa\bm k^2\rho(\bm k, t)+\sqrt{2D\bm k^4}\eta(\bm k, t).
\end{equation}
For underdamped active collision models, the velocity field reads (Eq.~\eqref{eq: hydro field 2}):
\begin{equation}
    \partial_t\bm u(\bm k, t)= -\gamma \bm u(\bm k, t) - \rho^{-1}_0 (i\bm k) p(\bm k, t)  +\rho^{-1}_0 (i\bm k)\cdot \bm \Pi^{\rm{rand}}
\end{equation}
In both cases, hyperuniformity arises from the violation of the fluctuation–dissipation theorem. Within field theory, such violations are straightforward to engineer: one simply imposes colored noise whose correlation scales differently in $|\bm k|$ than the deterministic linear term, as advocated in Ref.~\onlinecite{ikeda2023correlated}. For example, consider an order parameter $\phi$ that is not conserved by the deterministic averaged relaxation but is locally conserved by the noise:
\begin{align}
    \partial_t\phi(\bm k, t)&=-\tau\phi+\sqrt{2D\bm k^{2m}}\eta~~~ \text{if $m$ is even,}\label{eq: linear problems}\\
    \partial_t\phi(\bm k, t)&=-\tau\phi+\sqrt{2D\bm k^{2(m-1)}}(i\bm k) \cdot \bm\eta~~ \text{if $m$ is odd}\nonumber,
\end{align}
with $m\geq1$. The structure factor is found to be hyperuniform $\langle \phi(\bm k)\phi(-\bm k)\rangle\sim \bm k^{2m}$.

A natural question is whether this hyperuniformity survives the inclusion of non-linearities. To address this question, consider a non-equilibrium $\phi^4$ Allen-Cahn equation with divergence noise:
\begin{equation}
    \begin{split}
    \label{eq: non eq phi^4}
    \partial_t\phi(\bm k, t)&=-\dfrac{\delta F}{\delta \phi(-\bm k, t)}+\sqrt{2D}(i\bm k)\cdot \bm \eta(\bm k, t),\\
    F[\phi(\bm r)]&=\int \left(\dfrac{\tau}{2} \phi^2 + \dfrac{u}{4}\phi^4+\dfrac 1 2(\bm\nabla\phi)^2\right)d\bm r.
    \end{split}
\end{equation}
When $u=0$, the theory reduces to a linear equation equivalent to Eq.~\eqref{eq: linear problems}. For $u\neq 0$, non-linear couplings between Fourier modes appear:
\begin{equation}
    \begin{split}
        \partial_t\phi(\bm k)&= u\iint \phi(\bm q)\phi(\bm p)\phi(\bm k - \bm q - \bm p)\dfrac{d\bm qd\bm p}{(2\pi)^{2d}}\\
        &\quad-(\tau + \bm k^2)\phi(\bm k)+\sqrt{2D}(i\bm k)\cdot \bm \eta(\bm k, t),
    \end{split}
\end{equation}
producing non-trivial dynamics that must be analyzed perturbatively.

\begin{figure}
    \centering
    \includegraphics[width=0.999\linewidth]{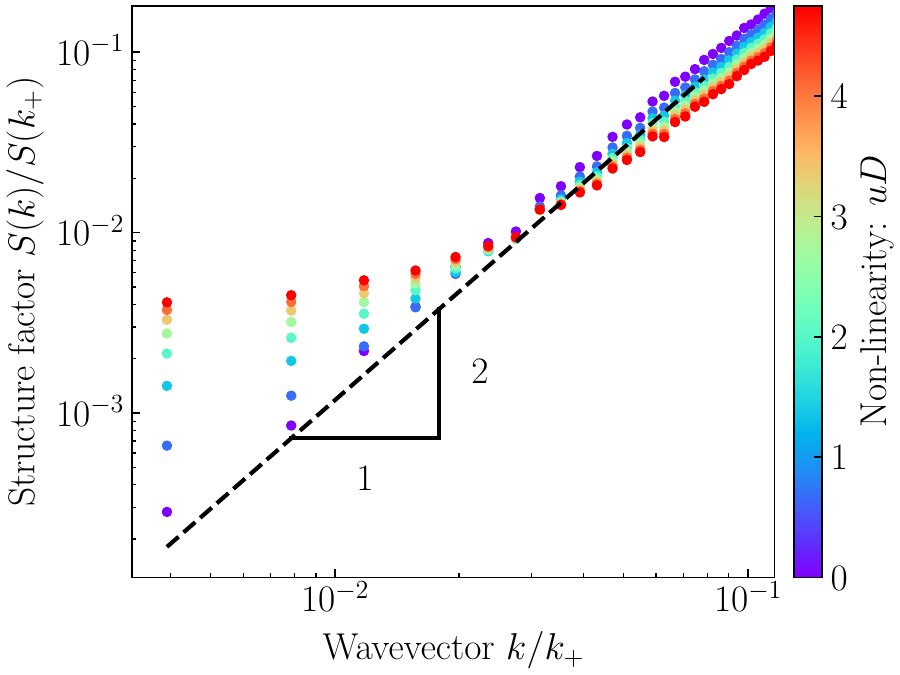}
    \caption{Structure factor of the non-equilibrium $\phi^4$ model (Eq.~\eqref{eq: non eq phi^4}) in a $d=2$ periodic box of size $L\times L$ driven by a conserved noise (divergence noise) but non-conserved deterministic relaxation, for various values of the non-linearity. Hyperuniformity is present only in the linear case and is destroyed by arbitrarily small non-linearities. Simulations are performed using a pseudo-spectral method~\cite{caballero2024cupss} with $\sqrt{\tau}dx=\sqrt{\tau}dy=1$, $\tau L^2=1024^2$ and $\tau dt= 0.025$.}
    \label{fig: modelA}
\end{figure}

We begin by simulating Eq.~\eqref{eq: non eq phi^4}, and present the results in Fig.~\ref{fig: modelA}. For $u=0$, the structure factor exhibits the expected hyperuniform scaling $S(k)\sim k^2$. However, for any finite amount of non-linearity, a plateau develops at small $k$: hyperuniformity is lost. Perturbative calculations of the non-linear corrections to $S(k)$ can capture this effect, but the underlying physics is more transparent in a renormalization-group framework, even if we remain far from the critical point.

Integrating out high-momentum modes of Eq.~\eqref{eq: non eq phi^4} yields an effective coarse-grained theory valid for $|\bm k|<|\bm k_*|$ (see Appendix~\ref{app: rg}):
\begin{align}
\partial_t\phi(\bm k, t)&=-\dfrac{\delta F_R}{\delta \phi(-\bm k, t)}+\sqrt{2D_R}(i\bm k)\cdot \bm \eta(\bm k, t)+\nonumber\\
&\quad\sqrt{2D'}\zeta(\bm k, t) \quad\text{for } |\bm k|<|\bm k_*|,\label{eq: non eq phi^4 renormalized}\\
\langle\zeta(\bm k, t)&\zeta(\bm k', t')\rangle=(2\pi)^d\delta(t-t')\delta(\bm k + \bm k')\nonumber,
\end{align}
with a renormalized free energy:
\begin{align}
F_R[\phi(\bm r)]&=\int \left(\dfrac{\tau_R}{2} \phi^2 + \dfrac{u_R}{4}\phi^4+\dfrac 1 2(\bm\nabla\phi)^2\right)d\bm r.
\end{align}
Besides scale dependent renormalized parameters $(\tau_R,u_R,D_R)$, the renormalization group procedure and the coupling between modes generate a new effective Gaussian white noise $\sqrt{2D'}\zeta$, similar to the one found at equilibrium for model A. This additional noise, absent in the bare dynamics and effectively acting on large scales, destroys hyperuniformity at small wavevectors, just as any simple Gaussian \textit{white} noise would in the linear regime.

More generally, when the deterministic dynamics obey conservation laws as well, we show in Appendix~\ref{app: rg} that if the noise conserves multipole moments up to the same order as the deterministic terms, the linear prediction for the structure factor remains valid, as new noises, qualitatively different from the bare one, are not generated. However, if the bare noise conserves higher multipoles than the deterministic part, the renormalization group generates a noise that obeys the same (lower) conservation laws as the deterministic dynamics. Thus, the effective large-scale behavior is governed by the conservation laws of the deterministic term. An analogous result holds for hyperuniformity induced by temporally correlated noise~\cite{ikeda2023correlated, ikeda2024continuous} (see Appendix~\ref{app: rg}). Consequently, hyperuniformity arising solely from spatio-temporally correlated noise, as in Ref.~\onlinecite{ikeda2023correlated}, is generally unstable to non-linearities in systems lacking conservation laws. This aligns with recent studies on the breakdown of the Mermin–Wagner theorem in non-equilibrium systems, which found that within their framework, a conserved mode is necessary to stabilize non-equilibrium–induced long-range order~\cite{diessel2025stabilization}. 

In the underdamped examples discussed in Sec.~\ref{sec: recap}, the evolution equation for the momentum field resembles the linear regime of Eq.~\eqref{eq: non eq phi^4}. We therefore expect the same reasoning presented here to carry over. Indeed, as shown in Appendix~\ref{app: nonlinearvelocity}, hyperuniformity in our molecular dynamics simulations emerges only under linear damping. By contrast, introducing a nonlinear damping term of the form $-\bm v |\bm v|^n$ with $n>0$ destroys hyperuniformity, for reasons analogous to those discussed above. This further complicates the experimental observation of such hyperuniformity.

One subtlety deserves comment. In Sec.~\ref{sec: recap} we argued that hyperuniformity reflects energy injection at short scales combined with dissipation across all scales, leading to a depletion of long-wavelength fluctuations. At first sight, this reasoning should also apply here, where deterministic dynamics do not conserve the order parameter. Yet hyperuniformity is absent. The resolution is that the energy-transfer mechanism must be local in $\bm k$-space for the argument to hold. Without a conservation law, nothing prevents non-local transfers: energy injected at high $\bm k$ can directly leak into small-$\bm k$ modes. This leakage provides the physical intuition for the emergence of an effective noise under renormalization.

\section{Conclusion}

Our results establish that conservation laws are decisive for stabilizing hyperuniformity in non-equilibrium systems, whereas mechanisms relying on partially conserved dynamics are intrinsically fragile. Using these conservation laws, we can generate configurations with arbitrarily strong hyperuniformity through short-range interactions alone and without any fine-tuning. 
This work opens intriguing directions, such as the study of dissipative Hamiltonian fractons (Eq.~\ref{eq: hamiltonian fracton}), as well as new lines of inquiry within the broader landscape of self-organized criticality. 

We only investigated the emergence of generic hyperuniformity via the interplay between conservation laws and a non-equilibrium density field. Other directions of research would include the effects of coupling the density field to other slow fields, a deeper look at the effect of temporally correlated noise, or non-equilibrium mechanisms leading to a generation of an effective mass. Theoretically, these systems are well-suited for investigation via non-equilibrium stochastic field theories and gradient expansions. A linear theory may often suffice, provided that nonlinearities do not generate additional noise absent at the bare level. The main challenge, however, lies in the fact that while constructing such field theories is relatively straightforward, identifying physical realizations of them is difficult.

Another approach to achieving generic hyperuniformity is to utilize long-range interactions. At equilibrium, liquid-state theory can often predict the emergence of hyperuniformity given an interaction potential~\cite{hansen2013theory}; far from equilibrium, however, each system must be analyzed individually to assess hyperuniformity, as a general liquid-state theory remains, for now, far from reach and a herculean task.

The systems considered here were homogeneous, and hyperuniformity emerged from fluctuations on top of this uniform background. This contrasts with systems in which the average density field itself is hyperuniform. Such hyperuniformity arise, for example, during phase separation. With the recent surge of interest in active matter and its tendency to form complex patterns, these systems may offer a promising platform for generating hyperuniform states. Such mechanisms are particularly intriguing because they could sustain hyperuniformity even in the presence of thermal fluctuations.

Finally, the emergence of hyperuniformity in finely tuned systems is also of considerable interest. In such cases, hyperuniformity is not generic and may arise only through non-linear effects. A notable example is conserved directed percolation, which is hyperuniform only at its non-trivial critical fixed point and not at its Gaussian one~\cite{ma2025hyperuniformity}. This distinction is significant because most mechanisms for hyperuniformity are understandable within linear theory, making these non-linear instances a considerable theoretical challenge. An interesting question is whether the critical point of the field theory proposed in Eq.~\eqref{eq: cdp mth multipole} is hyperuniform.

These open theoretical questions underscore the vast and promising landscape of hyperuniformity and its relation to other fields, offering a fertile ground for future exploration.
\section*{NOTES}
While finalizing this manuscript, we became aware of Refs.~\onlinecite{liu2023local, liu2025hyperuniform}, which employ self-rotating underdamped dumbbells to generate hyperuniformity; a system closely related to our self-rotating ellipses.
\begin{acknowledgments}
This work benefited greatly from discussions with Giuseppe Foffi, Frank Smallenburg, Andrea Plati, Yuta Kuroda, Ludovic Berthier, Leonardo Galliano and Andrew Lucas. 
\end{acknowledgments}

\appendix

\section{Methods and details on the models}\label{app: details}
\subsection{Chiral active particles}\label{app: chiral active}

We consider a system of $N$ particles in a 2D periodic box of size $L\times L$, of mass $m$ with position $\bm r_i$ undergoing the following underdamped dynamics:
\begin{equation}
    \begin{split}
    m\ddot{\bm r}_i=&-m\gamma\dot{\bm r}_i-\sum_{j\neq i} \left(\partial_{\bm r_i}U_{\textrm{WCA}}(|\bm r_i-\bm r_j|)+\right.\\ &\left.\partial_{\bm r_i}U_{\textrm{chiral}}(|\bm r_i-\bm r_j|)\times \hat{\bm z}\right).
    \end{split}
\end{equation}
$U_\textrm{WCA}$ is a repulsive WCA potential:
\begin{equation}
    \label{eq: WCA}
    U_{\text{WCA}}(r)=
\begin{cases}
4\varepsilon\!\left[\left(\dfrac{\sigma}{r}\right)^{12} - \left(\dfrac{\sigma}{r}\right)^{6}\right] + \varepsilon, & r \le r_c,\\[8pt]
0, & r > r_c,
\end{cases}
\end{equation}
with $r_c = 2^{1/6}\sigma$. We chose $U_{\textrm{chiral}}=\omega{r_c}/{r}$ but numerically cut off the force at $|\bm r_i - \bm r_j|=5r_c$, $\hat z$ is the unit vector perpendicular to the $xy$ plane. Note the absence of external noise: $U_{\textrm{chiral}}$ injects the energy that is retrieved by the viscous damping $\gamma$.

We define the natural time unit: $\tau \equiv \sigma\sqrt{m/\varepsilon}$.  Simulations are performed using a velocity-Verlet algorithm with time discretized as $dt/\tau=0.01$. The physical parameters are set to $\omega/\varepsilon=1$, $\gamma\tau=1$, $\phi = N\pi\sigma^2/(4L^2)=0.4$ and $N=20000$.

\subsection{Underdamped dissipative particle dynamics}
\label{app: dpd}

We consider a system of $N$ particles in a 2D periodic box of size $L\times L$,  with position $\bm r_i$  undergoing underdamped dynamics with a dissipative particle dynamics noise:
\begin{align}
m\dot{\bm v}_i=&-\gamma m\bm v_i- \sum_{j\neq i}\Big[\partial_{\bm r_i}U(\bm r_{ij})+\Gamma w(r_{ij}) \left( \hat{\bm{r}}_{ij} \cdot \bm{v}_{ij} \right) \hat{\bm{r}}_{ij}\nonumber\\&\qquad + \sqrt{2\Gamma T w( r_{ij})}\theta_{ij} \hat{\bm r}_{ij}\Big],\label{eq: dpd examples}
\end{align}
with $\theta$ a random Gaussian noise:
\begin{equation}
    \langle \theta_{ij}(t) \rangle = 0, \quad \langle \theta_{ij}(t)\theta_{kl}(t') \rangle = (\delta_{ik}\delta_{jl} + \delta_{il}\delta_{jk})\delta(t-t'),
\end{equation}
and $w(r)=\Theta(r/\sigma_w<1)$, with $\Theta(\mathtt{x})=1$ when $\mathtt{x}$ is satisfied, and $0$ otherwise, making the force local.  $U_{\textrm{WCA}}$ is defined in Eq.~\eqref{eq: WCA}. Note that the noise conserves momentum and is similar to a random stress for a coarse velocity field, associated to the viscosity at equilibrium.

We define the natural time unit: $\tau \equiv \sigma\sqrt{m/\varepsilon}$ and solve the dynamics using a velocity-Verlet integrator with $dt/\tau=0.0003$ (the potential large number of neighbors can make the noise variance large). The physical parameters are fixed to $T/\epsilon=1$, $\gamma/\tau = 10$, $\sigma_w/\sigma=4.5$, $\phi=N \sigma^2\pi/L^2=0.6$, $\Gamma/\tau = 1$  and $N = 20000$.

\subsection{Self-rotating active rods}\label{app: rods}

We consider $N$ rod-like particles with positions $\bm r_i$ and orientations 
$\hat{\bm u}_i=(\cos\theta_i,\sin\theta_i)$ in 2D. The dynamics is overdamped with a constant active rotation rate $\omega_0$:
\begin{equation}
    \begin{split}
    m\gamma \dot{\bm r}_i &=-\sum_{j\neq i} \partial_{\bm r_i} U^{\rm GB}_{\rm WCA}(\bm r_{ij},\hat{\bm u}_i,\hat{\bm u}_j), 
    \\
    \gamma_r \dot{\theta}_i &= \gamma_r\omega_0 -\sum_{j\neq i} \partial_{\theta_i} U_{\rm WCA}^{\rm GB}(\bm r_{ij},\hat{\bm u}_i,\hat{\bm u}_j),
    \end{split}
\end{equation}
The pair interaction is given by the purely repulsive Gay-Berne potential, a smooth anisotropic generalization of the WCA potential:
\begin{equation}
    \begin{split}
    &U_{\rm GB}(\bm r_{ij},\hat{\bm u}_i,\hat{\bm u}_j) =\\
       & \begin{cases}
        4\varepsilon 
        \left[ \left( \dfrac{\sigma_{\rm min}}{r^\theta_{ij}} \right)^{12} 
        - \left( \dfrac{\sigma_{\rm min}}{r^\theta_{ij}} \right)^{6} \right] + \varepsilon, 
        & r_{ij} \leq r_c,\\[6pt]
        0, & r_{ij} > r_c,
        \end{cases}
    \end{split}
\end{equation}
where $r^\theta_{ij}\equiv r_{ij}-\sigma(\hat{\bm r}_{ij},\hat{\bm u}_i,\hat{\bm u}_j)+\sigma$ and:
\begin{equation}
    \begin{split}
        \sigma(\hat{\bm r}_{ij},\hat{\bm u}_i,\hat{\bm u}_j) =& \sigma 
        \left[ 1 - \frac{\chi}{2} \left(
        \frac{(\hat{\bm r}_{ij}\!\cdot\!\hat{\bm u}_i+\hat{\bm r}_{ij}\!\cdot\!\hat{\bm u}_j)^2}{1+\chi\,\hat{\bm u}_i\!\cdot\!\hat{\bm u}_j} 
        \right.\right.+\\& \left.\left.\frac{(\hat{\bm r}_{ij}\!\cdot\!\hat{\bm u}_i-\hat{\bm r}_{ij}\!\cdot\!\hat{\bm u}_j)^2}{1-\chi\,\hat{\bm u}_i\!\cdot\!\hat{\bm u}_j} 
        \right) \right]^{-1/2},
    \end{split}
\end{equation}
with $\chi = (\kappa^2 - 1)/(\kappa^2 + 1)$. $r_c = 2^{1/6}\sigma_{\rm min}$ is the ellipse width and $\kappa=\sigma_{\rm max}/\sigma_{\rm min}$ is its aspect ratio. We note that for full consistency, we should use an orientation dependent potential well $\varepsilon(\hat{\bm r}_{ij},\hat{\bm u}_i,\hat{\bm u}_j) = \varepsilon_0 
\left[ 1 - \chi'^2(\hat{\bm u}_i\!\cdot\!\hat{\bm u}_j)^2 \right]^{-1/2}$, with $\chi' = (\kappa'^{1/2} - 1)/(\kappa'^{1/2} + 1)$ and $\kappa'$ a parameter controlling the energy anisotropy~\cite{cleaver1996extension}. However, our choice is particularly simple, and widely used in the literature~\cite{allen2006expressions}. Once again, no translational or rotational noise is included and the activity of the system is fully provided by the active angular drive $\omega_0$.

Simulations are performed using HOOMD-Blue~\cite{anderson2020hoomd} with the \texttt{Brownian} integrator~\cite{snook2006langevin}. We define the natural time unit: $\tau \equiv m\gamma \sigma_{\rm max}^2/\varepsilon$. The timestep is set to $dt/\tau=0.0005$. The physical parameters are set to $\omega_0\tau=0.25$, $\kappa=4$, $\gamma_r/\epsilon\tau=2$, $\phi = N\pi\sigma_{\rm min}\sigma_{\rm max}/(4L^2)=0.6$ and $N=100000$.

\subsection{Active oscillating radius}\label{app: oscillating radius}

We consider a system of $N$ hard-disks in a periodic box of size $L\times L$, with position $\bm r_i$ of mass $m$ and whose diameter $\sigma(t)$ varies periodically in time, following a triangular pattern between $\sigma_{\text{min}}$ and $\sigma_{\text{max}}$ at frequency $\omega$:
\begin{equation}
    \sigma(t) = \sigma_{\min} + \frac{\sigma_{\max} - \sigma_{\min}}{2} \left[ 1 + \frac{2}{\pi} \arcsin\left( \sin(\omega t) \right) \right].
\end{equation}
All particles have the same radius at each time. When two particles $i$ and $j$ are located at a distance $|\bm r_i-\bm r_j|=\sigma(t)$, they undergo the following collision~\cite{Lubachevsky_Stillinger_1990}:
\begin{equation}
    \begin{split}
    \bm v_i'&= \bm v_i+\left(\bm v_{ij}\cdot \hat{\bm r}_{ij}+\dfrac{d\sigma}{dt}\right)\hat{\bm r}_{ij},  \\
        \bm v_j'&= \bm v_j - \left(\bm v_{ij}\cdot \hat{\bm r}_{ij}+\dfrac{d\sigma}{dt}\right)\hat{\bm r}_{ij} ,
        \end{split}
\end{equation}
where $\bm v'_i$ and $\bm v_i$ are the post- and pre-collisional velocities of the particle $i$, respectively. $\bm v_{ij}\cdot \hat{\bm r}_{ij}$ is the translational velocity difference between $i$ and $j$ projected along the collision axis, which must be supplemented by the additional velocity of the radius increase or decrease. Particles are subject to a viscous damping $\gamma$ during their free flight:
\begin{equation}
    \bm v(t) = \bm v(0)e^{-\gamma t}.
     \label{eq: damping}
\end{equation}
Note the absence of external white noise usually associated to the global damping, required at equilibrium for consistency with the fluctuation-dissipation theorem.

The simulations are performed using an event-driven molecular dynamics algorithm~\cite{smallenburg2022efficient} with parameters fixed to $\omega/\gamma=2/\pi$, $\sigma_{\rm min}/\sigma_{\rm max}=0.95$, $\phi_{\rm max}=N\pi\sigma_{\rm max}^2/4L^2=0.4$ and $N=100000$. The dynamics can be solved exactly without damping with a piece wise linearly increasing or decreasing $\sigma$ or with damping and a constant $\sigma$. However, it cannot be solved exactly with both damping and changing radius, therefore, we timestep the event driven algorithm and apply a discrete damping $ \bm v \to \bm v e^{-\gamma \Delta t}$, every $\Delta t$ that we set approximately equal to $0.25$ times the \textit{average} frequency of collision.

\section{Dean's derivation for underdamped dissipative particle dynamics}\label{app: dean}

In this appendix, we derive a fluctuating hydrodynamics theory of an underdamped dissipative particle dynamics system using Dean's method~\cite{dean1996langevin, brossollet2025entropy}. We start from the microscopic equations of motion:
\begin{equation}
    \begin{split}
        \dot{\bm r}_i &= \frac{\bm{p}_i}{m}, \\
        \dot{\bm{p}}_i &= -\gamma\bm p_i + \sum_{j \neq i} \left( \bm{F}^{\rm cons}_{ij} + \bm{F}^{\rm diss}_{ij} + \bm{F}^{\rm rand}_{ij} \right),
    \end{split}
\end{equation}
with:
\begin{equation}
    \begin{split}
        \bm{F}^{\rm cons}_{ij} &= -\partial_{\bm r_i} U(r_{ij}), \\
        \bm{F}^{\rm diss}_{ij} &= -\Gamma w(r_{ij}) \left( \hat{\bm{r}}_{ij} \cdot \bm{v}_{ij} \right) \hat{\bm{r}}_{ij}, \\
        \bm{F}^{\rm rand}_{ij} &= \sqrt{2\Gamma T w(r_{ij})} \theta_{ij}(t) \hat{\bm{r}}_{ij},\\
            \langle \theta_{ij}(t) \theta_{kl}(t') \rangle &= (\delta_{ik}\delta_{jl} + \delta_{il}\delta_{jk}) \delta(t - t').
    \end{split}
\end{equation}
We define the empirical density and momentum:
\begin{equation}
    \begin{split}
    \tilde\rho(\bm r, t) &\equiv \sum_{i=1}^{N} \delta(\bm r - \bm r_i(t)),\\
    \tilde{\bm{g}}(\bm r, t) &\equiv \sum_{i=1}^{N} \bm{p}_i(t)\, \delta(\bm r - \bm r_i(t)).
\end{split}
\end{equation}
We could also define the empirical full distribution function $\tilde f(\bm r, \bm p)=\sum_{i=1}^N\delta(\bm r-\bm{r}_i)\delta(\bm p-\bm{p}_i)$ \textit{à la} Klimontovich and take its various moment~\cite{klimontovich2012statistical}. See also Refs.~\onlinecite{zhang2023pulsating, Barré_Chétrite_Muratori_Peruani_2014, farrell2012pattern, solon2015pressure}. We do not take this route and continue directly with our two fields~\cite{nakamura2009derivation,  Cornalba_Shardlow_Zimmer_2019, Lutsko_2012, perezbastías2025twofieldtheoryphasecoexistence}. Taking the derivative of the empirical density field directly yields:
\begin{equation}
     \partial_t\tilde\rho(\bm r, t) = -\bm\nabla \cdot \left( \frac{\bm{g}(\bm r, t)}{m} \right).
\end{equation}
We also have:
\begin{equation}
    \label{eq: medium vel}
    \begin{split}
        \partial_t \bm{\tilde g} =& \sum_{i}\left[\bm p_i\dot{\bm r}_i\cdot\partial_{\bm r_i}\delta(\bm r - \bm r_i)+\dot{\bm p}_i\delta(\bm r - \bm r_i)\right]\\
        =&-\bm \nabla\cdot \left(\sum_i  \dfrac{\bm p_i \otimes \bm p_i}{m}\delta(\bm r - \bm r_i)\right)-\gamma\bm{\tilde g}+\\
        &\dfrac 1 2 \sum_{i}\sum_{j \neq i} \left( \bm{F}^{\rm cons}_{ij} + \bm{F}^{\rm diss}_{ij} + \bm{F}^{\rm rand}_{ij} \right)\delta(\bm r - \bm r_i).
    \end{split}
\end{equation}

We will now do each force term in Eq.~\eqref{eq: medium vel} separately. The contribution arising from interparticle interaction yields: 
\begin{align}
(1)&=\sum_{i}\sum_{j \neq i}  \bm{F}^{\rm cons}_{ij}(\bm r_i-\bm r_j)\delta(\bm r - \bm r_i)\nonumber\\
&=-\sum_{i}\sum_{j \neq i}  \partial_{\bm r_i} U (\bm r_{i} - \bm r_{j})\delta(\bm r - \bm r_i)\int \delta(\bm r_j-\bm y)d\bm y\nonumber\\
&=-\int\sum_{i}\sum_{j \neq i}  \partial_{\bm r_i} U (\bm r - \bm y)\delta(\bm r - \bm r_i)\delta(\bm r_j-\bm y) d\bm y\nonumber\\
&=-\int  \tilde\rho(\bm r)\partial_{\bm r_i} U (\bm r - \bm y)\tilde\rho(\bm y)d\bm y.
\end{align}
We proceed similarly with the dissipative contribution:
\begin{equation}
    \label{eq: second term}
    \begin{split}
    (2)&=\sum_{i}\sum_{j \neq i}  \bm{F}^{\rm diss}_{ij}(\bm r_i-\bm r_j)\delta(\bm r - \bm r_i)\\
    &=-\dfrac{\Gamma}{m}\int \sum_{i}\sum_{j \neq i}  w(r_{ij}) \left( \hat{\bm{r}}_{ij} \cdot \bm{p}_{ij} \right) \hat{\bm{r}}_{ij}\times\\&\qquad\delta(\bm r - \bm r_i)\delta(\bm r_j-\bm y)d\bm y\\
    &=-\frac{\Gamma}{m}\int \sum_{i}\sum_{j \neq i}  w(\bm r - \bm y) \frac{(\bm p_i-\bm p_j)\cdot (\bm r- \bm y) }{(\bm r - \bm y)^2}\times\\&\qquad (\bm r - \bm y)\delta(\bm r - \bm r_i)\delta(\bm r_j-\bm y)d\bm y\\
    &=-\dfrac{\Gamma}{m}\int   w(\bm r - \bm y)(\bm r - \bm y)\times\\&\qquad \dfrac{(\rho(\bm y)\bm g(\bm r)-\rho(\bm r)\bm g(\bm y))\cdot (\bm r- \bm y)  }{(\bm r - \bm y)^2}d\bm y \\
    &=- \Gamma \int   \frac{w(|\bm{y}|)\bm y}{|\bm y|^2}\rho(\bm y)\rho(\bm r - \bm y)\times\\&\qquad [ \tilde{\bm v}(\bm r - \bm{y}) - \tilde{\bm{v}}(\bm r)] \cdot \bm{y}d\bm{y},
    \end{split}
\end{equation}
where we defined $m\tilde{\rho}(\bm r)\tilde{\bm v}(r) = \tilde{\bm g}(\bm r)$ the empirical velocity field. Since $w(\bm y)$ vanishes quickly away from zero, it is safe to expand the fields around $\bm y = 0$:
\begin{align}
    \label{eq: expansion}
    \tilde\rho(\bm r - \bm{y}) = \tilde\rho(\bm r) - y^\beta \partial_\beta \tilde\rho(\bm r) + \dots\\
    \tilde{\bm{v}}(\bm r - \bm{y}) = \tilde{\bm{v}}(\bm r) - y^\alpha \partial_\alpha\tilde{\bm{v}}(\bm r) + \dots
\end{align}
For simplicity, we will neglect the density gradients which introduce cumbersome coupling between velocity and density gradients, and therefore, coupling between reversible and conservative terms. Using the expansion Eq.~\eqref{eq: expansion} into Eq.~\eqref{eq: second term}, we obtain:
\begin{align}
    \tilde\rho(\bm {r})\tilde\rho(\bm r + \bm y) \left[ \tilde{\bm {v}}(\bm {r} - \bm {y}) - \tilde{\bm {v}}(\bm {r}) \right] =& -\tilde\rho^2(\bm r) y^i \partial_i \tilde{\bm {v}}(\bm r)+\nonumber\\
    & \frac{1}{2} \tilde\rho^2(\bm r)  y^i y^j \partial_i \partial_j \tilde{\bm {v}}(\bm r)+\nonumber\\
    & \dots\label{eq: damping dean}
\end{align}
The first order term vanishes after integration by isotropy, and we finally obtain:
\begin{equation}
    \begin{split}
(2)_l&= \frac{\Gamma \rho^2 B}{2} \left( \delta_{li} \delta_{jk} + \delta_{lj} \delta_{ik} + \delta_{lk} \delta_{ij} \right)  \partial_i \partial_j {\tilde v^k}\\
&=\frac{\Gamma \rho^2 B}{2} \left[ 2\bm \nabla (\bm \nabla \cdot \tilde{\bm{v}}) +  \bm \nabla^2 \tilde{\bm{v}} \right]_l,
\end{split}
\end{equation}
with:
\begin{equation}
    \begin{split}
    \int  w(|\bm{y}|) \frac{y_l y_i y_j y_k}{|\bm{y}|^2}d\bm{y} = B \left( \delta_{li} \delta_{jk} + \delta_{lj} \delta_{ik} + \delta_{lk} \delta_{ij} \right),
    \end{split}
\end{equation}
and $\Gamma \tilde\rho^2 B$ related to a contribution to the bulk and shear viscosities of the fluid. From the dissipative term, we obtain as expected the viscous dissipative contributions typical in Navier-Stokes fluids.

The last term is the noise. We 
 define the noise: $\bm \zeta=\sum_{i}\sum_{j \neq i}  \bm{F}^{\rm rand}_{ij}(\bm r_i-\bm r_j)\delta(\bm r - \bm r_i)$ and obtain its correlation $\langle \zeta_i(\bm r, t)\zeta_j(\bm r', t')\rangle=2\Gamma\tilde\rho(\bm r) TC_{ij}(\bm r, \bm r')\delta(t-t')$, following Ref.~\onlinecite{anand2025emergent}:
\begin{align}
    C_{ij}&=\delta(\bm r- \bm r')\int\tilde\rho(\bm y)w(\bm r - \bm y)\dfrac{(\bm r - \bm y)^i(\bm r - \bm y)^j}{(\bm r - \bm y)^2}d\bm{y}\nonumber-\\
    &\quad\tilde\rho(\bm r')w(\bm r - \bm r')\dfrac{(\bm r - \bm r')^i(\bm r - \bm r')^j}{(\bm r - \bm r')^2}\nonumber\\
    &=\int \tilde\rho(\bm y)w(\bm r - \bm y)\dfrac{(\bm r - \bm y)^i(\bm r - \bm y)^j}{(\bm r - \bm y)^2}\times\nonumber\\
    &\quad\big(\delta(\bm r- \bm r') - \delta(\bm y- \bm r')\big)d\bm y\label{eq: correlation dean}\\
    &=\tilde\rho(\bm r)\int w(\bm y)\dfrac{y^iy^j}{\bm y^2}\left(\delta(\bm r- \bm r') - \delta(\bm r - \bm r'- \bm y)\right)d\bm y+\nonumber\\
    &\quad\mathcal{O}(\partial_{\bm r}\tilde\rho(\bm r))\nonumber\\
    &\simeq \dfrac{\tilde\rho(\bm r)}{2}\int w(\bm y)\dfrac{y^iy^jy^ky^l}{\bm y^2}d\bm y\partial_k\partial_l\delta(\bm r- \bm r')\nonumber\\
    &= \dfrac{\tilde\rho(\bm r)B}{2}\left( \delta_{li} \delta_{jk} + \delta_{lj} \delta_{ik} + \delta_{lk} \delta_{ij} \right)\partial_k\partial_l\delta(\bm r- \bm r').\nonumber
\end{align}
We expanded around $\bm y=0$ since $w(\bm y)$ vanishes rapidly away from 0. The zeroth order term vanishes identically, while the first order term vanished after integration by isotropy. Only the second order contribution remains, corresponding to a second derivative of a Dirac delta. Because the dissipative viscous forces $\bm F^{\rm diss}$ and the noises $\bm F^{\rm rand}$ are related by the fluctuation-dissipation theorem, it is unsurprising that Eq.~\eqref{eq: correlation dean} and~\eqref{eq: damping dean} also satisfy a fluctuation-dissipation theorem. The global damping however, violates the total fluctuation-dissipation theorem. Ref.~\onlinecite{anand2025emergent} argued that the noise derived from Dean’s method in the random organization model is fundamentally distinct from a Laplacian noise. Our analysis shows that this distinction is not essential: the Laplacian noise naturally arises when focusing on hydrodynamic scales via a gradient expansion. This could be expected from the series expansion in $\bm k$ of its autocorrelation as correlations fully determine Gaussian noise or from the fact that at equilibrium its associated dissipation corresponds to a viscosity~\cite{groot1997dissipative}.

Collecting all terms, we obtain the following field equations:
\begin{align}
    \partial_t\tilde\rho(\bm r, t) =& -\bm\nabla \cdot \left( \tilde\rho \tilde {\bm v} \right),\label{eq: final dean equations}\\
    \partial_t \bm{\tilde v}(\bm r, t) + \tilde{\bm{v}}\cdot \bm\nabla \tilde{\bm{v}} =&-\dfrac{1}{m}\int \partial_{\bm r_i} U (\bm r - \bm y)\tilde\rho(\bm y)d\bm y+\nonumber\\
    &\quad\eta \left[ 2\bm \nabla (\bm \nabla \cdot \tilde{\bm{v}}) +  \bm \nabla^2 \tilde{\bm{v}} \right] -\gamma\tilde{\bm v}+\nonumber\\
    & \quad\bm \nabla \cdot \tilde{\bm \Pi}^{\rm rand},\nonumber
\end{align}
with $\eta = {\Gamma \tilde\rho B}/{2}$ and $\tilde{\bm \Pi}^{\rm rand}$ a Gaussian random stress with correlations:
\begin{equation}
    \begin{split}
    \langle\tilde{\Pi}^{\rm rand}_{ij}\tilde{\Pi}^{\rm rand}_{kl} \rangle&=2\eta T\left( \delta_{li} \delta_{jk} + \delta_{lj} \delta_{ik} + \delta_{lk} \delta_{ij} \right)\times\\&\qquad\delta(t-t')\delta(\bm r- \bm r').
    \end{split}
\end{equation}
These equations closely resemble the hydrodynamic equations for active hard disks Eqs.~\eqref{eq: hydro field}. The spatial integral of the interaction force contributes to the reversible stress, while the velocity gradients generate viscous stresses associated with $\bm F^{\rm diss}$. The stochastic term enters as the divergence of a stress tensor, consistent with momentum-conserving noise. Linearization of Eqs.~\eqref{eq: final dean equations}, indeed produces $S(\bm k)\sim \bm k^2$ for a local potential $U$. Adiabatic elimination of $\tilde{\bm v}$ recovers the hyperuniform model B dynamics, Eq.~\eqref{eq: model B HU}.

Despite these similarities with the equations given in the main text, Eqs.~\eqref{eq: final dean equations} differ in a crucial aspect: they govern the microscopic, non–coarse-grained density and velocity fields. No coarse-graining length or timescale has been introduced, nor have fast degrees of freedom been eliminated~\cite{illien2024dean, te2020classical,archer2004dynamical}. Without the gradient expansion used to obtain local dissipative terms and noises, the equations would be formally exact. Consequently, the transport coefficients lack contributions from the reversible interactions: for example, $U$ does not appear to contribute to the viscosities, even after a gradient expansion. Such effects emerge only after a proper coarse-graining. Similarly, the noise in Eq.~\eqref{eq: final dean equations} is the microscopic external noise inserted at the level of the particle dynamics. It should not be interpreted as an emergent mesoscopic internal noise arising from interactions at intermediate scales -- something Dean’s method cannot capture, since it is blind to mesoscopic physics.

This subtlety is easily overlooked. For instance, applying Dean’s method to a Hamiltonian system without noise or damping simply yields the Klimontovich equation for $\tilde f(\bm r, \bm v)$. At the microscopic level, the only stochasticity originates from initial conditions, and the method correctly predicts the (trivial) absence of noise or dissipation. On the mesoscopic scale, however, ensemble averaging induce dissipative terms which demand the inclusion of fluctuations, as described by Landau–Lifshitz fluctuating hydrodynamics. At equilibrium, projection-operator techniques recover the correct mesoscopic noise and dissipation, which include the effect of the reversible interactions~\cite{espanol1995hydrodynamics}. Out of equilibrium, the situation is more subtle. Typically, in active matter for example, external noise is introduced at the microscopic level. At low densities, Dean’s method is appropriate, as dissipative contributions and the emergent mesoscopic noise can be neglected~\cite{Bouchet_Gawȩdzki_Nardini_2016, dawsont1987large, illien2024dean, dinelli2024fluctuating, bon2025non}. Furthermore, the mesoscopic noise often resembles the externally imposed noise, which can lead to quantitatively inaccurate results, though the qualitative behavior is usually preserved. However, this resemblance is not guaranteed.

To show how problems related to the lack of dissipation in Dean's equation can arise, we will follow Ref.~\onlinecite{kuroda2023microscopic} and consider 2D self-propelled, chiral particles:
\begin{equation}
    \begin{split}
    \gamma\dot{\bm r}_i =& \gamma u_0 \bm n - \sum_i\partial_{\bm{r}_i}U(\bm r_i - \bm r_j)   , \\
    \gamma_r\dot{\bm \theta}_i=&\gamma_r\omega+\sqrt{2\gamma_r T}\zeta_i,
    \end{split}
\end{equation}
with $\bm n=(\cos(\theta), \sin(\theta))$ and $u_0$ the free self-propulsion speed. The active angular drive $\omega$ induces circular trajectories, perturbed by the noise. Dean’s method yields~\cite{kuroda2023microscopic}:
\begin{align}
    \gamma\partial_t\tilde\rho(\bm r, t) &= \bm \nabla\cdot\left(\tilde\rho \bm \nabla\dfrac{\delta \tilde F}{\delta \tilde\rho}+\sqrt{u_0^2\tilde\rho/2}\eta(\bm r, t)\right),\nonumber\\
    \langle\eta(\bm r, t)\eta(\bm r', t')\rangle &= e^{-T|t-t'|/\gamma_r}\cos(\omega(t-t'))\delta(\bm r-\bm r'),\label{eq: noise kuroda}
\end{align}
with $\tilde F$ a microscopic free energy, depending on $U$. Remarkably, as Ref.~\onlinecite{kuroda2023microscopic} found out, even in the limit $T\to 0$, the density field is still driven by a non-zero noise (Eq.~\eqref{eq: noise kuroda}) accounting for the external active angular drive. Yet in this limit, Dean’s predictions become pathological. In the linear regime, the intermediate scattering function is:
\begin{equation}
    \begin{split}
    S(\bm k, t)&=\langle \tilde\rho(\bm k, t)\tilde\rho(-\bm k, t')\rangle\\
    &=\dfrac{1}{2}\dfrac{u_0^2\rho_0\bm k^2}{\kappa^2\bm k^4 + \gamma^2\omega^2 }\cos(\omega (t-t')).
    \end{split}
\end{equation}
$\kappa$ is a coefficient arising from the linearization of $\tilde F$. Although $S(\bm k, 0)$ correctly exhibits hyperuniformity, decorrelation is never achieved since $S(\bm k, t\to\infty)\neq0$. This limitation reflects the microscopic nature of Dean’s description: it cannot generate dissipative terms. The $T\to0$ noise is non-dissipative, encoding only the active angular drive. Interestingly, simply adding a Laplacian mesoscopic noise (uncorrelated from Dean's noise) arising from interparticle collisions is insufficient to restore full decorrelation. To accurately recover the mesoscopic physics from Dean's equation, it would be necessary to separate the solution $\tilde\rho$ into two components: one averaged over initial conditions and a second fluctuating part. Both components would need to be determined for a fixed external realization of noise~\cite{bixon1969boltzmann,klimontovich2012statistical}. To our knowledge, this is a complex endeavor that has not yet been attempted.

\section{Details on the lattice dynamics}

\subsection{Derivation of a field theory for \texorpdfstring{$\bm\rho$}{ρ} from the lattice dynamics}
\label{app: detail lattice}

To better capture the emergent large-scale dynamics in the lattice, and to derive the associated correlation functions, it is useful to formulate an evolution equation for the coarse-grained density field~\cite{han2024scaling}.

Let $a$ be the lattice spacing, with total system size $L = aN$. In the continuum limit $a \to 0$ with $L$ fixed, it is straightforward to derive the deterministic evolution equation for $n(x,t)$. The stochastic equation, however, requires more care, as we now describe.

Following Ref.~\onlinecite{lefevre2007dynamics}, we introduce the moment generating function for the local increment $\Delta n_x(t) \equiv n_x(t+dt) - n_x(t)$:
\begin{equation}
\begin{split}
Z[\bm{\hat{n}}(t)] &= \left\langle \exp \left[ \sum_{x,t} \hat{n}_x(t) \Delta n_x(t) \right] \right\rangle\\&=\prod_{x,t} \left\langle \exp \left[ \hat{n}_x(t) \Delta n_x(t) \right] \right\rangle,
\end{split}
\end{equation}
where the average $\langle \cdots\rangle$ is over realizations of $\Delta \bm{n}(t)$ and $\hat{\bm{n}}$ is conjugate to $\Delta\bm{n}$.

Using the rates defined above, we perform the average and obtain in the limit $dt\to0$:
\begin{equation}
\begin{split}
Z[\bm{\hat{n}}(t)] =& \prod_{x,t} \left[
    dt\, \mathcal{R}^{\{\alpha\}}_{\mathcal{L}_x} 
    e^{\hat{\bm{n}} \cdot (\mathcal{L}_x \bm{n} - \bm{n})} 
    + dt\, \mathcal{R}^{\{\beta\}}_{\mathcal{L}_x^{-1}} 
    e^{\hat{\bm{n}} \cdot (\mathcal{L}_x^{-1} \bm{n} - \bm{n})}\right.\\&\left. 
    + 1 - dt \left( \mathcal{R}^{\{\alpha\}}_{\mathcal{L}_x} + \mathcal{R}^{\{\beta\}}_{\mathcal{L}_x^{-1}} \right)\right]\\
    =&\exp\left[a^{-1} \iint \mathcal{R}^{\{\alpha\}}_{\mathcal{L}_x} 
    \left(e^{\hat{\bm{n}} \cdot (\mathcal{L}_x \bm{n} - \bm{n})} -1\right) \right. \\ &\left.
    +\mathcal{R}^{\{\beta\}}_{\mathcal{L}_x^{-1}} \left(
    e^{\hat{\bm{n}} \cdot (\mathcal{L}_x^{-1} \bm{n} - \bm{n})}  -1\right)dtdx\right].
\end{split}
    \label{eq: to refer}
\end{equation}
Nonlocal products such as $n_x n_{x+2}$ are expanded in $a$ using Taylor series, e.g. $n_xn_{x+2}\simeq n(x)(n(x)+2an'(x)+(2a)^2n''(x)/2+\dots)$, which yields a local field theory for the occupation number.

We define the cumulant generating function by: $W[\bm{\hat{n}}(t)]\equiv \log\left(Z[\bm{\hat{n}}(t)]\right)$. Its functional derivative leads to various cumulant of $\partial_t n$:
\begin{equation}
    \begin{split}
    \left.\dfrac{\delta W}{\delta\hat n(x, t)}\right|_{\hat n=0}=&\langle\partial_tn(x, t)\rangle,\\
    \left.\dfrac{\delta^2 W}{\delta\hat n(x, t)\delta\hat n(x', t')}\right|_{\hat n=0}=&\langle\partial_{t'}n(x', t')\partial_{t}n(x, t)\rangle-\\&\langle\partial_tn(x, t)\rangle\langle\partial_{t'}n(x', t')\rangle.
    \label{eq: cumulant}
    \end{split}
\end{equation}
Using the non-anticipating property of the noise in It\=o discretization, we can obtain the associated Langevin equation.

As an instructive example, consider the process $\bm{\mathcal L}=[-1,4, -6, 4, -1]$ which conserves all multipole of order below or equal to $3$. We will consider the time reversed event too. We obtain:
\begin{equation}
\begin{split}
    \hat{\bm{n}} \cdot (\mathcal{L}_x^{\pm 1} \bm{n} - \bm{n})
&= \mp\hat n_{x-2} \pm 4\hat n_{x-1} \mp 6\hat n_x \pm 4\hat n_{x+1} \mp \hat n_{x+2} \\
&\simeq \mp a^4 \partial_x^4 \hat n(x)
\end{split}
\end{equation}
\begin{equation}
\begin{split}
\mathcal{R}^{\{\alpha\}}_{\mathcal{L}_x}
&= \alpha n_{x-2} n_x (n_x - 1)(n_x - 2)(n_x - 3)(n_x - 4)\times\\
&\quad(n_x - 5) n_{x+2} \\
&\simeq \alpha n_{x-2} n_{x+2} n_x^6 \\
&\simeq n(x)^6 \Bigg( n(x)^2 + 4 a^2 n(x)^2 \partial_x^2 \log(n(x)) +\\
&\quad  \frac{4}{3} a^4 \Big( 3 (\partial_x^2 n(x))^2 - 4 (\partial_x n(x)) (\partial_x^3 n(x)) +\\
&\quad  n(x) \partial_x^4 n(x) \Big) \Bigg),
\end{split}
\end{equation}
\begin{equation}
\begin{split}
\mathcal{R}^{\{\beta\}}_{\mathcal{L}_x^{-1}}
&= \beta n_{x-1} (n_{x-1}-1)(n_{x-1}-2)(n_{x-1}-3) \times\\
&\quad n_{x+1} (n_{x+1}-1)(n_{x+1}-2)(n_{x+1}-3) \\
&\simeq \beta n_{x-1}^4 n_{x+1}^4 \\
&\simeq \beta n(x)^6 \Bigg( n(x)^2 + 4 a^2 n(x)^2 \partial_x^2 \log(n(x))+ \\
&\quad  a^4 \frac{1}{3} n(x)^{-2} \Big( 18 (\partial_x n(x))^4 - 36 n(x) \times\\
&\quad (\partial_x n(x))^2\partial_x^2 n(x) - 4 n(x)^2 (\partial_x n(x)) (\partial_x^3 n(x))+ \\
&\quad n(x)^2 \big( 21 (\partial_x^2 n(x))^2 + n(x) \partial_x^4 n(x) \big) \Big) \Bigg).
\end{split}
\end{equation}

This yields the cumulant generating function:
\begin{equation}
    \begin{split}
    W[\bm{\hat{n}}(t)]= a^{-1}\iint &\mathcal{R}^{\{\alpha\}}_{\mathcal{L}_x}\left(e^{-a^4\partial_x^4\hat n} -1\right)+\\&\mathcal{R}^{\{\beta\}}_{\mathcal{L}_x^{-1}}\left(e^{a^4\partial_x^4\hat n} -1\right)dtdx,\\
    &
    \end{split}
\end{equation}
and its functional derivatives:
\begin{align}
    \frac{\delta W}{\delta \hat{n}(x,t)} &= -a^3 \, \partial_x^4 \Big(  \mathcal{R}^{\{\alpha\}}_{\mathcal{L}_x}e^{-a^4 \partial_x^4 \hat{n}(x,t)}-\nonumber\\
    &\quad\mathcal{R}^{\{\beta\}}_{\mathcal{L}_x^{-1}} \, e^{a^4 \partial_x^4 \hat{n}(x,t)} \Big),\\
    \frac{\delta^2 W}{\delta \hat{n}(x', t') \delta \hat{n}(x, t)} 
    &= a^7 \, \partial_x^4 \partial_{x'}^4 \Big[\Big(  \mathcal{R}^{\{\alpha\}}_{\mathcal{L}_x} e^{-a^4 \partial_x^4 \hat{n}(x,t)}+\nonumber\\
    &\quad\mathcal{R}^{\{\beta\}}_{\mathcal{L}_x^{-1}}  e^{a^7 \partial_x^4 \hat{n}(x,t)} \Big)  \delta(x - x') \delta(t - t') \Big]\nonumber.
\end{align}

When detailed balance holds, $\alpha=\beta$, the leading contribution to the deterministic part cancels at zeroth order in gradients and the first nonzero deterministic term appears at higher order in gradients. We find, to leading linear order:
\begin{align}
    \left.\frac{\delta W}{\delta \hat{n}(x,t)}\right|_{\hat n= 0}&\simeq -a^7 n_0^7\partial_x^8 \delta n(x, t),\\
    \left.\frac{\delta^2 W}{\delta \hat{n}(x,t)\delta \hat{n}(x',t')}\right|_{\hat n= 0}&\simeq 2a^7n_0^8\partial_x^4\partial_{x'}^4\delta(x-x')\delta(t-t').\nonumber
\end{align}
We linearized the occupation field as: $n(x, t)=n_0+\delta n(x, t)$. From Eq.~\eqref{eq: cumulant}, these cumulants correspond to the linear Langevin equation

\begin{equation}
    \partial_t\delta n(x, t)=-a^7 n_0^7\partial_x^8 \delta n(x, t)+\sqrt{2a^7n_0^8}\partial_x^4\eta(x, t).
\end{equation}
We readily verify that this dynamics yields the expected equilibrium result for the structure factor:
\begin{equation}
    S(k)\equiv \lim_{t\to\infty}\langle \delta n(k, t)\delta n(-k, t)\rangle=n_0,
\end{equation}
as expected for the dynamics that satisfies detailed balance~\cite{han2024scaling}.

We now consider a case in which detailed balance is broken: $\beta=0$. The lower-order gradient terms can survive in the deterministic drift. Evaluating the functional derivatives in this case and linearizing yields:
\begin{align}
        \left.\frac{\delta W}{\delta \hat{n}(x,t)}\right|_{\hat n= 0}&\simeq -8a^3n_0^7\partial_x^4 \delta n(x, t),\\
        \left.\frac{\delta^2 W}{\delta \hat{n}(x,t)\delta \hat{n}(x',t')}\right|_{\hat n= 0}&\simeq a^7n_0^8\partial_x^4\partial_{x'}^4\delta(x-x')\delta(t-t').\nonumber
\end{align}
Hence, the linear Langevin equation becomes:
\begin{equation}
    \partial_t\delta n(x, t)=-8a^3 n_0^7\partial_x^4\delta n(x, t)+\sqrt{a^7n_0^8}\partial_x^4\eta(x, t),
\end{equation}
which leads a strongly non-equilibrium and hyperuniform result:
\begin{equation}
    S(k)\equiv \lim_{t\to\infty}\langle \delta n(k, t)\delta n(-k, t)\rangle=n_0k^4/(8a^4).
\end{equation}
We note the dependence of the final result on the lattice spacing $a$.

The example above generalizes naturally. We want to find a general equation for the local occupation number written as:
\begin{equation}
    \begin{split}
    \label{eq: starting point}
   \partial_t n(x, t) &= \left.\frac{\delta W}{\delta \hat{n}(x,t)}\right|_{\hat n = 0}+\eta(x, t),\\
    \langle\eta(x, t)\eta(x', t')\rangle&=\left.\frac{\delta^2 W}{\delta \hat{n}(x', t') \delta \hat{n}(x, t)} \right|_{\hat n = 0}.
    \end{split}
\end{equation}
Suppose the events locally conserve all multipole of order equal or below $M$ with the $M-$th multipole denoted $\mathcal Q_{M}=\sum_jj^{M}n_j$.  In the continuum limit, it implies the following:
\begin{equation}
    \begin{split}
    \label{eq: general condition 1}
    \hat{\bm{n}} \cdot (\mathcal{L}_x^{\pm 1} \bm{n} - \bm{n})&=\pm\sum_{j=-m'}^m\delta_j \hat n_{x + j}\\
    &=\pm\sum_{k=0}^\infty\dfrac{a^k}{k!}\partial_x^k\hat n(x)\overbrace{\sum_{j=-m'}^m\delta_j j^k}^{=0 \textrm { if } k \leq M}\\
    &=\pm\gamma\partial_x^{M+1}\hat n(x)+\mathcal O (\partial_x^{M+2}\hat n),
    \end{split}
\end{equation}
where $\gamma=a^{M+1}/k!\sum_j\delta_jj^{M+1}$. The vanishing of the $k-$th sum comes from the fact that it is equal to the change of the $k-$th multipole after an event $\sum_{j=-m'}^m\delta_j j^k=\bm{\mathcal L}^\pm \mathcal Q_k - \mathcal Q_k$. From Eqs.~\eqref{eq: to refer} and~\eqref{eq: general condition 1}, we find:
\begin{align}
\left.\frac{\delta W}{\delta \hat{n}(x,t)}\right|_{\hat n = 0} =& \dfrac{\gamma}{a} \, \partial_x^{M+1} \left(  \mathcal{R}^{\{\alpha\}}_{\mathcal{L}_x} -\mathcal{R}^{\{\beta\}}_{\mathcal{L}_x^{-1}}  \right),\nonumber\\
\left.\frac{\delta^2 W}{\delta \hat{n}(x', t') \delta \hat{n}(x, t)} \right|_{\hat n = 0}=& \dfrac{\gamma^2}{a} \, \partial_x^{M+1} \partial_{x'}^{M+1} \left[\left(  \mathcal{R}^{\{\alpha\}}_{\mathcal{L}_x} +\mathcal{R}^{\{\beta\}}_{\mathcal{L}_x^{-1}}  \right) \right.\nonumber\\
&\left.\delta(x - x') \delta(t - t')\vphantom{\mathcal{R}^{\{\beta\}}_{\mathcal{L}_x^{-1}}} \right].\label{eq: general case app}
\end{align}
The second functional derivative is straightforward to obtain as at lowest order, we must find:
\begin{equation}
        \label{eq: eq1 WHOA}
        \mathcal{R}^{\{\alpha/\beta\}}_{\mathcal{L}_x^{\pm 1}}=\kappa_{\alpha/\beta} n(x)^\Delta + \mathcal{O}(\partial_x n),
\end{equation}
with $\kappa_{\alpha/\beta}$ a \textit{positive} constant and $\Delta=\sum_i \delta_i \Theta(\delta_i>0)=-\sum_i \delta_i \Theta(\delta_i<0)$ the number of particles that are moved by the event. Therefore, from Eqs.~\eqref{eq: general case app} and~\eqref{eq: eq1 WHOA}, we deduce that in the linear regime, the noise can always be written as a derivative of order $M+1$. We rewrite Eq.~\eqref{eq: starting point} as:
\begin{equation}
    \label{eq: almost there}
   \partial_t \delta n(x, t) =  \partial_x^{M+1} \left[  \Gamma\left(\mathcal{R}^{\{\alpha\}}_{\mathcal{L}_x} -\mathcal{R}^{\{\beta\}}_{\mathcal{L}_x^{-1}}\right)  +\sqrt{2D}\zeta(x, t)\right],
\end{equation}
with:
\begin{equation}
    \langle\zeta(x, t)\zeta(x', t')\rangle= \delta(x - x') \delta(t - t'),
\end{equation}
where $\Gamma=a^{-1}\gamma$ and $D$ is a constant. It is now left to us to obtain $\mathcal{R}^{\{\alpha\}}_{\mathcal{L}_x} -\mathcal{R}^{\{\beta\}}_{\mathcal{L}_x^{-1}}$. We will show that:
\begin{equation}
    \label{eq: a bit complicated}
    \mathcal{R}^{\{\alpha\}}_{\mathcal{L}_x}=\dfrac \alpha \beta\mathcal{R}^{\{\beta\}}_{\bm{\mathcal L}^{{-1}}_x}+\mathcal O(\partial_x^{M+1}\delta n).
\end{equation}
That is, at equilibrium, when $\alpha=\beta$: $\mathcal{R}^{\{\alpha\}}_{\mathcal{L}_x} -\mathcal{R}^{\{\beta\}}_{\mathcal{L}_x^{-1}}=\mathcal O(\partial_x^{M+1}\delta n)$. We start the proof of Eq.~\eqref{eq: a bit complicated} by taking the large density limit, in this case Eq.~\eqref{eq: rate} simplifies:
\begin{equation}
    \label{eq: rate simple}
    R_{\mathcal L_x }^{\{\alpha\}}= \alpha \prod_{j=-m'}^{m} n_{x+j}^{\underline{\delta_j}}\simeq \alpha \prod_{j=-m'}^{m} n_{x+j}^{\max(0, \delta_j)}.
\end{equation}
We also recall that the reversed process has rate:
\begin{equation}
    R_{\mathcal L_x^{-1} }^{\{\beta\}}\simeq \beta \prod_{j=-m'}^{m} n_{x+j}^{\max(0, -\delta_j)}.
\end{equation}
Since for any number $x$: $x = \max(0, x)-\max(0, -x)$, we have:
\begin{equation}
        \begin{split}
        \frac{\mathcal R_{\mathcal L_x}}{\mathcal R_{\mathcal L_x^{-1}}}&=\dfrac \alpha \beta \prod_{j=-m'}^mn_{x+j}^{\underline{\delta_j}-\underline{-\delta_j}}=\dfrac \alpha \beta\prod_{j=-m'}^m n_{x+j}^{\delta_j}\\
        &=\dfrac \alpha \beta\exp\left(\sum_{j=-m'}^m\delta_j\log(n_{x+j})\right).
        \end{split}
\end{equation}
We now continue by performing a Taylor expansion:
\begin{equation}
    \log\frac{\beta \mathcal R_{\mathcal L_x}}{\alpha \mathcal R_{\mathcal L_x^{-1}}}=\sum_{k=0}^\infty\frac{a^k}{k!}\partial_x^{k}\log(n(x))\overbrace{\sum_{j=-m'}^m\delta_j j^k}^{=0 \textrm { if } k \leq M},
\end{equation}
where the sum vanishes, again due to multipole conservation. We exponentiate both side to find Eq.~\eqref{eq: a bit complicated}:
\begin{equation}
    \begin{split}
        \mathcal R_{\mathcal L_x}^\alpha &= \mathcal R_{\mathcal L_x^{-1}}^\beta\frac{\alpha}{\beta}e^{\left(\frac{a^M \left(\sum_{j=-m'}^m\delta_j j^{M+1}\right)}{(M+1)!}\partial_x^{M+1}\log n(x)\right)}\\
        & = \mathcal R_{\mathcal L_x^{-1}}\frac{\alpha}{\beta}+\mathcal O(\partial_x^{M+1} \delta n).    
    \end{split}
\end{equation}
This concludes the derivation of the field theory as we now have all the required information to analyze Eq.~\eqref{eq: almost there}.
\\
If $\beta=\alpha$ (equilibrium/detailed balance), the leading deterministic contribution is of order $\partial_x^{2(M+1)}\delta n$:
\begin{equation}
    \begin{split}
    \partial_t\delta n(x, t)=&(-1)^{M}\Gamma_{\textrm{eq}}\partial_x^{2(M+1)}\delta n(x, t)+\\
    &\quad\sqrt{2D_{\textrm{eq}}}\partial_x^{M+1}\eta(x, t).        
    \end{split}
\end{equation}
The derived structure factor is flat as expected from equilibrium fluctuations.
\\
If $\beta=0$, the deterministic drive generically contains a $\partial_x^{M}\delta n$ term:
\begin{equation}
    \begin{split}
    \partial_t\delta n(x, t)&= (-1)^{M/2}\Gamma_{\textrm{neq}}\partial_x^{M+1}\delta n(x, t)+\\&\sqrt{2D_{\text {neq}}}\partial_x^{M+1}\eta(x, t),
    \end{split}
\end{equation}
with $M$ odd. It produces a non-equilibrium scaling of the structure factor $S(k)\sim k^{M+1}$. With $M$ even, we need to consider the next to leading term in derivatives to properly compute the structure factor, as the leading order term is non-dissipative:
\begin{equation}
    \begin{split}
    \partial_t\delta n(x, t)&= (v\partial_x^{M+1} +(-1)^{(M+1)/2} \Gamma_{\textrm{neq}}\partial_x^{M+2})\delta n(x, t)\\&+\sqrt{2D_{\text {neq}}}\partial_x^{M+1}\eta(x, t).
    \end{split}
\end{equation}
This yields $S(k)\sim k^{M}$. 

\subsection{Non-equilibrium quantum lattice dynamics}\label{app: quantum}

For systems of indistinguishable quantum particles, transition rates must incorporate Bose or Fermi statistics~\cite{han2024scaling}.
\\
For fermions, each single-particle state can be occupied by at most one particle. Introducing a degeneracy $g$ (necessary to avoid falling too easily into an absorbing phase), the classical rate in Eq.~\eqref{eq: rate} must therefore be multiplied by $\prod_{j=-m'}^m (g-n_{x+j})^{\underline{-\delta_j}}$ which enforces the Pauli exclusion constraint at the arrival sites.
\\
For bosons, the presence of particles at the arrival sites enhances the rate (Bose stimulation). The classical rate is multiplied by $\prod_{j=-m'}^m (1+n_{x+j})^{\overline{-\delta_j}}$ where $n^{\overline{\delta}} = n (n+1) \dots (n+|\delta|-1)$ if $\delta<0$ and $1$ otherwise. In principle, a similar construction extends to anyonic statistics~\cite{riccardo2019multiple} and parastatistics~\cite{wang2025particle}.

These modifications respect detailed balance whenever $\alpha=\beta$. Consequently, the equilibrium hydrodynamic equation for the density field retains the same form and derivative order as in the classical case. Out of equilibrium when $\alpha\neq\beta$, the rates are altered numerically relative to the classical dynamics, but the conclusions of Sec.~\ref{app: detail lattice} remain unchanged, notably Eq.~\eqref{eq: a bit complicated} still holds and hyperuniformity is expected.

\subsection{Structure of kernels conserving multipole moments}
\label{app: proof}
We now demonstrate how to construct kernels that conserve multipole moments up to order $M$.

We recall that a kernel $\bm{\mathcal L} = [\delta_1, \delta_2, \dots, \delta_n]$ is a vector constructed from the particle changes $\delta_j$ at a given lattice site during an event. Conservation of all moments $m\leq M$ requires:
\begin{equation}
    \label{eq: contraint equ}
    \sum_{j=1}^{n} \delta_j j^m = 0, \qquad m=0,1,\dots,M,
\end{equation}
which forms a system of $M+1$ linear equations for $n$ unknowns $\{\delta_j\}$.

Equivalently, these constraints state that the kernel is orthogonal to the vector of powers $(1^m,2^m,\dots,n^m)$ for all $m\leq M$.

To construct such kernels, we invoke the forward finite-difference operator of order $M+1$, defined for any vector $\bm s$ by:
\begin{equation}
    \Delta^{M+1} s_j = \sum_{i=0}^{M+1} (-1)^{M+1-i} \binom{M+1}{i} s_{j+i}.
\end{equation}
This operator acts as a discrete derivative. For instance: $\Delta^1 s_j=s_{j+1}-s_j$, $\Delta^2 s_j = s_j - 2 s_{j+1}+ s_{j+2}$, \dots.  Since derivatives of order higher than the degree of a polynomial annihilate it, we have $\Delta^{M+1}(1^m,2^m,\dots)=0$. This is similar to the assertion that continuous derivative of order higher than a given polynomial annihilates it. It follows that the kernel $\delta_{j}=(-1)^{M+1-j}\binom{M+1}{j}$ automatically satisfies Eq.~\eqref{eq: contraint equ}. The fact that the change of particle at a given lattice site can be written as a discrete derivative of order $M+1$ explains why, in the continuum, the time evolution of the density field is given by a derivative of order $M+1$ in Eq.~\eqref{eq: general conservation equation}.

More generally, any kernel of length $n > M+1$ that conserves all moments up to order $M$ can be constructed from shifted copies of the finite difference operator along the lattice and their linear combinations. Concretely, if $c_0, \dots, c_{n-M-2}$ are arbitrary coefficients, the kernel
\begin{equation}
    \delta_j = \sum_{r=0}^{n-M-2} c_r\, k^{(M+1)}_{j-r}, \qquad j=1,\dots,n,
\end{equation}
satisfies all the constraints, with $k^{(M+1)}_i=0$ if $i<0$ or $i>M+1$ and $k^{(M+1)}_i=(-1)^{M+1-i}\binom{M+1}{i}$ otherwise. 

For example, $\bm{\mathcal L} = [1,-2,1]$ conserves all moments up to $M=1$. The most general solution with $n=5$ and $M+1$ can be obtained by:
\begin{equation}
    \bm{\mathcal L} = c_0[1,-2,1, 0, 0] + c_1[0,1,-2,1,0] + c_2[0,0,1,-2,1].
\end{equation}

\subsection{Dynamics conserving up to an even multipole moment}\label{app: even}
\begin{figure}
    \centering
    \includegraphics[width=0.999\linewidth]{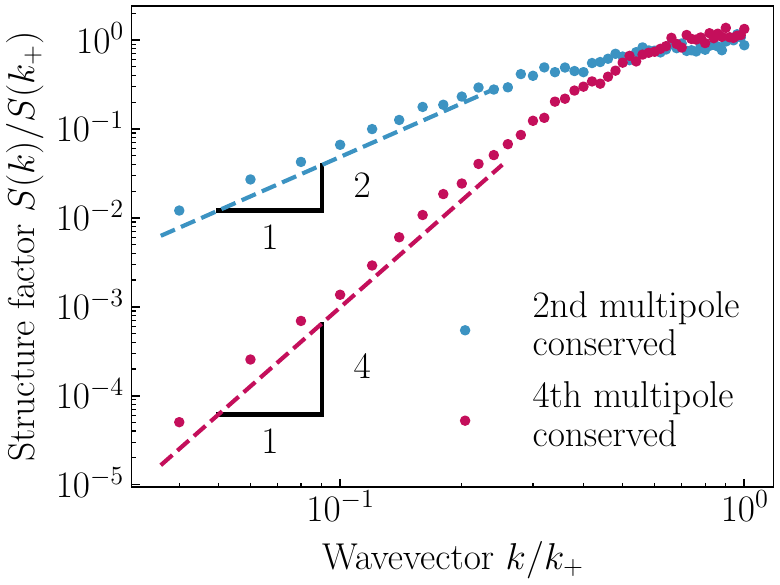}
    \caption{Structure factor of a 1D lattice simulation with kernels $\bm{\mathcal L}=[-1,  1,  3, -5,  2]$ and $\bm{\mathcal L}=[  1,  -3,   0,  10, -15,   9,  -2]$, conserving up to the second and fourth multipole moments, respectively. The dynamics is non-equilibrium with $\beta=0$.}
    \label{fig: evenMoments}
\end{figure}

In Fig.~\ref{fig: evenMoments}, we present simulation of two lattices with kernels $\bm{\mathcal L}=[-1,  1,  3, -5,  2]$ and $\bm{\mathcal L}=[  1,  -3,   0,  10, -15,   9,  -2]$ constructed using the method of Sec.~\ref{app: proof} and conserving, respectively, up to the 2nd and 4th multipole moment. We set $\beta=0$ to enforce a non-equilibrium dynamics from which we indeed observe a hyperuniform scaling: $S(k)\sim k^M$ with $M$ the highest order multipole conserved, in agreement with Eq.~\eqref{eq: main message structure factor}. 

The doubtful reader will rightfully remark that the kernel used are not the simplest ones that conserve up to 2nd and 4th multipole moments: $\bm{\mathcal L}=[-1,  3,  -3, 1]$ and $\bm{\mathcal L}=[ 1, -5, 10, -10, 5, -1]$, respectively.
We simulated these cases and found a flat structure factor, indicating that the lowest-order dissipative term in the density evolution are those expected for an equilibrium system, despite the dynamics being non-equilibrium.  Using the method of Sec.~\ref{app: detail lattice}, we correctly find a leading non-dissipative term of order $\partial_x^{M+1}$ but also a non-zero dissipative term of order $\partial_x^{M+2}$ instead of the numerically obtained $\partial_x^{2(M+1)}$. This suggests that our theory does not correctly capture next-to-leading-order terms.  

A possible explanation is that simple kernels such as $[-1,3,-3,1]$ lack sufficient dissipative character, since they correspond essentially to permutations of particles. In contrast, kernels like $[1,-2,1]$ represent particle splitting and redistribution, which generate genuine dissipation. Similarly, events such as $[1,-2,0,2,-1]$ or $[1,1,-8,8,-1,-1]$ (all conserving multipoles up to order 2) also yield flat structure factors, consistent with the interpretation that pure rearrangements without redistribution fail to produce effective dissipative terms.  

This phenomenon remains to be fully understood, but it appears to concern only a very restricted class of multipole-conserving kernels -- albeit a natural one to consider.

\section{Power counting of the modified conserved directed percolation equations}\label{app: power counting}

In this appendix, we assess the relevance of the density field in the modified conserved directed percolation equations Eq.~\eqref{eq: cdp mth multipole}.

We consider the scale transformations: $t\to l^zt$, $x\to lx$, $\rho_a\to l^{\chi_a}\rho_a$ and $\delta\rho\to l^{\chi}\delta\rho$. We want to find $z$, $\chi_a$ and $\chi$ such that the linear terms in Eq.~\eqref{eq: cdp} are scale invariant to place ourselves in the vincinity of the Gaussian fixed point. To this end, we set $a=0$ and assume that all coefficients ($b$, $\kappa$, \dots) are scale invariant, as expected at the Gaussian fixed point. This yields:
\begin{equation}
    \label{eq: cdp power counting}
    \begin{split}
        b^{\chi_a - z}\partial_t\rho_a &= l^{\chi_a+\chi}b\rho\rho_a - l^{2\chi_a}c\rho_a^2 + l^{-2 +\chi_a}\kappa\bm \nabla^{2}\rho_a +\\&\quad l^{(-d-2+\chi_a)/2}\sqrt{2D\rho_a}\eta,\\
        b^{\chi - z}\partial_t\delta\rho &= l^{-(M+1) +\chi_a}(-1)^{\frac{M-1}{2}}\kappa'\bm\nabla^{M+1}\rho_a.
    \end{split}
\end{equation}
Scale invariance of the linear terms require $z=2$, $\chi_a = 2 - d$, $\chi = 3 - M - d$. The term $c\rho_a^2$ is therefore irrelevant near the Gaussian fixed point for $d>4$, while the non-linearity coupling the two fields $b\rho_a\delta\rho$ is irrelevant for $d>5-M$. 

\section{Renormalization group generation of a lower order noise}\label{app: rg}
We now perform a partial renormalization group analysis, focusing on the renormalization of the noise term which arises at two loops in $\phi^4$ theories. We consider the general theory:
\begin{align}
    \partial_t\phi(\bm k, t) &= -|\bm k|^{2M}\dfrac{\delta F}{\delta \phi(-\bm k, t)}+\eta(\bm k, t),\nonumber\\
        F[\phi(\bm r)]&=\int \left(\dfrac{\tau}{2} \phi^2 + \dfrac{u}{4}\phi^4+\dfrac 1 2(\bm\nabla\phi)^2\right)d\bm r,\nonumber\\
    \dfrac{\langle \eta(\bm k, w)\eta(\bm k', w')\rangle}{(2\pi)^{d+1}} &= D|\bm k|^{2M'}|w|^{2\theta}\delta(\bm k + \bm k')\delta(w + w').
\end{align}
We introduce the bare propagator $G_0$, the bare vertex $g_0$, and the bare correlation function $C_0$:
\begin{equation}
    \begin{split}
        G_0(\bm k, w)&=\dfrac{1}{-iw +| \bm k|^{2M}(\tau+\bm k^2)},\\
        g_0(\bm k)&=-u|\bm k|^{2M},\\
        C_0(\bm k, w)&=2D_0(\bm k, w) |G_0(\bm k, w)|^2,\\
        D_0(\bm k, w) &=D|\bm k|^{2M'}|w|^{2\theta}.
    \end{split}    
\end{equation}
The noise in the correlator is renormalized at two loops by a sunset Feynman diagram made of bare correlator~\cite{tauber2014critical}:
\begin{align}
    \label{eq: renormalization correlator}
    C_R&(\bm k, w) = C_0(\bm k, w) + 12 g_0^2(\bm k)|G_0(\bm k, w)|^2\iint C_0(\bm q, w_{\bm q})\nonumber\\
   &\times C_0(\bm p, w_{\bm p})C_0(\bm k-\bm q - \bm p, w - w_{\bm q}-w_{\bm p})d \tilde{\bm q} d\tilde{\bm p}+ \mathcal G,
\end{align}
with $d\tilde{\bm p}=d\bm p dw_{\bm p}/(2\pi)^{d+1}$ and $\mathcal G$ denotes additional diagrams contributing only to the propagator renormalization at one and two loops. We define the renormalized correlator as $C_R(\bm k, w)=2\tilde D_R(\bm k, w)|G_R(\bm k, w)|^2$ with $G_R$ the renormalized propagator and $\tilde D_R$ proportional to the autocorrelation of the renormalized noise. From Eq.~\eqref{eq: renormalization correlator}, we obtain:
\begin{align}
    \label{eq: noise}
    &\tilde D_R(\bm k, w)= D_0(\bm k, w) + 48 u^2 |\bm k|^{4M}\iint D_0(\bm q, w_{\bm q})\times\nonumber\\
   & \quad D_0(\bm p, w_{\bm p})D_0(\bm k-\bm q - \bm p, w - w_{\bm q}-w_{\bm p}) \times\nonumber\\
   &\quad\dfrac{1}{w_{\bm q}^2 + |\bm q|^{2M}(\tau+\bm q^2)}\dfrac{1}{w_{\bm p}^2 + \bm p^{2M}(\tau+\bm p^2)} \\
   &\dfrac{d\tilde{\bm q} d\tilde{\bm p}}{(w-w_{\bm p}-w_{\bm q})^2 + |\bm k-\bm p-\bm q|^{2M}(\tau+ (\bm k-\bm p-\bm q)^2)}\nonumber.
\end{align}
In a first time, we consider the simpler case $\theta=0$ which corresponds to a noise delta correlated in time. The double integral in Eq.~\eqref{eq: noise} over $w_{\bm q}$ and $w_{\bm p}$ that we call $I$, is easily done. We also take the limit $w\to 0$ and obtain:
\begin{align}
    I(\bm k,0)=D^3\iint\frac{|\bm q|^{2M'}|\bm p|^{2M'}|\bm r|^{2M'}}{4 A_{\bm q}A_{\bm p} A_{\bm r}(A_{\bm q}+A_{\bm p}+A_{\bm r})}\frac{d \bm q}{(2\pi)^d}\frac{d \bm p}{(2\pi)^d},
\end{align}
with $A_{\bm q}=|\bm q|^{2M}(1+\bm q^2)$, and $\bm r = \bm k - \bm q - \bm p$. $I(\bm k\to 0, 0)$ has a well-defined non-zero limit with an integration region over large wavelengths, as usual in renormalization group computation. It might be necessary to include higher power in the denominator (corresponding to higher power of $(\bm \nabla \phi)^2$ in the free-energy) to make the integral well-behaved in the UV, in any case, we use a UV cut-off. Therefore, we find at lowest $\bm k$ order the renormalized noise:
\begin{equation}
    \label{eq: renormalized noise k only}
    \tilde D_R(\bm k, 0)= D|\bm k|^{2M'} + 12 u^2 D^3\left(I(0, 0) + \mathcal{O}(\bm k^2)\right)|\bm k|^{4M}.
\end{equation}
We first recall the equilibrium results $M=M'$. Model A is $M=0$ and the bare noise as well as the perturbative contribution both have leading order terms in $\bm k^0$, implying that the noise intensity $D$ simply gets renormalized. In model $B$, $M=1$ and the bare noise has a leading order term in $\bm k^2$, while the perturbative contribution has a leading order term in $\bm k^4$. The latter being less relevant than the former, we usually ignore this contribution if we are interested on the large scale behavior and $D$ gets no perturbative contribution, $D_0$ is therefore not renormalized~\cite{tauber2014critical}. However, in the case simulated in the main text with $M=0$ and $M'\leq 1$, an effective noise with correlation $\propto\bm k^0$ is generated which eventually completely changes the analysis of the linearized system as it prevents the hyperuniformity predicted in the linear regime since it acts as a white noise.

To correctly predict the structure factor, we should also check that the functional shape of the linear deterministic term is not changed and therefore that $\tau$ is simply renormalized. To check that, we use the renormalization of the propagator~\cite{tauber2014critical}:
\begin{align}
    G_R(\bm k, 0)&=G_0(\bm k, 0)-3u|\bm k|^{2M}G_0(\bm k, 0)^2\int C_0(\bm q,\bm w)d\tilde{\bm q}\nonumber\\
    &=G_0(\bm k, 0)-3u|\bm k|^{2M}G_0(\bm k, 0)^2\int S_0(\bm q) \dfrac{d\bm q}{(2\pi)^d},\nonumber\\
\end{align}
with $S_0$ the bare structure factor, the integral $I'=\int S_0(\bm q) {d\bm q}/{(2\pi)^d}$ over a ring is finite. Therefore, we obtain that the linear term is renormalized to:
\begin{equation}
    \tau_r=\tau+3uI'/\tau.
\end{equation}
No lower terms, such as $\tau_r = \tau+3uI'/\tau + \tau' \bm k^{-2}$, get generated.

With the noise correlation qualitatively modified in Eq.~\eqref{eq: noise 2} and the linear term only quantitatively modified, we find a structure factor in:
\begin{equation}
    S(\bm k)\sim |\bm k|^{2(\min(2M, M')-M)}.
\end{equation}
We note that if $M\leq M'\leq 2M$, the noise conserves lower or equal multipole moment that the deterministic term (as in equilibrium), then $S(k)=k^{2(M'-M)}$ which coincides with the prediction of linear theory. However, when $2M\leq M'$, the noise conserves higher moments than the deterministic term and the obtained structure factor is: $S(\bm k)\sim k^{2M}$, in disagreement with linear theory. In the linear regime, this would correspond to a situation in which both the noise and the deterministic term obey the same conservation law.

Let's go back to our original problem, with the full expression for the bare noise including temporal correlations:
\begin{align}
    \label{eq: noise 2}
    \tilde D_R(\bm k, w)&= D_0(\bm k, w) + 48 D^3 u^2 |\bm k|^{4M}\iint \dfrac{|\bm q|^{2M'}|w_{\bm{q}}|^{2\theta}}{w_{\bm q}^2 + A_{\bm q}}\times\nonumber\\
    &\quad\dfrac{|\bm p|^{2M'}|w_{\bm{p}}|^{2\theta}}{w_{\bm p}^2 + A_{\bm p}}\dfrac{|\bm r|^{2M'}|w_{\bm{r}}|^{2\theta}}{w_{\bm r}^2 + A_{\bm r}} d\tilde{\bm q} d\tilde{\bm p}.
\end{align}
In the limit $w\to 0$, Power counting tells us that the integral converges when $\theta<2/3$, moreover, the limit $\bm k\to 0$ of the integral over internal momenta has no reason to be non-zero. We therefore find again a result equivalent to Eq.~\eqref{eq: renormalized noise k only}, which holds  now, \textit{a priori} only for $\theta<2/3$. In this, case, a noise non-correlated in time is generated which can destroy the time correlated induced hyperuniformity found in the linear regime. We note that, once again, conservation laws can protect the bare noise to be renormalized, for example if $M'=M$ and $\theta<2/3$.

\section{Non-linear damping for the momentum field and loss of hyperuniformity}\label{app: nonlinearvelocity}

\begin{figure}
    \centering
    \includegraphics[width=0.999\linewidth]{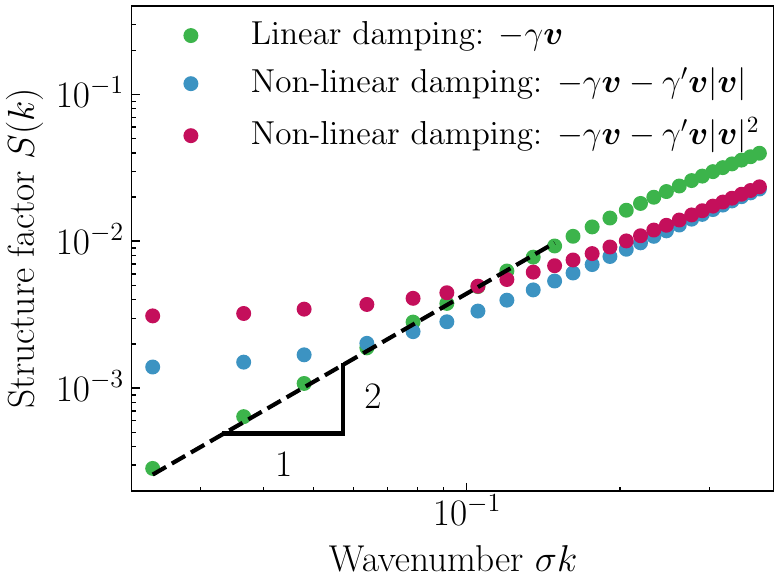}
    \caption{Structure factor from the molecular dynamics simulations of Eqs.~\eqref{eq: ran ni} and~\eqref{eq: non-linear damping} with $n=0$ (linear damping), $n=1$ and $n=2$. When the damping is non-linear, hyperuniformity is lost. $N = 100000$, $\phi=N\pi\sigma^2/(4L^2) = 0.4$, $\Delta E/(m\sigma^2\gamma^2) = 0.5$. We set $\gamma'\sigma=1$ for $n=1$ and $\gamma'(\sigma^2\gamma)=1$ for $n=2$.}
    \label{fig: speed}
\end{figure}

We simulate the model of Lei and Ni~\cite{lei2019hydrodynamics}, in which a collision between two particles $i$ and $j$ of mass $m$ and diameter $\sigma$ injects energy according to
\begin{equation}
\label{eq: ran ni}
\frac{1}{2}m \bm v_i'^2 + \frac{1}{2}m \bm v_j'^2 = \frac{1}{2}m \bm v_i^2 + \frac{1}{2}m \bm v_j^2 + \Delta E.
\end{equation}
To this model, we introduce a modification in the dissipative free-flight dynamics by including a nonlinear damping term:
\begin{equation}
\label{eq: non-linear damping}
\dot{\bm v}_i = -\gamma \bm v_i - \gamma' \bm v_i |\bm v_i|^n.
\end{equation}
When $n=0$, the damping is purely linear. For $n>0$, however, the damping couples modes in a way that breaks momentum-conservation (unlike, for example, the nonlinear advective term). As demonstrated in Fig.~\ref{fig: speed}, this nonlinear damping eliminates hyperuniformity.

A comprehensive theoretical analysis of this behavior would require a perturbative treatment of the compressible Navier–Stokes equations. However, the mechanism underlying the loss of hyperuniformity is most likely the same as that responsible for the emergence of the renormalized white noise in Eq.~\eqref{eq: non eq phi^4 renormalized}, we do not attempt the perturbative calculation.

\bibliography{bib}

\begin{thebibliography}{183}%
\makeatletter
\providecommand \@ifxundefined [1]{%
 \@ifx{#1\undefined}
}%
\providecommand \@ifnum [1]{%
 \ifnum #1\expandafter \@firstoftwo
 \else \expandafter \@secondoftwo
 \fi
}%
\providecommand \@ifx [1]{%
 \ifx #1\expandafter \@firstoftwo
 \else \expandafter \@secondoftwo
 \fi
}%
\providecommand \natexlab [1]{#1}%
\providecommand \enquote  [1]{``#1''}%
\providecommand \bibnamefont  [1]{#1}%
\providecommand \bibfnamefont [1]{#1}%
\providecommand \citenamefont [1]{#1}%
\providecommand \href@noop [0]{\@secondoftwo}%
\providecommand \href [0]{\begingroup \@sanitize@url \@href}%
\providecommand \@href[1]{\@@startlink{#1}\@@href}%
\providecommand \@@href[1]{\endgroup#1\@@endlink}%
\providecommand \@sanitize@url [0]{\catcode `\\12\catcode `\$12\catcode `\&12\catcode `\#12\catcode `\^12\catcode `\_12\catcode `\%12\relax}%
\providecommand \@@startlink[1]{}%
\providecommand \@@endlink[0]{}%
\providecommand \url  [0]{\begingroup\@sanitize@url \@url }%
\providecommand \@url [1]{\endgroup\@href {#1}{\urlprefix }}%
\providecommand \urlprefix  [0]{URL }%
\providecommand \Eprint [0]{\href }%
\providecommand \doibase [0]{https://doi.org/}%
\providecommand \selectlanguage [0]{\@gobble}%
\providecommand \bibinfo  [0]{\@secondoftwo}%
\providecommand \bibfield  [0]{\@secondoftwo}%
\providecommand \translation [1]{[#1]}%
\providecommand \BibitemOpen [0]{}%
\providecommand \bibitemStop [0]{}%
\providecommand \bibitemNoStop [0]{.\EOS\space}%
\providecommand \EOS [0]{\spacefactor3000\relax}%
\providecommand \BibitemShut  [1]{\csname bibitem#1\endcsname}%
\let\auto@bib@innerbib\@empty
\bibitem [{\citenamefont {Cavagna}\ \emph {et~al.}(2017)\citenamefont {Cavagna}, \citenamefont {Conti}, \citenamefont {Creato}, \citenamefont {Del~Castello}, \citenamefont {Giardina}, \citenamefont {Grigera}, \citenamefont {Melillo}, \citenamefont {Parisi},\ and\ \citenamefont {Viale}}]{cavagna2017dynamic}%
  \BibitemOpen
  \bibfield  {author} {\bibinfo {author} {\bibfnamefont {A.}~\bibnamefont {Cavagna}}, \bibinfo {author} {\bibfnamefont {D.}~\bibnamefont {Conti}}, \bibinfo {author} {\bibfnamefont {C.}~\bibnamefont {Creato}}, \bibinfo {author} {\bibfnamefont {L.}~\bibnamefont {Del~Castello}}, \bibinfo {author} {\bibfnamefont {I.}~\bibnamefont {Giardina}}, \bibinfo {author} {\bibfnamefont {T.~S.}\ \bibnamefont {Grigera}}, \bibinfo {author} {\bibfnamefont {S.}~\bibnamefont {Melillo}}, \bibinfo {author} {\bibfnamefont {L.}~\bibnamefont {Parisi}},\ and\ \bibinfo {author} {\bibfnamefont {M.}~\bibnamefont {Viale}},\ }\bibfield  {title} {\enquote {\bibinfo {title} {Dynamic scaling in natural swarms},}\ }\href {https://doi.org/10.1038/nphys4153} {\bibfield  {journal} {\bibinfo  {journal} {Nature Physics}\ }\textbf {\bibinfo {volume} {13}},\ \bibinfo {pages} {914--918} (\bibinfo {year} {2017})}\BibitemShut {NoStop}%
\bibitem [{\citenamefont {Mandelbrot}\ and\ \citenamefont {Hudson}(2007)}]{mandelbrot2007misbehavior}%
  \BibitemOpen
  \bibfield  {author} {\bibinfo {author} {\bibfnamefont {B.}~\bibnamefont {Mandelbrot}}\ and\ \bibinfo {author} {\bibfnamefont {R.~L.}\ \bibnamefont {Hudson}},\ }\href {https://www.hachettebookgroup.com/titles/benoit-mandelbrot/the-misbehavior-of-markets/9780465043576/} {\emph {\bibinfo {title} {The Misbehavior of Markets: A fractal view of financial turbulence}}}\ (\bibinfo  {publisher} {Basic books},\ \bibinfo {year} {2007})\BibitemShut {NoStop}%
\bibitem [{\citenamefont {Friedman}\ \emph {et~al.}(2012)\citenamefont {Friedman}, \citenamefont {Ito}, \citenamefont {Brinkman}, \citenamefont {Shimono}, \citenamefont {DeVille}, \citenamefont {Dahmen}, \citenamefont {Beggs},\ and\ \citenamefont {Butler}}]{friedman2012universal}%
  \BibitemOpen
  \bibfield  {author} {\bibinfo {author} {\bibfnamefont {N.}~\bibnamefont {Friedman}}, \bibinfo {author} {\bibfnamefont {S.}~\bibnamefont {Ito}}, \bibinfo {author} {\bibfnamefont {B.~A.}\ \bibnamefont {Brinkman}}, \bibinfo {author} {\bibfnamefont {M.}~\bibnamefont {Shimono}}, \bibinfo {author} {\bibfnamefont {R.~L.}\ \bibnamefont {DeVille}}, \bibinfo {author} {\bibfnamefont {K.~A.}\ \bibnamefont {Dahmen}}, \bibinfo {author} {\bibfnamefont {J.~M.}\ \bibnamefont {Beggs}},\ and\ \bibinfo {author} {\bibfnamefont {T.~C.}\ \bibnamefont {Butler}},\ }\bibfield  {title} {\enquote {\bibinfo {title} {Universal critical dynamics in high resolution neuronal avalanche data},}\ }\href {https://doi.org/10.1103/PhysRevLett.108.208102} {\bibfield  {journal} {\bibinfo  {journal} {Physical review letters}\ }\textbf {\bibinfo {volume} {108}},\ \bibinfo {pages} {208102} (\bibinfo {year} {2012})}\BibitemShut {NoStop}%
\bibitem [{\citenamefont {Chaikin}, \citenamefont {Lubensky},\ and\ \citenamefont {Witten}(1995)}]{chaikin1995principles}%
  \BibitemOpen
  \bibfield  {author} {\bibinfo {author} {\bibfnamefont {P.~M.}\ \bibnamefont {Chaikin}}, \bibinfo {author} {\bibfnamefont {T.~C.}\ \bibnamefont {Lubensky}},\ and\ \bibinfo {author} {\bibfnamefont {T.~A.}\ \bibnamefont {Witten}},\ }\href {https://doi.org/10.1063/1.2808258} {\emph {\bibinfo {title} {Principles of condensed matter physics}}},\ Vol.~\bibinfo {volume} {10}\ (\bibinfo  {publisher} {Cambridge university press Cambridge},\ \bibinfo {year} {1995})\BibitemShut {NoStop}%
\bibitem [{\citenamefont {Bak}, \citenamefont {Tang},\ and\ \citenamefont {Wiesenfeld}(1988)}]{bak1988self}%
  \BibitemOpen
  \bibfield  {author} {\bibinfo {author} {\bibfnamefont {P.}~\bibnamefont {Bak}}, \bibinfo {author} {\bibfnamefont {C.}~\bibnamefont {Tang}},\ and\ \bibinfo {author} {\bibfnamefont {K.}~\bibnamefont {Wiesenfeld}},\ }\bibfield  {title} {\enquote {\bibinfo {title} {Self-organized criticality},}\ }\href {https://doi.org/10.1016/S1474-6670(17)46153-2} {\bibfield  {journal} {\bibinfo  {journal} {Physical review A}\ }\textbf {\bibinfo {volume} {38}},\ \bibinfo {pages} {364} (\bibinfo {year} {1988})}\BibitemShut {NoStop}%
\bibitem [{\citenamefont {Lei}\ and\ \citenamefont {Ni}(2024)}]{lei2024non}%
  \BibitemOpen
  \bibfield  {author} {\bibinfo {author} {\bibfnamefont {Y.}~\bibnamefont {Lei}}\ and\ \bibinfo {author} {\bibfnamefont {R.}~\bibnamefont {Ni}},\ }\bibfield  {title} {\enquote {\bibinfo {title} {Non-equilibrium dynamic hyperuniform states},}\ }\href {https://doi.org/10.1088/1361-648X/ad83a0} {\bibfield  {journal} {\bibinfo  {journal} {Journal of Physics: Condensed Matter}\ }\textbf {\bibinfo {volume} {37}},\ \bibinfo {pages} {023004} (\bibinfo {year} {2024})}\BibitemShut {NoStop}%
\bibitem [{\citenamefont {Torquato}(2018)}]{torquato2018hyperuniform}%
  \BibitemOpen
  \bibfield  {author} {\bibinfo {author} {\bibfnamefont {S.}~\bibnamefont {Torquato}},\ }\bibfield  {title} {\enquote {\bibinfo {title} {Hyperuniform states of matter},}\ }\href {https://doi.org/10.1016/j.physrep.2018.03.001} {\bibfield  {journal} {\bibinfo  {journal} {Physics Reports}\ }\textbf {\bibinfo {volume} {745}},\ \bibinfo {pages} {1--95} (\bibinfo {year} {2018})}\BibitemShut {NoStop}%
\bibitem [{\citenamefont {Salvalaglio}\ \emph {et~al.}(2024)\citenamefont {Salvalaglio}, \citenamefont {Skinner}, \citenamefont {Dunkel},\ and\ \citenamefont {Voigt}}]{salvalaglio2024persistent}%
  \BibitemOpen
  \bibfield  {author} {\bibinfo {author} {\bibfnamefont {M.}~\bibnamefont {Salvalaglio}}, \bibinfo {author} {\bibfnamefont {D.~J.}\ \bibnamefont {Skinner}}, \bibinfo {author} {\bibfnamefont {J.}~\bibnamefont {Dunkel}},\ and\ \bibinfo {author} {\bibfnamefont {A.}~\bibnamefont {Voigt}},\ }\bibfield  {title} {\enquote {\bibinfo {title} {Persistent homology and topological statistics of hyperuniform point clouds},}\ }\href {https://doi.org/10.1103/PhysRevResearch.6.023107} {\bibfield  {journal} {\bibinfo  {journal} {Phys. Rev. Res.}\ }\textbf {\bibinfo {volume} {6}},\ \bibinfo {pages} {023107} (\bibinfo {year} {2024})}\BibitemShut {NoStop}%
\bibitem [{\citenamefont {Kim}\ and\ \citenamefont {Torquato}(2018)}]{kim2018effect}%
  \BibitemOpen
  \bibfield  {author} {\bibinfo {author} {\bibfnamefont {J.}~\bibnamefont {Kim}}\ and\ \bibinfo {author} {\bibfnamefont {S.}~\bibnamefont {Torquato}},\ }\bibfield  {title} {\enquote {\bibinfo {title} {Effect of imperfections on the hyperuniformity of many-body systems},}\ }\href {https://doi.org/10.1103/PhysRevB.97.054105} {\bibfield  {journal} {\bibinfo  {journal} {Physical Review B}\ }\textbf {\bibinfo {volume} {97}},\ \bibinfo {pages} {054105} (\bibinfo {year} {2018})}\BibitemShut {NoStop}%
\bibitem [{\citenamefont {Hansen}\ and\ \citenamefont {McDonald}(2013)}]{hansen2013theory}%
  \BibitemOpen
  \bibfield  {author} {\bibinfo {author} {\bibfnamefont {J.-P.}\ \bibnamefont {Hansen}}\ and\ \bibinfo {author} {\bibfnamefont {I.~R.}\ \bibnamefont {McDonald}},\ }\href {https://www.sciencedirect.com/book/9780123870322/theory-of-simple-liquids} {\emph {\bibinfo {title} {Theory of simple liquids: with applications to soft matter}}}\ (\bibinfo  {publisher} {Academic press},\ \bibinfo {year} {2013})\BibitemShut {NoStop}%
\bibitem [{\citenamefont {Hexner}\ and\ \citenamefont {Levine}(2017)}]{hexner2017noise}%
  \BibitemOpen
  \bibfield  {author} {\bibinfo {author} {\bibfnamefont {D.}~\bibnamefont {Hexner}}\ and\ \bibinfo {author} {\bibfnamefont {D.}~\bibnamefont {Levine}},\ }\bibfield  {title} {\enquote {\bibinfo {title} {Noise, diffusion, and hyperuniformity},}\ }\href {https://doi.org/10.1103/PhysRevLett.118.020601} {\bibfield  {journal} {\bibinfo  {journal} {Physical review letters}\ }\textbf {\bibinfo {volume} {118}},\ \bibinfo {pages} {020601} (\bibinfo {year} {2017})}\BibitemShut {NoStop}%
\bibitem [{\citenamefont {Ma}\ and\ \citenamefont {Torquato}(2019)}]{PhysRevE.99.022115}%
  \BibitemOpen
  \bibfield  {author} {\bibinfo {author} {\bibfnamefont {Z.}~\bibnamefont {Ma}}\ and\ \bibinfo {author} {\bibfnamefont {S.}~\bibnamefont {Torquato}},\ }\bibfield  {title} {\enquote {\bibinfo {title} {Hyperuniformity of generalized random organization models},}\ }\href {https://doi.org/10.1103/PhysRevE.99.022115} {\bibfield  {journal} {\bibinfo  {journal} {Phys. Rev. E}\ }\textbf {\bibinfo {volume} {99}},\ \bibinfo {pages} {022115} (\bibinfo {year} {2019})}\BibitemShut {NoStop}%
\bibitem [{\citenamefont {Bertrand}, \citenamefont {Chatenay},\ and\ \citenamefont {Voituriez}(2019)}]{bertrand2019nonlinear}%
  \BibitemOpen
  \bibfield  {author} {\bibinfo {author} {\bibfnamefont {T.}~\bibnamefont {Bertrand}}, \bibinfo {author} {\bibfnamefont {D.}~\bibnamefont {Chatenay}},\ and\ \bibinfo {author} {\bibfnamefont {R.}~\bibnamefont {Voituriez}},\ }\bibfield  {title} {\enquote {\bibinfo {title} {Nonlinear diffusion and hyperuniformity from poisson representation in systems with interaction mediated dynamics},}\ }\href {https://doi.org/https://doi.org/10.1088/1367-2630/ab5f17} {\bibfield  {journal} {\bibinfo  {journal} {New Journal of Physics}\ }\textbf {\bibinfo {volume} {21}},\ \bibinfo {pages} {123048} (\bibinfo {year} {2019})}\BibitemShut {NoStop}%
\bibitem [{\citenamefont {Mukherjee}\ \emph {et~al.}(2024)\citenamefont {Mukherjee}, \citenamefont {Tapader}, \citenamefont {Hazra},\ and\ \citenamefont {Pradhan}}]{mukherjee2024anomalous}%
  \BibitemOpen
  \bibfield  {author} {\bibinfo {author} {\bibfnamefont {A.}~\bibnamefont {Mukherjee}}, \bibinfo {author} {\bibfnamefont {D.}~\bibnamefont {Tapader}}, \bibinfo {author} {\bibfnamefont {A.}~\bibnamefont {Hazra}},\ and\ \bibinfo {author} {\bibfnamefont {P.}~\bibnamefont {Pradhan}},\ }\bibfield  {title} {\enquote {\bibinfo {title} {Anomalous relaxation and hyperuniform fluctuations in center-of-mass conserving systems with broken time-reversal symmetry},}\ }\href {https://doi.org/10.1103/PhysRevE.110.024119} {\bibfield  {journal} {\bibinfo  {journal} {Physical Review E}\ }\textbf {\bibinfo {volume} {110}},\ \bibinfo {pages} {024119} (\bibinfo {year} {2024})}\BibitemShut {NoStop}%
\bibitem [{\citenamefont {Hazra}, \citenamefont {Mukherjee},\ and\ \citenamefont {Pradhan}(2025)}]{hazra2025hyperuniformity}%
  \BibitemOpen
  \bibfield  {author} {\bibinfo {author} {\bibfnamefont {A.}~\bibnamefont {Hazra}}, \bibinfo {author} {\bibfnamefont {A.}~\bibnamefont {Mukherjee}},\ and\ \bibinfo {author} {\bibfnamefont {P.}~\bibnamefont {Pradhan}},\ }\bibfield  {title} {\enquote {\bibinfo {title} {Hyperuniformity in mass transport processes with center-of-mass conservation: some exact results},}\ }\href {https://doi.org/10.1088/1742-5468/ada88c} {\bibfield  {journal} {\bibinfo  {journal} {Journal of Statistical Mechanics: Theory and Experiment}\ }\textbf {\bibinfo {volume} {2025}},\ \bibinfo {pages} {023201} (\bibinfo {year} {2025})}\BibitemShut {NoStop}%
\bibitem [{\citenamefont {Dandekar}(2020)}]{dandekar2020exact}%
  \BibitemOpen
  \bibfield  {author} {\bibinfo {author} {\bibfnamefont {R.}~\bibnamefont {Dandekar}},\ }\bibfield  {title} {\enquote {\bibinfo {title} {Exact hyperuniformity exponents and entropy cusps in models of active-absorbing transition},}\ }\href {https://doi.org/10.1209/0295-5075/132/10008} {\bibfield  {journal} {\bibinfo  {journal} {Europhysics Letters}\ }\textbf {\bibinfo {volume} {132}},\ \bibinfo {pages} {10008} (\bibinfo {year} {2020})}\BibitemShut {NoStop}%
\bibitem [{\citenamefont {Kuroda}\ and\ \citenamefont {Miyazaki}(2023)}]{kuroda2023microscopic}%
  \BibitemOpen
  \bibfield  {author} {\bibinfo {author} {\bibfnamefont {Y.}~\bibnamefont {Kuroda}}\ and\ \bibinfo {author} {\bibfnamefont {K.}~\bibnamefont {Miyazaki}},\ }\bibfield  {title} {\enquote {\bibinfo {title} {Microscopic theory for hyperuniformity in two-dimensional chiral active fluid},}\ }\href {https://doi.org/10.1088/1742-5468/ad0639} {\bibfield  {journal} {\bibinfo  {journal} {Journal of Statistical Mechanics: Theory and Experiment}\ }\textbf {\bibinfo {volume} {2023}},\ \bibinfo {pages} {103203} (\bibinfo {year} {2023})}\BibitemShut {NoStop}%
\bibitem [{\citenamefont {Kuroda}, \citenamefont {Kawasaki},\ and\ \citenamefont {Miyazaki}(2025{\natexlab{a}})}]{kuroda2025singulardensitycorrelationschiral}%
  \BibitemOpen
  \bibfield  {author} {\bibinfo {author} {\bibfnamefont {Y.}~\bibnamefont {Kuroda}}, \bibinfo {author} {\bibfnamefont {T.}~\bibnamefont {Kawasaki}},\ and\ \bibinfo {author} {\bibfnamefont {K.}~\bibnamefont {Miyazaki}},\ }\href {https://arxiv.org/abs/2507.08770} {\enquote {\bibinfo {title} {Singular density correlations in chiral active fluids in three dimensions},}\ } (\bibinfo {year} {2025}{\natexlab{a}})\BibitemShut {NoStop}%
\bibitem [{\citenamefont {Keta}\ and\ \citenamefont {Henkes}(2025)}]{keta2024long}%
  \BibitemOpen
  \bibfield  {author} {\bibinfo {author} {\bibfnamefont {Y.-E.}\ \bibnamefont {Keta}}\ and\ \bibinfo {author} {\bibfnamefont {S.}~\bibnamefont {Henkes}},\ }\bibfield  {title} {\enquote {\bibinfo {title} {Long-range order in two-dimensional systems with fluctuating active stresses},}\ }\href {https://doi.org/10.1039/d5sm00208g} {\bibfield  {journal} {\bibinfo  {journal} {Soft Matter}\ ,\ \bibinfo {pages} {20}} (\bibinfo {year} {2025})}\BibitemShut {NoStop}%
\bibitem [{\citenamefont {Lei}\ and\ \citenamefont {Ni}(2019)}]{lei2019hydrodynamics}%
  \BibitemOpen
  \bibfield  {author} {\bibinfo {author} {\bibfnamefont {Q.-L.}\ \bibnamefont {Lei}}\ and\ \bibinfo {author} {\bibfnamefont {R.}~\bibnamefont {Ni}},\ }\bibfield  {title} {\enquote {\bibinfo {title} {Hydrodynamics of random-organizing hyperuniform fluids},}\ }\href {https://doi.org/10.1073/pnas.1911596116} {\bibfield  {journal} {\bibinfo  {journal} {Proceedings of the National Academy of Sciences}\ }\textbf {\bibinfo {volume} {116}},\ \bibinfo {pages} {22983--22989} (\bibinfo {year} {2019})}\BibitemShut {NoStop}%
\bibitem [{\citenamefont {Lei}, \citenamefont {Ciamarra},\ and\ \citenamefont {Ni}(2019)}]{lei2019nonequilibrium}%
  \BibitemOpen
  \bibfield  {author} {\bibinfo {author} {\bibfnamefont {Q.-L.}\ \bibnamefont {Lei}}, \bibinfo {author} {\bibfnamefont {M.~P.}\ \bibnamefont {Ciamarra}},\ and\ \bibinfo {author} {\bibfnamefont {R.}~\bibnamefont {Ni}},\ }\bibfield  {title} {\enquote {\bibinfo {title} {Nonequilibrium strongly hyperuniform fluids of circle active particles with large local density fluctuations},}\ }\href {https://doi.org/10.1126/sciadv.aau7423} {\bibfield  {journal} {\bibinfo  {journal} {Science advances}\ }\textbf {\bibinfo {volume} {5}},\ \bibinfo {pages} {eaau7423} (\bibinfo {year} {2019})}\BibitemShut {NoStop}%
\bibitem [{\citenamefont {Lei}, \citenamefont {Zheng},\ and\ \citenamefont {Ni}(2023)}]{lei2023random}%
  \BibitemOpen
  \bibfield  {author} {\bibinfo {author} {\bibfnamefont {Y.}~\bibnamefont {Lei}}, \bibinfo {author} {\bibfnamefont {N.}~\bibnamefont {Zheng}},\ and\ \bibinfo {author} {\bibfnamefont {R.}~\bibnamefont {Ni}},\ }\bibfield  {title} {\enquote {\bibinfo {title} {Random organization and non-equilibrium hyperuniform fluids on a sphere},}\ }\href {https://doi.org/10.1063/5.0165527} {\bibfield  {journal} {\bibinfo  {journal} {The Journal of Chemical Physics}\ }\textbf {\bibinfo {volume} {159}} (\bibinfo {year} {2023})}\BibitemShut {NoStop}%
\bibitem [{\citenamefont {Maire}\ \emph {et~al.}(2025{\natexlab{a}})\citenamefont {Maire}, \citenamefont {Plati}, \citenamefont {Smallenburg},\ and\ \citenamefont {Foffi}}]{maire2025dynamical}%
  \BibitemOpen
  \bibfield  {author} {\bibinfo {author} {\bibfnamefont {R.}~\bibnamefont {Maire}}, \bibinfo {author} {\bibfnamefont {A.}~\bibnamefont {Plati}}, \bibinfo {author} {\bibfnamefont {F.}~\bibnamefont {Smallenburg}},\ and\ \bibinfo {author} {\bibfnamefont {G.}~\bibnamefont {Foffi}},\ }\bibfield  {title} {\enquote {\bibinfo {title} {Dynamical and structural properties of an absorbing phase transition: a case study from granular systems},}\ }\href {https://doi.org/10.48550/arXiv.2507.06083} {\bibfield  {journal} {\bibinfo  {journal} {arXiv:2507.06083}\ } (\bibinfo {year} {2025}{\natexlab{a}})}\BibitemShut {NoStop}%
\bibitem [{\citenamefont {Li}, \citenamefont {Lei},\ and\ \citenamefont {Ma}(2025)}]{li2025fluidization}%
  \BibitemOpen
  \bibfield  {author} {\bibinfo {author} {\bibfnamefont {Z.-Q.}\ \bibnamefont {Li}}, \bibinfo {author} {\bibfnamefont {Q.-L.}\ \bibnamefont {Lei}},\ and\ \bibinfo {author} {\bibfnamefont {Y.-Q.}\ \bibnamefont {Ma}},\ }\bibfield  {title} {\enquote {\bibinfo {title} {Fluidization and anomalous density fluctuations in 2d voronoi cell tissues with pulsating activity},}\ }\href {https://doi.org/10.1073/pnas.2421518122} {\bibfield  {journal} {\bibinfo  {journal} {Proceedings of the National Academy of Sciences}\ }\textbf {\bibinfo {volume} {122}},\ \bibinfo {pages} {e2421518122} (\bibinfo {year} {2025})}\BibitemShut {NoStop}%
\bibitem [{\citenamefont {Gao}\ \emph {et~al.}(2025)\citenamefont {Gao}, \citenamefont {Shang}, \citenamefont {Hu}, \citenamefont {Ma},\ and\ \citenamefont {Lei}}]{gao2025liquidgascriticalityhyperuniformfluids}%
  \BibitemOpen
  \bibfield  {author} {\bibinfo {author} {\bibfnamefont {S.}~\bibnamefont {Gao}}, \bibinfo {author} {\bibfnamefont {H.}~\bibnamefont {Shang}}, \bibinfo {author} {\bibfnamefont {H.}~\bibnamefont {Hu}}, \bibinfo {author} {\bibfnamefont {Y.-Q.}\ \bibnamefont {Ma}},\ and\ \bibinfo {author} {\bibfnamefont {Q.-L.}\ \bibnamefont {Lei}},\ }\href {https://arxiv.org/abs/2507.06023} {\enquote {\bibinfo {title} {Liquid-gas criticality of hyperuniform fluids},}\ } (\bibinfo {year} {2025})\BibitemShut {NoStop}%
\bibitem [{\citenamefont {Liu}\ \emph {et~al.}(2023)\citenamefont {Liu}, \citenamefont {Gong}, \citenamefont {Yang},\ and\ \citenamefont {Chen}}]{liu2023local}%
  \BibitemOpen
  \bibfield  {author} {\bibinfo {author} {\bibfnamefont {R.}~\bibnamefont {Liu}}, \bibinfo {author} {\bibfnamefont {J.}~\bibnamefont {Gong}}, \bibinfo {author} {\bibfnamefont {M.}~\bibnamefont {Yang}},\ and\ \bibinfo {author} {\bibfnamefont {K.}~\bibnamefont {Chen}},\ }\bibfield  {title} {\enquote {\bibinfo {title} {Local rotational jamming and multi-stage hyperuniformities in an active spinner system},}\ }\href {https://doi.org/10.1088/0256-307x/40/12/126402} {\bibfield  {journal} {\bibinfo  {journal} {Chinese Physics Letters}\ }\textbf {\bibinfo {volume} {40}},\ \bibinfo {pages} {126402} (\bibinfo {year} {2023})}\BibitemShut {NoStop}%
\bibitem [{\citenamefont {Liu}, \citenamefont {Yang},\ and\ \citenamefont {Chen}(2025)}]{liu2025hyperuniform}%
  \BibitemOpen
  \bibfield  {author} {\bibinfo {author} {\bibfnamefont {R.}~\bibnamefont {Liu}}, \bibinfo {author} {\bibfnamefont {M.}~\bibnamefont {Yang}},\ and\ \bibinfo {author} {\bibfnamefont {K.}~\bibnamefont {Chen}},\ }\bibfield  {title} {\enquote {\bibinfo {title} {Hyperuniform mixing of binary active spinners},}\ }\href {https://doi.org/10.1039/D5SM00458F} {\bibfield  {journal} {\bibinfo  {journal} {arXiv:2504.17197}\ } (\bibinfo {year} {2025}),\ 10.1039/D5SM00458F}\BibitemShut {NoStop}%
\bibitem [{\citenamefont {De~Luca}\ \emph {et~al.}(2024)\citenamefont {De~Luca}, \citenamefont {Ma}, \citenamefont {Nardini},\ and\ \citenamefont {Cates}}]{de2024hyperuniformity}%
  \BibitemOpen
  \bibfield  {author} {\bibinfo {author} {\bibfnamefont {F.}~\bibnamefont {De~Luca}}, \bibinfo {author} {\bibfnamefont {X.}~\bibnamefont {Ma}}, \bibinfo {author} {\bibfnamefont {C.}~\bibnamefont {Nardini}},\ and\ \bibinfo {author} {\bibfnamefont {M.~E.}\ \bibnamefont {Cates}},\ }\bibfield  {title} {\enquote {\bibinfo {title} {Hyperuniformity in phase ordering: the roles of activity, noise, and non-constant mobility},}\ }\href {https://doi.org/10.1088/1361-648X/ad5b45} {\bibfield  {journal} {\bibinfo  {journal} {Journal of Physics: Condensed Matter}\ }\textbf {\bibinfo {volume} {36}},\ \bibinfo {pages} {405101} (\bibinfo {year} {2024})}\BibitemShut {NoStop}%
\bibitem [{\citenamefont {Padhan}\ and\ \citenamefont {Voigt}(2025{\natexlab{a}})}]{padhan2025suppression}%
  \BibitemOpen
  \bibfield  {author} {\bibinfo {author} {\bibfnamefont {N.~B.}\ \bibnamefont {Padhan}}\ and\ \bibinfo {author} {\bibfnamefont {A.}~\bibnamefont {Voigt}},\ }\bibfield  {title} {\enquote {\bibinfo {title} {Suppression of hyperuniformity in hydrodynamic scalar active field theories},}\ }\href {https://doi.org/10.1088/1361-648X/ada243} {\bibfield  {journal} {\bibinfo  {journal} {Journal of Physics: Condensed Matter}\ }\textbf {\bibinfo {volume} {37}},\ \bibinfo {pages} {105101} (\bibinfo {year} {2025}{\natexlab{a}})}\BibitemShut {NoStop}%
\bibitem [{\citenamefont {Padhan}\ and\ \citenamefont {Voigt}(2025{\natexlab{b}})}]{padhan2025hyperuniformity}%
  \BibitemOpen
  \bibfield  {author} {\bibinfo {author} {\bibfnamefont {N.~B.}\ \bibnamefont {Padhan}}\ and\ \bibinfo {author} {\bibfnamefont {A.}~\bibnamefont {Voigt}},\ }\bibfield  {title} {\enquote {\bibinfo {title} {Hyperuniformity in ternary fluid mixtures: the role of wetting and hydrodynamics},}\ }\href@noop {} {\bibfield  {journal} {\bibinfo  {journal} {arXiv:2506.22647}\ } (\bibinfo {year} {2025}{\natexlab{b}})}\BibitemShut {NoStop}%
\bibitem [{\citenamefont {Wilken}, \citenamefont {Chaderjian},\ and\ \citenamefont {Saleh}(2023)}]{wilken2023spatial}%
  \BibitemOpen
  \bibfield  {author} {\bibinfo {author} {\bibfnamefont {S.}~\bibnamefont {Wilken}}, \bibinfo {author} {\bibfnamefont {A.}~\bibnamefont {Chaderjian}},\ and\ \bibinfo {author} {\bibfnamefont {O.~A.}\ \bibnamefont {Saleh}},\ }\bibfield  {title} {\enquote {\bibinfo {title} {Spatial organization of phase-separated dna droplets},}\ }\href {https://doi.org/10.1103/PhysRevX.13.031014} {\bibfield  {journal} {\bibinfo  {journal} {Physical Review X}\ }\textbf {\bibinfo {volume} {13}},\ \bibinfo {pages} {031014} (\bibinfo {year} {2023})}\BibitemShut {NoStop}%
\bibitem [{\citenamefont {Zheng}, \citenamefont {Klatt},\ and\ \citenamefont {L{\"o}wen}(2024)}]{zheng2024universal}%
  \BibitemOpen
  \bibfield  {author} {\bibinfo {author} {\bibfnamefont {Y.}~\bibnamefont {Zheng}}, \bibinfo {author} {\bibfnamefont {M.~A.}\ \bibnamefont {Klatt}},\ and\ \bibinfo {author} {\bibfnamefont {H.}~\bibnamefont {L{\"o}wen}},\ }\bibfield  {title} {\enquote {\bibinfo {title} {Universal hyperuniformity in active field theories},}\ }\href {https://doi.org/10.1103/PhysRevResearch.6.L032056} {\bibfield  {journal} {\bibinfo  {journal} {Physical Review Research}\ }\textbf {\bibinfo {volume} {6}},\ \bibinfo {pages} {L032056} (\bibinfo {year} {2024})}\BibitemShut {NoStop}%
\bibitem [{\citenamefont {Salvalaglio}\ \emph {et~al.}(2020)\citenamefont {Salvalaglio}, \citenamefont {Bouabdellaoui}, \citenamefont {Bollani}, \citenamefont {Benali}, \citenamefont {Favre}, \citenamefont {Claude}, \citenamefont {Wenger}, \citenamefont {de~Anna}, \citenamefont {Intonti}, \citenamefont {Voigt} \emph {et~al.}}]{salvalaglio2020hyperuniform}%
  \BibitemOpen
  \bibfield  {author} {\bibinfo {author} {\bibfnamefont {M.}~\bibnamefont {Salvalaglio}}, \bibinfo {author} {\bibfnamefont {M.}~\bibnamefont {Bouabdellaoui}}, \bibinfo {author} {\bibfnamefont {M.}~\bibnamefont {Bollani}}, \bibinfo {author} {\bibfnamefont {A.}~\bibnamefont {Benali}}, \bibinfo {author} {\bibfnamefont {L.}~\bibnamefont {Favre}}, \bibinfo {author} {\bibfnamefont {J.-B.}\ \bibnamefont {Claude}}, \bibinfo {author} {\bibfnamefont {J.}~\bibnamefont {Wenger}}, \bibinfo {author} {\bibfnamefont {P.}~\bibnamefont {de~Anna}}, \bibinfo {author} {\bibfnamefont {F.}~\bibnamefont {Intonti}}, \bibinfo {author} {\bibfnamefont {A.}~\bibnamefont {Voigt}}, \emph {et~al.},\ }\bibfield  {title} {\enquote {\bibinfo {title} {Hyperuniform monocrystalline structures by spinodal solid-state dewetting},}\ }\href {https://doi.org/10.1103/PhysRevLett.125.126101} {\bibfield  {journal} {\bibinfo  {journal} {Physical review letters}\ }\textbf {\bibinfo {volume} {125}},\ \bibinfo {pages} {126101} (\bibinfo {year}
  {2020})}\BibitemShut {NoStop}%
\bibitem [{\citenamefont {Oppenheimer}\ \emph {et~al.}(2022)\citenamefont {Oppenheimer}, \citenamefont {Stein}, \citenamefont {Zion},\ and\ \citenamefont {Shelley}}]{oppenheimer2022hyperuniformity}%
  \BibitemOpen
  \bibfield  {author} {\bibinfo {author} {\bibfnamefont {N.}~\bibnamefont {Oppenheimer}}, \bibinfo {author} {\bibfnamefont {D.~B.}\ \bibnamefont {Stein}}, \bibinfo {author} {\bibfnamefont {M.~Y.~B.}\ \bibnamefont {Zion}},\ and\ \bibinfo {author} {\bibfnamefont {M.~J.}\ \bibnamefont {Shelley}},\ }\bibfield  {title} {\enquote {\bibinfo {title} {Hyperuniformity and phase enrichment in vortex and rotor assemblies},}\ }\href {https://doi.org/10.1038/s41467-022-28375-9} {\bibfield  {journal} {\bibinfo  {journal} {Nature communications}\ }\textbf {\bibinfo {volume} {13}},\ \bibinfo {pages} {804} (\bibinfo {year} {2022})}\BibitemShut {NoStop}%
\bibitem [{\citenamefont {Huang}\ \emph {et~al.}(2021)\citenamefont {Huang}, \citenamefont {Hu}, \citenamefont {Yang}, \citenamefont {Liu},\ and\ \citenamefont {Zhang}}]{huang2021circular}%
  \BibitemOpen
  \bibfield  {author} {\bibinfo {author} {\bibfnamefont {M.}~\bibnamefont {Huang}}, \bibinfo {author} {\bibfnamefont {W.}~\bibnamefont {Hu}}, \bibinfo {author} {\bibfnamefont {S.}~\bibnamefont {Yang}}, \bibinfo {author} {\bibfnamefont {Q.-X.}\ \bibnamefont {Liu}},\ and\ \bibinfo {author} {\bibfnamefont {H.}~\bibnamefont {Zhang}},\ }\bibfield  {title} {\enquote {\bibinfo {title} {Circular swimming motility and disordered hyperuniform state in an algae system},}\ }\href {https://doi.org/10.1073/pnas.2100493118} {\bibfield  {journal} {\bibinfo  {journal} {Proceedings of the National Academy of Sciences}\ }\textbf {\bibinfo {volume} {118}},\ \bibinfo {pages} {e2100493118} (\bibinfo {year} {2021})}\BibitemShut {NoStop}%
\bibitem [{\citenamefont {Levesque}, \citenamefont {Weis},\ and\ \citenamefont {Lebowitz}(2000)}]{levesque2000charge}%
  \BibitemOpen
  \bibfield  {author} {\bibinfo {author} {\bibfnamefont {D.}~\bibnamefont {Levesque}}, \bibinfo {author} {\bibfnamefont {J.-J.}\ \bibnamefont {Weis}},\ and\ \bibinfo {author} {\bibfnamefont {J.}~\bibnamefont {Lebowitz}},\ }\bibfield  {title} {\enquote {\bibinfo {title} {Charge fluctuations in the two-dimensional one-component plasma},}\ }\href {https://doi.org/10.1023/A:1018643829340} {\bibfield  {journal} {\bibinfo  {journal} {Journal of Statistical Physics}\ }\textbf {\bibinfo {volume} {100}},\ \bibinfo {pages} {209--222} (\bibinfo {year} {2000})}\BibitemShut {NoStop}%
\bibitem [{\citenamefont {Ganguly}\ and\ \citenamefont {Sarkar}(2020)}]{ganguly2020ground}%
  \BibitemOpen
  \bibfield  {author} {\bibinfo {author} {\bibfnamefont {S.}~\bibnamefont {Ganguly}}\ and\ \bibinfo {author} {\bibfnamefont {S.}~\bibnamefont {Sarkar}},\ }\bibfield  {title} {\enquote {\bibinfo {title} {Ground states and hyperuniformity of the hierarchical coulomb gas in all dimensions},}\ }\href {https://doi.org/10.1007/s00440-019-00955-9} {\bibfield  {journal} {\bibinfo  {journal} {Probability Theory and Related Fields}\ }\textbf {\bibinfo {volume} {177}},\ \bibinfo {pages} {621--675} (\bibinfo {year} {2020})}\BibitemShut {NoStop}%
\bibitem [{\citenamefont {Yashunsky}\ \emph {et~al.}(2024)\citenamefont {Yashunsky}, \citenamefont {Pearce}, \citenamefont {Ariel},\ and\ \citenamefont {Be’er}}]{yashunsky2024topological}%
  \BibitemOpen
  \bibfield  {author} {\bibinfo {author} {\bibfnamefont {V.}~\bibnamefont {Yashunsky}}, \bibinfo {author} {\bibfnamefont {D.~J.}\ \bibnamefont {Pearce}}, \bibinfo {author} {\bibfnamefont {G.}~\bibnamefont {Ariel}},\ and\ \bibinfo {author} {\bibfnamefont {A.}~\bibnamefont {Be’er}},\ }\bibfield  {title} {\enquote {\bibinfo {title} {Topological defects in multi-layered swarming bacteria},}\ }\href {https://doi.org/10.1039/D4SM00038B} {\bibfield  {journal} {\bibinfo  {journal} {Soft Matter}\ }\textbf {\bibinfo {volume} {20}},\ \bibinfo {pages} {4237--4245} (\bibinfo {year} {2024})}\BibitemShut {NoStop}%
\bibitem [{\citenamefont {Backofen}\ \emph {et~al.}(2024)\citenamefont {Backofen}, \citenamefont {Altawil}, \citenamefont {Salvalaglio},\ and\ \citenamefont {Voigt}}]{backofen2024nonequilibrium}%
  \BibitemOpen
  \bibfield  {author} {\bibinfo {author} {\bibfnamefont {R.}~\bibnamefont {Backofen}}, \bibinfo {author} {\bibfnamefont {A.~Y.}\ \bibnamefont {Altawil}}, \bibinfo {author} {\bibfnamefont {M.}~\bibnamefont {Salvalaglio}},\ and\ \bibinfo {author} {\bibfnamefont {A.}~\bibnamefont {Voigt}},\ }\bibfield  {title} {\enquote {\bibinfo {title} {Nonequilibrium hyperuniform states in active turbulence},}\ }\href {https://doi.org/10.1073/pnas.2320719121} {\bibfield  {journal} {\bibinfo  {journal} {Proceedings of the National Academy of Sciences}\ }\textbf {\bibinfo {volume} {121}},\ \bibinfo {pages} {e2320719121} (\bibinfo {year} {2024})}\BibitemShut {NoStop}%
\bibitem [{\citenamefont {Nizam}\ \emph {et~al.}(2021)\citenamefont {Nizam}, \citenamefont {Makey}, \citenamefont {Barbier}, \citenamefont {Kahraman}, \citenamefont {Demir}, \citenamefont {Shafigh}, \citenamefont {Galioglu}, \citenamefont {Vahabli}, \citenamefont {H{\"u}sn{\"u}gil}, \citenamefont {G{\"u}ne{\c{s}}} \emph {et~al.}}]{nizam2021dynamic}%
  \BibitemOpen
  \bibfield  {author} {\bibinfo {author} {\bibfnamefont {{\"U}.~S.}\ \bibnamefont {Nizam}}, \bibinfo {author} {\bibfnamefont {G.}~\bibnamefont {Makey}}, \bibinfo {author} {\bibfnamefont {M.}~\bibnamefont {Barbier}}, \bibinfo {author} {\bibfnamefont {S.~S.}\ \bibnamefont {Kahraman}}, \bibinfo {author} {\bibfnamefont {E.}~\bibnamefont {Demir}}, \bibinfo {author} {\bibfnamefont {E.~E.}\ \bibnamefont {Shafigh}}, \bibinfo {author} {\bibfnamefont {S.}~\bibnamefont {Galioglu}}, \bibinfo {author} {\bibfnamefont {D.}~\bibnamefont {Vahabli}}, \bibinfo {author} {\bibfnamefont {S.}~\bibnamefont {H{\"u}sn{\"u}gil}}, \bibinfo {author} {\bibfnamefont {M.~H.}\ \bibnamefont {G{\"u}ne{\c{s}}}}, \emph {et~al.},\ }\bibfield  {title} {\enquote {\bibinfo {title} {Dynamic evolution of hyperuniformity in a driven dissipative colloidal system},}\ }\href {https://doi.org/10.1088/1361-648X/abf9b8} {\bibfield  {journal} {\bibinfo  {journal} {Journal of Physics: Condensed Matter}\ }\textbf {\bibinfo {volume} {33}},\ \bibinfo {pages}
  {304002} (\bibinfo {year} {2021})}\BibitemShut {NoStop}%
\bibitem [{\citenamefont {Thambi}\ and\ \citenamefont {Uspal}(2025)}]{thambi2025clustering}%
  \BibitemOpen
  \bibfield  {author} {\bibinfo {author} {\bibfnamefont {A.~G.}\ \bibnamefont {Thambi}}\ and\ \bibinfo {author} {\bibfnamefont {W.~E.}\ \bibnamefont {Uspal}},\ }\bibfield  {title} {\enquote {\bibinfo {title} {Clustering and emergent hyperuniformity by breaking microswimmer shape and actuation symmetries},}\ }\href {https://doi.org/10.48550/arXiv.2506.12293} {\bibfield  {journal} {\bibinfo  {journal} {arXiv:2506.12293}\ } (\bibinfo {year} {2025})}\BibitemShut {NoStop}%
\bibitem [{\citenamefont {Hexner}\ and\ \citenamefont {Levine}(2015)}]{hexner2015hyperuniformity}%
  \BibitemOpen
  \bibfield  {author} {\bibinfo {author} {\bibfnamefont {D.}~\bibnamefont {Hexner}}\ and\ \bibinfo {author} {\bibfnamefont {D.}~\bibnamefont {Levine}},\ }\bibfield  {title} {\enquote {\bibinfo {title} {Hyperuniformity of critical absorbing states},}\ }\href {https://doi.org/10.1103/PhysRevLett.114.110602} {\bibfield  {journal} {\bibinfo  {journal} {Physical review letters}\ }\textbf {\bibinfo {volume} {114}},\ \bibinfo {pages} {110602} (\bibinfo {year} {2015})}\BibitemShut {NoStop}%
\bibitem [{\citenamefont {Mitra}\ \emph {et~al.}(2021)\citenamefont {Mitra}, \citenamefont {Parmar}, \citenamefont {Leishangthem}, \citenamefont {Sastry},\ and\ \citenamefont {Foffi}}]{mitra2021hyperuniformity}%
  \BibitemOpen
  \bibfield  {author} {\bibinfo {author} {\bibfnamefont {S.}~\bibnamefont {Mitra}}, \bibinfo {author} {\bibfnamefont {A.~D.}\ \bibnamefont {Parmar}}, \bibinfo {author} {\bibfnamefont {P.}~\bibnamefont {Leishangthem}}, \bibinfo {author} {\bibfnamefont {S.}~\bibnamefont {Sastry}},\ and\ \bibinfo {author} {\bibfnamefont {G.}~\bibnamefont {Foffi}},\ }\bibfield  {title} {\enquote {\bibinfo {title} {Hyperuniformity in cyclically driven glasses},}\ }\href {https://doi.org/10.1088/1742-5468/abdeb0} {\bibfield  {journal} {\bibinfo  {journal} {Journal of Statistical Mechanics: Theory and Experiment}\ }\textbf {\bibinfo {volume} {2021}},\ \bibinfo {pages} {033203} (\bibinfo {year} {2021})}\BibitemShut {NoStop}%
\bibitem [{\citenamefont {Anand}, \citenamefont {Zhang},\ and\ \citenamefont {Martiniani}(2025)}]{anand2025emergent}%
  \BibitemOpen
  \bibfield  {author} {\bibinfo {author} {\bibfnamefont {S.}~\bibnamefont {Anand}}, \bibinfo {author} {\bibfnamefont {G.}~\bibnamefont {Zhang}},\ and\ \bibinfo {author} {\bibfnamefont {S.}~\bibnamefont {Martiniani}},\ }\bibfield  {title} {\enquote {\bibinfo {title} {Emergent universal long-range structure in random-organizing systems},}\ }\href {https://doi.org/10.48550/arXiv.2505.22933} {\bibfield  {journal} {\bibinfo  {journal} {arXiv:2505.22933}\ } (\bibinfo {year} {2025})}\BibitemShut {NoStop}%
\bibitem [{\citenamefont {Chen}\ \emph {et~al.}(2024)\citenamefont {Chen}, \citenamefont {Lei}, \citenamefont {Xiang}, \citenamefont {Duan}, \citenamefont {Peng},\ and\ \citenamefont {Zhang}}]{chen2024emergent}%
  \BibitemOpen
  \bibfield  {author} {\bibinfo {author} {\bibfnamefont {J.}~\bibnamefont {Chen}}, \bibinfo {author} {\bibfnamefont {X.}~\bibnamefont {Lei}}, \bibinfo {author} {\bibfnamefont {Y.}~\bibnamefont {Xiang}}, \bibinfo {author} {\bibfnamefont {M.}~\bibnamefont {Duan}}, \bibinfo {author} {\bibfnamefont {X.}~\bibnamefont {Peng}},\ and\ \bibinfo {author} {\bibfnamefont {H.}~\bibnamefont {Zhang}},\ }\bibfield  {title} {\enquote {\bibinfo {title} {Emergent chirality and hyperuniformity in an active mixture with nonreciprocal interactions},}\ }\href {https://doi.org/10.1103/PhysRevLett.132.118301} {\bibfield  {journal} {\bibinfo  {journal} {Physical Review Letters}\ }\textbf {\bibinfo {volume} {132}},\ \bibinfo {pages} {118301} (\bibinfo {year} {2024})}\BibitemShut {NoStop}%
\bibitem [{\citenamefont {Wilken}\ \emph {et~al.}(2021)\citenamefont {Wilken}, \citenamefont {Guerra}, \citenamefont {Levine},\ and\ \citenamefont {Chaikin}}]{wilken2021random}%
  \BibitemOpen
  \bibfield  {author} {\bibinfo {author} {\bibfnamefont {S.}~\bibnamefont {Wilken}}, \bibinfo {author} {\bibfnamefont {R.~E.}\ \bibnamefont {Guerra}}, \bibinfo {author} {\bibfnamefont {D.}~\bibnamefont {Levine}},\ and\ \bibinfo {author} {\bibfnamefont {P.~M.}\ \bibnamefont {Chaikin}},\ }\bibfield  {title} {\enquote {\bibinfo {title} {Random close packing as a dynamical phase transition},}\ }\href {https://doi.org/10.1103/PhysRevLett.127.038002} {\bibfield  {journal} {\bibinfo  {journal} {Phys. Rev. Lett.}\ }\textbf {\bibinfo {volume} {127}},\ \bibinfo {pages} {038002} (\bibinfo {year} {2021})}\BibitemShut {NoStop}%
\bibitem [{\citenamefont {Wilken}\ \emph {et~al.}(2020)\citenamefont {Wilken}, \citenamefont {Guerra}, \citenamefont {Pine},\ and\ \citenamefont {Chaikin}}]{wilken2020hyperuniform}%
  \BibitemOpen
  \bibfield  {author} {\bibinfo {author} {\bibfnamefont {S.}~\bibnamefont {Wilken}}, \bibinfo {author} {\bibfnamefont {R.~E.}\ \bibnamefont {Guerra}}, \bibinfo {author} {\bibfnamefont {D.~J.}\ \bibnamefont {Pine}},\ and\ \bibinfo {author} {\bibfnamefont {P.~M.}\ \bibnamefont {Chaikin}},\ }\bibfield  {title} {\enquote {\bibinfo {title} {Hyperuniform structures formed by shearing colloidal suspensions},}\ }\href {https://doi.org/10.1103/PhysRevLett.125.148001} {\bibfield  {journal} {\bibinfo  {journal} {Physical Review Letters}\ }\textbf {\bibinfo {volume} {125}},\ \bibinfo {pages} {148001} (\bibinfo {year} {2020})}\BibitemShut {NoStop}%
\bibitem [{\citenamefont {Weijs}\ \emph {et~al.}(2015)\citenamefont {Weijs}, \citenamefont {Jeanneret}, \citenamefont {Dreyfus},\ and\ \citenamefont {Bartolo}}]{weijs2015emergent}%
  \BibitemOpen
  \bibfield  {author} {\bibinfo {author} {\bibfnamefont {J.~H.}\ \bibnamefont {Weijs}}, \bibinfo {author} {\bibfnamefont {R.}~\bibnamefont {Jeanneret}}, \bibinfo {author} {\bibfnamefont {R.}~\bibnamefont {Dreyfus}},\ and\ \bibinfo {author} {\bibfnamefont {D.}~\bibnamefont {Bartolo}},\ }\bibfield  {title} {\enquote {\bibinfo {title} {Emergent hyperuniformity in periodically driven emulsions},}\ }\href {https://doi.org/10.1103/PhysRevLett.115.108301} {\bibfield  {journal} {\bibinfo  {journal} {Physical review letters}\ }\textbf {\bibinfo {volume} {115}},\ \bibinfo {pages} {108301} (\bibinfo {year} {2015})}\BibitemShut {NoStop}%
\bibitem [{\citenamefont {Boltz}\ and\ \citenamefont {Ihle}(2024)}]{boltz2024hyperuniformity}%
  \BibitemOpen
  \bibfield  {author} {\bibinfo {author} {\bibfnamefont {H.-H.}\ \bibnamefont {Boltz}}\ and\ \bibinfo {author} {\bibfnamefont {T.}~\bibnamefont {Ihle}},\ }\bibfield  {title} {\enquote {\bibinfo {title} {Hyperuniformity in deterministic anti-aligning active matter},}\ }\href {https://doi.org/10.48550/arXiv.2402.19451} {\bibfield  {journal} {\bibinfo  {journal} {arXiv:2402.19451}\ } (\bibinfo {year} {2024})}\BibitemShut {NoStop}%
\bibitem [{\citenamefont {Castillo}\ \emph {et~al.}(2019)\citenamefont {Castillo}, \citenamefont {Mujica}, \citenamefont {Sepúlveda}, \citenamefont {Sobarzo}, \citenamefont {Guzmán},\ and\ \citenamefont {Soto}}]{Castillo_2019}%
  \BibitemOpen
  \bibfield  {author} {\bibinfo {author} {\bibfnamefont {G.}~\bibnamefont {Castillo}}, \bibinfo {author} {\bibfnamefont {N.}~\bibnamefont {Mujica}}, \bibinfo {author} {\bibfnamefont {N.}~\bibnamefont {Sepúlveda}}, \bibinfo {author} {\bibfnamefont {J.~C.}\ \bibnamefont {Sobarzo}}, \bibinfo {author} {\bibfnamefont {M.}~\bibnamefont {Guzmán}},\ and\ \bibinfo {author} {\bibfnamefont {R.}~\bibnamefont {Soto}},\ }\bibfield  {title} {\enquote {\bibinfo {title} {Hyperuniform states generated by a critical friction field},}\ }\href {http://dx.doi.org/10.1103/PhysRevE.100.032902} {\bibfield  {journal} {\bibinfo  {journal} {Physical Review E}\ }\textbf {\bibinfo {volume} {100}} (\bibinfo {year} {2019})}\BibitemShut {NoStop}%
\bibitem [{\citenamefont {Hexner}, \citenamefont {Chaikin},\ and\ \citenamefont {Levine}(2017)}]{hexner2017enhanced}%
  \BibitemOpen
  \bibfield  {author} {\bibinfo {author} {\bibfnamefont {D.}~\bibnamefont {Hexner}}, \bibinfo {author} {\bibfnamefont {P.~M.}\ \bibnamefont {Chaikin}},\ and\ \bibinfo {author} {\bibfnamefont {D.}~\bibnamefont {Levine}},\ }\bibfield  {title} {\enquote {\bibinfo {title} {Enhanced hyperuniformity from random reorganization},}\ }\href {https://doi.org/10.1073/pnas.1619260114} {\bibfield  {journal} {\bibinfo  {journal} {Proceedings of the National Academy of Sciences}\ }\textbf {\bibinfo {volume} {114}},\ \bibinfo {pages} {4294--4299} (\bibinfo {year} {2017})}\BibitemShut {NoStop}%
\bibitem [{\citenamefont {Shang}\ \emph {et~al.}(2025)\citenamefont {Shang}, \citenamefont {Wang}, \citenamefont {Pan}, \citenamefont {Jin},\ and\ \citenamefont {Zhang}}]{shang2025jamming}%
  \BibitemOpen
  \bibfield  {author} {\bibinfo {author} {\bibfnamefont {J.}~\bibnamefont {Shang}}, \bibinfo {author} {\bibfnamefont {Y.}~\bibnamefont {Wang}}, \bibinfo {author} {\bibfnamefont {D.}~\bibnamefont {Pan}}, \bibinfo {author} {\bibfnamefont {Y.}~\bibnamefont {Jin}},\ and\ \bibinfo {author} {\bibfnamefont {J.}~\bibnamefont {Zhang}},\ }\bibfield  {title} {\enquote {\bibinfo {title} {Jamming as a topological satisfiability transition with contact number hyperuniformity and criticality},}\ }\href {https://doi.org/10.48550/arXiv.2506.16474} {\bibfield  {journal} {\bibinfo  {journal} {arXiv:2506.16474}\ } (\bibinfo {year} {2025})}\BibitemShut {NoStop}%
\bibitem [{\citenamefont {Atkinson}\ \emph {et~al.}(2016)\citenamefont {Atkinson}, \citenamefont {Zhang}, \citenamefont {Hopkins},\ and\ \citenamefont {Torquato}}]{atkinson2016critical}%
  \BibitemOpen
  \bibfield  {author} {\bibinfo {author} {\bibfnamefont {S.}~\bibnamefont {Atkinson}}, \bibinfo {author} {\bibfnamefont {G.}~\bibnamefont {Zhang}}, \bibinfo {author} {\bibfnamefont {A.~B.}\ \bibnamefont {Hopkins}},\ and\ \bibinfo {author} {\bibfnamefont {S.}~\bibnamefont {Torquato}},\ }\bibfield  {title} {\enquote {\bibinfo {title} {Critical slowing down and hyperuniformity on approach to jamming},}\ }\href {https://doi.org/10.1103/PhysRevE.94.012902} {\bibfield  {journal} {\bibinfo  {journal} {Physical Review E}\ }\textbf {\bibinfo {volume} {94}},\ \bibinfo {pages} {012902} (\bibinfo {year} {2016})}\BibitemShut {NoStop}%
\bibitem [{\citenamefont {Wang}\ \emph {et~al.}(2025)\citenamefont {Wang}, \citenamefont {Qian}, \citenamefont {Tong},\ and\ \citenamefont {Tanaka}}]{wang2025hyperuniform}%
  \BibitemOpen
  \bibfield  {author} {\bibinfo {author} {\bibfnamefont {Y.}~\bibnamefont {Wang}}, \bibinfo {author} {\bibfnamefont {Z.}~\bibnamefont {Qian}}, \bibinfo {author} {\bibfnamefont {H.}~\bibnamefont {Tong}},\ and\ \bibinfo {author} {\bibfnamefont {H.}~\bibnamefont {Tanaka}},\ }\bibfield  {title} {\enquote {\bibinfo {title} {Hyperuniform disordered solids with crystal-like stability},}\ }\href {https://doi.org/10.1038/s41467-025-56283-1} {\bibfield  {journal} {\bibinfo  {journal} {Nature Communications}\ }\textbf {\bibinfo {volume} {16}},\ \bibinfo {pages} {1398} (\bibinfo {year} {2025})}\BibitemShut {NoStop}%
\bibitem [{\citenamefont {Wilken}\ \emph {et~al.}(2023)\citenamefont {Wilken}, \citenamefont {Guo}, \citenamefont {Levine},\ and\ \citenamefont {Chaikin}}]{wilken2023dynamical}%
  \BibitemOpen
  \bibfield  {author} {\bibinfo {author} {\bibfnamefont {S.}~\bibnamefont {Wilken}}, \bibinfo {author} {\bibfnamefont {A.~Z.}\ \bibnamefont {Guo}}, \bibinfo {author} {\bibfnamefont {D.}~\bibnamefont {Levine}},\ and\ \bibinfo {author} {\bibfnamefont {P.~M.}\ \bibnamefont {Chaikin}},\ }\bibfield  {title} {\enquote {\bibinfo {title} {Dynamical approach to the jamming problem},}\ }\href {https://doi.org/10.1103/PhysRevLett.131.238202} {\bibfield  {journal} {\bibinfo  {journal} {Physical review letters}\ }\textbf {\bibinfo {volume} {131}},\ \bibinfo {pages} {238202} (\bibinfo {year} {2023})}\BibitemShut {NoStop}%
\bibitem [{\citenamefont {Hexner}, \citenamefont {Liu},\ and\ \citenamefont {Nagel}(2018)}]{hexner2018two}%
  \BibitemOpen
  \bibfield  {author} {\bibinfo {author} {\bibfnamefont {D.}~\bibnamefont {Hexner}}, \bibinfo {author} {\bibfnamefont {A.~J.}\ \bibnamefont {Liu}},\ and\ \bibinfo {author} {\bibfnamefont {S.~R.}\ \bibnamefont {Nagel}},\ }\bibfield  {title} {\enquote {\bibinfo {title} {Two diverging length scales in the structure of jammed packings},}\ }\href {https://doi.org/10.1103/PhysRevLett.121.115501} {\bibfield  {journal} {\bibinfo  {journal} {Physical review letters}\ }\textbf {\bibinfo {volume} {121}},\ \bibinfo {pages} {115501} (\bibinfo {year} {2018})}\BibitemShut {NoStop}%
\bibitem [{\citenamefont {Ozawa}, \citenamefont {Berthier},\ and\ \citenamefont {Coslovich}(2017)}]{ozawa2017exploring}%
  \BibitemOpen
  \bibfield  {author} {\bibinfo {author} {\bibfnamefont {M.}~\bibnamefont {Ozawa}}, \bibinfo {author} {\bibfnamefont {L.}~\bibnamefont {Berthier}},\ and\ \bibinfo {author} {\bibfnamefont {D.}~\bibnamefont {Coslovich}},\ }\bibfield  {title} {\enquote {\bibinfo {title} {Exploring the jamming transition over a wide range of critical densities},}\ }\href {https://doi.org/10.21468/scipostphys.3.4.027} {\bibfield  {journal} {\bibinfo  {journal} {SciPost Physics}\ }\textbf {\bibinfo {volume} {3}},\ \bibinfo {pages} {027} (\bibinfo {year} {2017})}\BibitemShut {NoStop}%
\bibitem [{\citenamefont {Ikeda}\ and\ \citenamefont {Berthier}(2015)}]{ikeda2015thermal}%
  \BibitemOpen
  \bibfield  {author} {\bibinfo {author} {\bibfnamefont {A.}~\bibnamefont {Ikeda}}\ and\ \bibinfo {author} {\bibfnamefont {L.}~\bibnamefont {Berthier}},\ }\bibfield  {title} {\enquote {\bibinfo {title} {Thermal fluctuations, mechanical response, and hyperuniformity in jammed solids},}\ }\href {https://doi.org/10.1103/physreve.92.012309} {\bibfield  {journal} {\bibinfo  {journal} {Physical Review E}\ }\textbf {\bibinfo {volume} {92}},\ \bibinfo {pages} {012309} (\bibinfo {year} {2015})}\BibitemShut {NoStop}%
\bibitem [{\citenamefont {Maher}, \citenamefont {Jiao},\ and\ \citenamefont {Torquato}(2023)}]{maher2023hyperuniformity}%
  \BibitemOpen
  \bibfield  {author} {\bibinfo {author} {\bibfnamefont {C.~E.}\ \bibnamefont {Maher}}, \bibinfo {author} {\bibfnamefont {Y.}~\bibnamefont {Jiao}},\ and\ \bibinfo {author} {\bibfnamefont {S.}~\bibnamefont {Torquato}},\ }\bibfield  {title} {\enquote {\bibinfo {title} {Hyperuniformity of maximally random jammed packings of hyperspheres across spatial dimensions},}\ }\href {https://doi.org/10.1103/PhysRevE.108.064602} {\bibfield  {journal} {\bibinfo  {journal} {Physical Review E}\ }\textbf {\bibinfo {volume} {108}},\ \bibinfo {pages} {064602} (\bibinfo {year} {2023})}\BibitemShut {NoStop}%
\bibitem [{\citenamefont {Dam}\ \emph {et~al.}(2025)\citenamefont {Dam}, \citenamefont {Kawasaki}, \citenamefont {Ikeda},\ and\ \citenamefont {Miyazaki}}]{dam2025hyperuniformity}%
  \BibitemOpen
  \bibfield  {author} {\bibinfo {author} {\bibfnamefont {D.~T.}\ \bibnamefont {Dam}}, \bibinfo {author} {\bibfnamefont {T.}~\bibnamefont {Kawasaki}}, \bibinfo {author} {\bibfnamefont {A.}~\bibnamefont {Ikeda}},\ and\ \bibinfo {author} {\bibfnamefont {K.}~\bibnamefont {Miyazaki}},\ }\bibfield  {title} {\enquote {\bibinfo {title} {Hyperuniformity near jamming transition over a wide range of bidispersity},}\ }\href {https://doi.org/10.48550/arXiv.2507.12738} {\bibfield  {journal} {\bibinfo  {journal} {arXiv:2507.12738}\ } (\bibinfo {year} {2025})}\BibitemShut {NoStop}%
\bibitem [{\citenamefont {Zhu}\ \emph {et~al.}(2023)\citenamefont {Zhu}, \citenamefont {Hallet}, \citenamefont {Sipos}, \citenamefont {Domokos},\ and\ \citenamefont {Liu}}]{zhu2023hyperuniformity}%
  \BibitemOpen
  \bibfield  {author} {\bibinfo {author} {\bibfnamefont {Z.}~\bibnamefont {Zhu}}, \bibinfo {author} {\bibfnamefont {B.}~\bibnamefont {Hallet}}, \bibinfo {author} {\bibfnamefont {A.~A.}\ \bibnamefont {Sipos}}, \bibinfo {author} {\bibfnamefont {G.}~\bibnamefont {Domokos}},\ and\ \bibinfo {author} {\bibfnamefont {Q.-X.}\ \bibnamefont {Liu}},\ }\bibfield  {title} {\enquote {\bibinfo {title} {Hyperuniformity on mars: Pebbles scattered on sand},}\ }\href {https://doi.org/10.48550/arXiv.2312.13818} {\bibfield  {journal} {\bibinfo  {journal} {arXiv:2312.13818}\ } (\bibinfo {year} {2023})}\BibitemShut {NoStop}%
\bibitem [{\citenamefont {Philcox}\ and\ \citenamefont {Torquato}(2023)}]{philcox2023disordered}%
  \BibitemOpen
  \bibfield  {author} {\bibinfo {author} {\bibfnamefont {O.~H.}\ \bibnamefont {Philcox}}\ and\ \bibinfo {author} {\bibfnamefont {S.}~\bibnamefont {Torquato}},\ }\bibfield  {title} {\enquote {\bibinfo {title} {Disordered heterogeneous universe: Galaxy distribution and clustering across length scales},}\ }\href {https://doi.org/10.1103/PhysRevX.13.011038} {\bibfield  {journal} {\bibinfo  {journal} {Physical Review X}\ }\textbf {\bibinfo {volume} {13}},\ \bibinfo {pages} {011038} (\bibinfo {year} {2023})}\BibitemShut {NoStop}%
\bibitem [{\citenamefont {Ezoe}, \citenamefont {Katori},\ and\ \citenamefont {Shirai}(2025)}]{ezoe2025weighted}%
  \BibitemOpen
  \bibfield  {author} {\bibinfo {author} {\bibfnamefont {A.}~\bibnamefont {Ezoe}}, \bibinfo {author} {\bibfnamefont {M.}~\bibnamefont {Katori}},\ and\ \bibinfo {author} {\bibfnamefont {T.}~\bibnamefont {Shirai}},\ }\bibfield  {title} {\enquote {\bibinfo {title} {Weighted point configurations with hyperuniformity: An ecological example and models},}\ }\href {https://doi.org/10.7566/JPSJ.94.064002} {\bibfield  {journal} {\bibinfo  {journal} {Journal of the Physical Society of Japan}\ }\textbf {\bibinfo {volume} {94}},\ \bibinfo {pages} {064002} (\bibinfo {year} {2025})}\BibitemShut {NoStop}%
\bibitem [{\citenamefont {Dong}(2023)}]{dong2023hyperuniform}%
  \BibitemOpen
  \bibfield  {author} {\bibinfo {author} {\bibfnamefont {L.}~\bibnamefont {Dong}},\ }\bibfield  {title} {\enquote {\bibinfo {title} {Hyperuniform organization in human settlements},}\ }\href {https://doi.org/10.48550/arXiv.2306.04149} {\bibfield  {journal} {\bibinfo  {journal} {arXiv:2306.04149}\ } (\bibinfo {year} {2023})}\BibitemShut {NoStop}%
\bibitem [{\citenamefont {Jiao}\ \emph {et~al.}(2014)\citenamefont {Jiao}, \citenamefont {Lau}, \citenamefont {Hatzikirou}, \citenamefont {Meyer-Hermann}, \citenamefont {Corbo},\ and\ \citenamefont {Torquato}}]{PhysRevE.89.022721}%
  \BibitemOpen
  \bibfield  {author} {\bibinfo {author} {\bibfnamefont {Y.}~\bibnamefont {Jiao}}, \bibinfo {author} {\bibfnamefont {T.}~\bibnamefont {Lau}}, \bibinfo {author} {\bibfnamefont {H.}~\bibnamefont {Hatzikirou}}, \bibinfo {author} {\bibfnamefont {M.}~\bibnamefont {Meyer-Hermann}}, \bibinfo {author} {\bibfnamefont {J.~C.}\ \bibnamefont {Corbo}},\ and\ \bibinfo {author} {\bibfnamefont {S.}~\bibnamefont {Torquato}},\ }\bibfield  {title} {\enquote {\bibinfo {title} {Avian photoreceptor patterns represent a disordered hyperuniform solution to a multiscale packing problem},}\ }\href {https://doi.org/10.1103/PhysRevE.89.022721} {\bibfield  {journal} {\bibinfo  {journal} {Phys. Rev. E}\ }\textbf {\bibinfo {volume} {89}},\ \bibinfo {pages} {022721} (\bibinfo {year} {2014})}\BibitemShut {NoStop}%
\bibitem [{\citenamefont {Ge}(2023)}]{Ge_2023}%
  \BibitemOpen
  \bibfield  {author} {\bibinfo {author} {\bibfnamefont {Z.}~\bibnamefont {Ge}},\ }\bibfield  {title} {\enquote {\bibinfo {title} {The hidden order of turing patterns in arid and semi‐arid vegetation ecosystems},}\ }\href {https://doi.org/10.1073/pnas.2306514120} {\bibfield  {journal} {\bibinfo  {journal} {Proceedings of the National Academy of Sciences}\ }\textbf {\bibinfo {volume} {120}} (\bibinfo {year} {2023})}\BibitemShut {NoStop}%
\bibitem [{\citenamefont {Liu}\ \emph {et~al.}(2024)\citenamefont {Liu}, \citenamefont {Chen}, \citenamefont {Tian}, \citenamefont {Xu},\ and\ \citenamefont {Jiao}}]{PhysRevLett.133.028401}%
  \BibitemOpen
  \bibfield  {author} {\bibinfo {author} {\bibfnamefont {Y.}~\bibnamefont {Liu}}, \bibinfo {author} {\bibfnamefont {D.}~\bibnamefont {Chen}}, \bibinfo {author} {\bibfnamefont {J.}~\bibnamefont {Tian}}, \bibinfo {author} {\bibfnamefont {W.}~\bibnamefont {Xu}},\ and\ \bibinfo {author} {\bibfnamefont {Y.}~\bibnamefont {Jiao}},\ }\bibfield  {title} {\enquote {\bibinfo {title} {Universal hyperuniform organization in looped leaf vein networks},}\ }\href {https://doi.org/10.1103/PhysRevLett.133.028401} {\bibfield  {journal} {\bibinfo  {journal} {Phys. Rev. Lett.}\ }\textbf {\bibinfo {volume} {133}},\ \bibinfo {pages} {028401} (\bibinfo {year} {2024})}\BibitemShut {NoStop}%
\bibitem [{\citenamefont {Tang}, \citenamefont {Li},\ and\ \citenamefont {Bi}(2024)}]{tang2024tunable}%
  \BibitemOpen
  \bibfield  {author} {\bibinfo {author} {\bibfnamefont {Y.}~\bibnamefont {Tang}}, \bibinfo {author} {\bibfnamefont {X.}~\bibnamefont {Li}},\ and\ \bibinfo {author} {\bibfnamefont {D.}~\bibnamefont {Bi}},\ }\bibfield  {title} {\enquote {\bibinfo {title} {Tunable hyperuniformity in cellular structures},}\ }\href {https://doi.org/10.48550/arXiv.2408.08976} {\bibfield  {journal} {\bibinfo  {journal} {arXiv:2408.08976}\ } (\bibinfo {year} {2024})}\BibitemShut {NoStop}%
\bibitem [{\citenamefont {Mayer}\ \emph {et~al.}(2015)\citenamefont {Mayer}, \citenamefont {Balasubramanian}, \citenamefont {Mora},\ and\ \citenamefont {Walczak}}]{mayer2015well}%
  \BibitemOpen
  \bibfield  {author} {\bibinfo {author} {\bibfnamefont {A.}~\bibnamefont {Mayer}}, \bibinfo {author} {\bibfnamefont {V.}~\bibnamefont {Balasubramanian}}, \bibinfo {author} {\bibfnamefont {T.}~\bibnamefont {Mora}},\ and\ \bibinfo {author} {\bibfnamefont {A.~M.}\ \bibnamefont {Walczak}},\ }\bibfield  {title} {\enquote {\bibinfo {title} {How a well-adapted immune system is organized},}\ }\href {https://doi.org/10.1073/pnas.1421827112} {\bibfield  {journal} {\bibinfo  {journal} {Proceedings of the National Academy of Sciences}\ }\textbf {\bibinfo {volume} {112}},\ \bibinfo {pages} {5950--5955} (\bibinfo {year} {2015})}\BibitemShut {NoStop}%
\bibitem [{\citenamefont {Emery}\ \emph {et~al.}(2024)\citenamefont {Emery}, \citenamefont {Bercegol}, \citenamefont {Jonqueres},\ and\ \citenamefont {Auma{\^\i}tre}}]{emery2024complex}%
  \BibitemOpen
  \bibfield  {author} {\bibinfo {author} {\bibfnamefont {E.}~\bibnamefont {Emery}}, \bibinfo {author} {\bibfnamefont {H.}~\bibnamefont {Bercegol}}, \bibinfo {author} {\bibfnamefont {N.}~\bibnamefont {Jonqueres}},\ and\ \bibinfo {author} {\bibfnamefont {S.}~\bibnamefont {Auma{\^\i}tre}},\ }\bibfield  {title} {\enquote {\bibinfo {title} {Complex network analysis of transmission networks preparing for the energy transition: application to the current french power grid},}\ }\href {https://doi.org/10.1140/epjb/s10051-024-00837-7} {\bibfield  {journal} {\bibinfo  {journal} {The European Physical Journal B}\ }\textbf {\bibinfo {volume} {97}},\ \bibinfo {pages} {201} (\bibinfo {year} {2024})}\BibitemShut {NoStop}%
\bibitem [{\citenamefont {Galliano}, \citenamefont {Cates},\ and\ \citenamefont {Berthier}(2023)}]{PhysRevLett.131.047101}%
  \BibitemOpen
  \bibfield  {author} {\bibinfo {author} {\bibfnamefont {L.}~\bibnamefont {Galliano}}, \bibinfo {author} {\bibfnamefont {M.~E.}\ \bibnamefont {Cates}},\ and\ \bibinfo {author} {\bibfnamefont {L.}~\bibnamefont {Berthier}},\ }\bibfield  {title} {\enquote {\bibinfo {title} {Two-dimensional crystals far from equilibrium},}\ }\href {https://doi.org/10.1103/PhysRevLett.131.047101} {\bibfield  {journal} {\bibinfo  {journal} {Phys. Rev. Lett.}\ }\textbf {\bibinfo {volume} {131}},\ \bibinfo {pages} {047101} (\bibinfo {year} {2023})}\BibitemShut {NoStop}%
\bibitem [{\citenamefont {Maire}\ and\ \citenamefont {Plati}(2024)}]{enhancing2024Maire}%
  \BibitemOpen
  \bibfield  {author} {\bibinfo {author} {\bibfnamefont {R.}~\bibnamefont {Maire}}\ and\ \bibinfo {author} {\bibfnamefont {A.}~\bibnamefont {Plati}},\ }\bibfield  {title} {\enquote {\bibinfo {title} {Enhancing (quasi-)long-range order in a two-dimensional driven crystal},}\ }\href {https://doi.org/10.1063/5.0217958} {\bibfield  {journal} {\bibinfo  {journal} {The Journal of Chemical Physics}\ }\textbf {\bibinfo {volume} {161}},\ \bibinfo {pages} {054902} (\bibinfo {year} {2024})}\BibitemShut {NoStop}%
\bibitem [{\citenamefont {Kuroda}, \citenamefont {Kawasaki},\ and\ \citenamefont {Miyazaki}(2025{\natexlab{b}})}]{kuroda2024long}%
  \BibitemOpen
  \bibfield  {author} {\bibinfo {author} {\bibfnamefont {Y.}~\bibnamefont {Kuroda}}, \bibinfo {author} {\bibfnamefont {T.}~\bibnamefont {Kawasaki}},\ and\ \bibinfo {author} {\bibfnamefont {K.}~\bibnamefont {Miyazaki}},\ }\bibfield  {title} {\enquote {\bibinfo {title} {Long-range translational order and hyperuniformity in two-dimensional chiral active crystal},}\ }\href {https://doi.org/10.1103/PhysRevResearch.7.L012048} {\bibfield  {journal} {\bibinfo  {journal} {Physical Review Research}\ }\textbf {\bibinfo {volume} {7}},\ \bibinfo {pages} {L012048} (\bibinfo {year} {2025}{\natexlab{b}})}\BibitemShut {NoStop}%
\bibitem [{\citenamefont {Ikeda}(2024)}]{ikeda2024harmonic}%
  \BibitemOpen
  \bibfield  {author} {\bibinfo {author} {\bibfnamefont {H.}~\bibnamefont {Ikeda}},\ }\bibfield  {title} {\enquote {\bibinfo {title} {Harmonic chain far from equilibrium: single-file diffusion, long-range order, and hyperuniformity},}\ }\href {https://doi.org/10.21468/SciPostPhys.17.4.103} {\bibfield  {journal} {\bibinfo  {journal} {SciPost Physics}\ }\textbf {\bibinfo {volume} {17}},\ \bibinfo {pages} {103} (\bibinfo {year} {2024})}\BibitemShut {NoStop}%
\bibitem [{\citenamefont {Mkhonta}, \citenamefont {Huang},\ and\ \citenamefont {Elder}(2024)}]{mkhonta2024liquid}%
  \BibitemOpen
  \bibfield  {author} {\bibinfo {author} {\bibfnamefont {S.}~\bibnamefont {Mkhonta}}, \bibinfo {author} {\bibfnamefont {Z.-F.}\ \bibnamefont {Huang}},\ and\ \bibinfo {author} {\bibfnamefont {K.}~\bibnamefont {Elder}},\ }\bibfield  {title} {\enquote {\bibinfo {title} {Liquid-substrate fluctuation effects on crystal growth and disordered hyperuniformity of two-dimensional materials},}\ }\href {https://doi.org/10.1103/PhysRevMaterials.8.104002} {\bibfield  {journal} {\bibinfo  {journal} {Physical Review Materials}\ }\textbf {\bibinfo {volume} {8}},\ \bibinfo {pages} {104002} (\bibinfo {year} {2024})}\BibitemShut {NoStop}%
\bibitem [{\citenamefont {Lei}\ and\ \citenamefont {Ni}(2023)}]{Lei_Ni_2023}%
  \BibitemOpen
  \bibfield  {author} {\bibinfo {author} {\bibfnamefont {Y.}~\bibnamefont {Lei}}\ and\ \bibinfo {author} {\bibfnamefont {R.}~\bibnamefont {Ni}},\ }\bibfield  {title} {\enquote {\bibinfo {title} {How does a hyperuniform fluid freeze?}}\ }\href {https://doi.org/10.1073/pnas.2312866120} {\bibfield  {journal} {\bibinfo  {journal} {Proceedings of the National Academy of Sciences}\ }\textbf {\bibinfo {volume} {120}} (\bibinfo {year} {2023})}\BibitemShut {NoStop}%
\bibitem [{\citenamefont {Di~Santo}\ \emph {et~al.}(2016)\citenamefont {Di~Santo}, \citenamefont {Burioni}, \citenamefont {Vezzani},\ and\ \citenamefont {Munoz}}]{di2016self}%
  \BibitemOpen
  \bibfield  {author} {\bibinfo {author} {\bibfnamefont {S.}~\bibnamefont {Di~Santo}}, \bibinfo {author} {\bibfnamefont {R.}~\bibnamefont {Burioni}}, \bibinfo {author} {\bibfnamefont {A.}~\bibnamefont {Vezzani}},\ and\ \bibinfo {author} {\bibfnamefont {M.~A.}\ \bibnamefont {Munoz}},\ }\bibfield  {title} {\enquote {\bibinfo {title} {Self-organized bistability associated with first-order phase transitions},}\ }\href {https://doi.org/10.1103/PhysRevLett.116.240601} {\bibfield  {journal} {\bibinfo  {journal} {Physical review letters}\ }\textbf {\bibinfo {volume} {116}},\ \bibinfo {pages} {240601} (\bibinfo {year} {2016})}\BibitemShut {NoStop}%
\bibitem [{\citenamefont {Maire}\ \emph {et~al.}(2025{\natexlab{b}})\citenamefont {Maire}, \citenamefont {Galliano}, \citenamefont {Plati},\ and\ \citenamefont {Berthier}}]{maire2025hyperuniforminterfacesnonequilibriumphase}%
  \BibitemOpen
  \bibfield  {author} {\bibinfo {author} {\bibfnamefont {R.}~\bibnamefont {Maire}}, \bibinfo {author} {\bibfnamefont {L.}~\bibnamefont {Galliano}}, \bibinfo {author} {\bibfnamefont {A.}~\bibnamefont {Plati}},\ and\ \bibinfo {author} {\bibfnamefont {L.}~\bibnamefont {Berthier}},\ }\href {https://arxiv.org/abs/2507.03957} {\enquote {\bibinfo {title} {Hyperuniform interfaces in non-equilibrium phase coexistence},}\ } (\bibinfo {year} {2025}{\natexlab{b}})\BibitemShut {NoStop}%
\bibitem [{\citenamefont {Ikeda}(2023)}]{ikeda2023correlated}%
  \BibitemOpen
  \bibfield  {author} {\bibinfo {author} {\bibfnamefont {H.}~\bibnamefont {Ikeda}},\ }\bibfield  {title} {\enquote {\bibinfo {title} {Correlated noise and critical dimensions},}\ }\href {https://doi.org/10.1103/PhysRevE.108.064119} {\bibfield  {journal} {\bibinfo  {journal} {Physical Review E}\ }\textbf {\bibinfo {volume} {108}},\ \bibinfo {pages} {064119} (\bibinfo {year} {2023})}\BibitemShut {NoStop}%
\bibitem [{\citenamefont {Man}\ \emph {et~al.}(2013)\citenamefont {Man}, \citenamefont {Florescu}, \citenamefont {Williamson}, \citenamefont {He}, \citenamefont {Hashemizad}, \citenamefont {Leung}, \citenamefont {Liner}, \citenamefont {Torquato}, \citenamefont {Chaikin},\ and\ \citenamefont {Steinhardt}}]{man2013isotropic}%
  \BibitemOpen
  \bibfield  {author} {\bibinfo {author} {\bibfnamefont {W.}~\bibnamefont {Man}}, \bibinfo {author} {\bibfnamefont {M.}~\bibnamefont {Florescu}}, \bibinfo {author} {\bibfnamefont {E.~P.}\ \bibnamefont {Williamson}}, \bibinfo {author} {\bibfnamefont {Y.}~\bibnamefont {He}}, \bibinfo {author} {\bibfnamefont {S.~R.}\ \bibnamefont {Hashemizad}}, \bibinfo {author} {\bibfnamefont {B.~Y.}\ \bibnamefont {Leung}}, \bibinfo {author} {\bibfnamefont {D.~R.}\ \bibnamefont {Liner}}, \bibinfo {author} {\bibfnamefont {S.}~\bibnamefont {Torquato}}, \bibinfo {author} {\bibfnamefont {P.~M.}\ \bibnamefont {Chaikin}},\ and\ \bibinfo {author} {\bibfnamefont {P.~J.}\ \bibnamefont {Steinhardt}},\ }\bibfield  {title} {\enquote {\bibinfo {title} {Isotropic band gaps and freeform waveguides observed in hyperuniform disordered photonic solids},}\ }\href {https://doi.org/10.1073/pnas.1307879110} {\bibfield  {journal} {\bibinfo  {journal} {Proceedings of the National Academy of Sciences}\ }\textbf {\bibinfo {volume} {110}},\ \bibinfo
  {pages} {15886--15891} (\bibinfo {year} {2013})}\BibitemShut {NoStop}%
\bibitem [{\citenamefont {Florescu}, \citenamefont {Torquato},\ and\ \citenamefont {Steinhardt}(2009)}]{florescu2009designer}%
  \BibitemOpen
  \bibfield  {author} {\bibinfo {author} {\bibfnamefont {M.}~\bibnamefont {Florescu}}, \bibinfo {author} {\bibfnamefont {S.}~\bibnamefont {Torquato}},\ and\ \bibinfo {author} {\bibfnamefont {P.~J.}\ \bibnamefont {Steinhardt}},\ }\bibfield  {title} {\enquote {\bibinfo {title} {Designer disordered materials with large, complete photonic band gaps},}\ }\href {https://doi.org/10.1073/pnas.0907744106} {\bibfield  {journal} {\bibinfo  {journal} {Proceedings of the National Academy of Sciences}\ }\textbf {\bibinfo {volume} {106}},\ \bibinfo {pages} {20658--20663} (\bibinfo {year} {2009})}\BibitemShut {NoStop}%
\bibitem [{\citenamefont {Diego}\ \emph {et~al.}(2025)\citenamefont {Diego}, \citenamefont {Hardouin}, \citenamefont {Mazevet-Schargrod}, \citenamefont {Pirro}, \citenamefont {Kim}, \citenamefont {Anufriev},\ and\ \citenamefont {Nomura}}]{diego2025hypersonic}%
  \BibitemOpen
  \bibfield  {author} {\bibinfo {author} {\bibfnamefont {M.}~\bibnamefont {Diego}}, \bibinfo {author} {\bibfnamefont {J.}~\bibnamefont {Hardouin}}, \bibinfo {author} {\bibfnamefont {G.}~\bibnamefont {Mazevet-Schargrod}}, \bibinfo {author} {\bibfnamefont {M.}~\bibnamefont {Pirro}}, \bibinfo {author} {\bibfnamefont {B.}~\bibnamefont {Kim}}, \bibinfo {author} {\bibfnamefont {R.}~\bibnamefont {Anufriev}},\ and\ \bibinfo {author} {\bibfnamefont {M.}~\bibnamefont {Nomura}},\ }\bibfield  {title} {\enquote {\bibinfo {title} {Hypersonic acoustic wave control via stealthy hyperuniform phononic nanostructures},}\ }\href {https://doi.org/10.1126/sciadv.adw7205} {\bibfield  {journal} {\bibinfo  {journal} {Science Advances}\ }\textbf {\bibinfo {volume} {11}} (\bibinfo {year} {2025})}\BibitemShut {NoStop}%
\bibitem [{\citenamefont {Milo{\v{s}}evi{\'c}}\ \emph {et~al.}(2019)\citenamefont {Milo{\v{s}}evi{\'c}}, \citenamefont {Man}, \citenamefont {Nahal}, \citenamefont {Steinhardt}, \citenamefont {Torquato}, \citenamefont {Chaikin}, \citenamefont {Amoah}, \citenamefont {Yu}, \citenamefont {Mullen},\ and\ \citenamefont {Florescu}}]{milovsevic2019hyperuniform}%
  \BibitemOpen
  \bibfield  {author} {\bibinfo {author} {\bibfnamefont {M.~M.}\ \bibnamefont {Milo{\v{s}}evi{\'c}}}, \bibinfo {author} {\bibfnamefont {W.}~\bibnamefont {Man}}, \bibinfo {author} {\bibfnamefont {G.}~\bibnamefont {Nahal}}, \bibinfo {author} {\bibfnamefont {P.~J.}\ \bibnamefont {Steinhardt}}, \bibinfo {author} {\bibfnamefont {S.}~\bibnamefont {Torquato}}, \bibinfo {author} {\bibfnamefont {P.~M.}\ \bibnamefont {Chaikin}}, \bibinfo {author} {\bibfnamefont {T.}~\bibnamefont {Amoah}}, \bibinfo {author} {\bibfnamefont {B.}~\bibnamefont {Yu}}, \bibinfo {author} {\bibfnamefont {R.~A.}\ \bibnamefont {Mullen}},\ and\ \bibinfo {author} {\bibfnamefont {M.}~\bibnamefont {Florescu}},\ }\bibfield  {title} {\enquote {\bibinfo {title} {Hyperuniform disordered waveguides and devices for near infrared silicon photonics},}\ }\href {https://doi.org/10.1038/s41598-019-56692-5} {\bibfield  {journal} {\bibinfo  {journal} {Scientific reports}\ }\textbf {\bibinfo {volume} {9}},\ \bibinfo {pages} {20338} (\bibinfo {year}
  {2019})}\BibitemShut {NoStop}%
\bibitem [{\citenamefont {Barsukova}\ \emph {et~al.}(2025)\citenamefont {Barsukova}, \citenamefont {Zhang}, \citenamefont {Gould}, \citenamefont {Sadri}, \citenamefont {Rosiek}, \citenamefont {Stobbe}, \citenamefont {Karcher},\ and\ \citenamefont {Rechtsman}}]{barsukova2025stealthy}%
  \BibitemOpen
  \bibfield  {author} {\bibinfo {author} {\bibfnamefont {M.}~\bibnamefont {Barsukova}}, \bibinfo {author} {\bibfnamefont {Z.}~\bibnamefont {Zhang}}, \bibinfo {author} {\bibfnamefont {B.}~\bibnamefont {Gould}}, \bibinfo {author} {\bibfnamefont {K.}~\bibnamefont {Sadri}}, \bibinfo {author} {\bibfnamefont {C.}~\bibnamefont {Rosiek}}, \bibinfo {author} {\bibfnamefont {S.}~\bibnamefont {Stobbe}}, \bibinfo {author} {\bibfnamefont {J.}~\bibnamefont {Karcher}},\ and\ \bibinfo {author} {\bibfnamefont {M.~C.}\ \bibnamefont {Rechtsman}},\ }\bibfield  {title} {\enquote {\bibinfo {title} {Stealthy-hyperuniform wave dynamics in two-dimensional photonic crystals},}\ }\href {https://doi.org/10.48550/arXiv.2507.05253} {\bibfield  {journal} {\bibinfo  {journal} {arXiv:2507.05253}\ } (\bibinfo {year} {2025})}\BibitemShut {NoStop}%
\bibitem [{\citenamefont {Aubry}\ \emph {et~al.}(2020)\citenamefont {Aubry}, \citenamefont {Froufe-P{\'e}rez}, \citenamefont {Kuhl}, \citenamefont {Legrand}, \citenamefont {Scheffold},\ and\ \citenamefont {Mortessagne}}]{aubry2020experimental}%
  \BibitemOpen
  \bibfield  {author} {\bibinfo {author} {\bibfnamefont {G.~J.}\ \bibnamefont {Aubry}}, \bibinfo {author} {\bibfnamefont {L.~S.}\ \bibnamefont {Froufe-P{\'e}rez}}, \bibinfo {author} {\bibfnamefont {U.}~\bibnamefont {Kuhl}}, \bibinfo {author} {\bibfnamefont {O.}~\bibnamefont {Legrand}}, \bibinfo {author} {\bibfnamefont {F.}~\bibnamefont {Scheffold}},\ and\ \bibinfo {author} {\bibfnamefont {F.}~\bibnamefont {Mortessagne}},\ }\bibfield  {title} {\enquote {\bibinfo {title} {Experimental tuning of transport regimes in hyperuniform disordered photonic materials},}\ }\href {https://doi.org/10.1103/PhysRevLett.125.127402} {\bibfield  {journal} {\bibinfo  {journal} {Physical review letters}\ }\textbf {\bibinfo {volume} {125}},\ \bibinfo {pages} {127402} (\bibinfo {year} {2020})}\BibitemShut {NoStop}%
\bibitem [{\citenamefont {Gkantzounis}, \citenamefont {Amoah},\ and\ \citenamefont {Florescu}(2017)}]{gkantzounis2017hyperuniform}%
  \BibitemOpen
  \bibfield  {author} {\bibinfo {author} {\bibfnamefont {G.}~\bibnamefont {Gkantzounis}}, \bibinfo {author} {\bibfnamefont {T.}~\bibnamefont {Amoah}},\ and\ \bibinfo {author} {\bibfnamefont {M.}~\bibnamefont {Florescu}},\ }\bibfield  {title} {\enquote {\bibinfo {title} {Hyperuniform disordered phononic structures},}\ }\href {https://doi.org/10.1103/PhysRevB.95.094120} {\bibfield  {journal} {\bibinfo  {journal} {Physical Review B}\ }\textbf {\bibinfo {volume} {95}},\ \bibinfo {pages} {094120} (\bibinfo {year} {2017})}\BibitemShut {NoStop}%
\bibitem [{\citenamefont {Hong}\ \emph {et~al.}(2024)\citenamefont {Hong}, \citenamefont {Nerse}, \citenamefont {Oberst},\ and\ \citenamefont {Saadatfar}}]{hong2024topological}%
  \BibitemOpen
  \bibfield  {author} {\bibinfo {author} {\bibfnamefont {S.}~\bibnamefont {Hong}}, \bibinfo {author} {\bibfnamefont {C.}~\bibnamefont {Nerse}}, \bibinfo {author} {\bibfnamefont {S.}~\bibnamefont {Oberst}},\ and\ \bibinfo {author} {\bibfnamefont {M.}~\bibnamefont {Saadatfar}},\ }\bibfield  {title} {\enquote {\bibinfo {title} {Topological mechanical states in geometry-driven hyperuniform materials},}\ }\href {https://doi.org/10.1093/pnasnexus/pgae510} {\bibfield  {journal} {\bibinfo  {journal} {PNAS nexus}\ }\textbf {\bibinfo {volume} {3}},\ \bibinfo {pages} {p. 510} (\bibinfo {year} {2024})}\BibitemShut {NoStop}%
\bibitem [{\citenamefont {Dale}\ \emph {et~al.}(2022)\citenamefont {Dale}, \citenamefont {Sartor}, \citenamefont {Dennis},\ and\ \citenamefont {Corwin}}]{dale2022hyperuniform}%
  \BibitemOpen
  \bibfield  {author} {\bibinfo {author} {\bibfnamefont {J.~R.}\ \bibnamefont {Dale}}, \bibinfo {author} {\bibfnamefont {J.~D.}\ \bibnamefont {Sartor}}, \bibinfo {author} {\bibfnamefont {R.~C.}\ \bibnamefont {Dennis}},\ and\ \bibinfo {author} {\bibfnamefont {E.~I.}\ \bibnamefont {Corwin}},\ }\bibfield  {title} {\enquote {\bibinfo {title} {Hyperuniform jammed sphere packings have anomalous material properties},}\ }\href {https://doi.org/10.1103/PhysRevE.106.024903} {\bibfield  {journal} {\bibinfo  {journal} {Physical Review E}\ }\textbf {\bibinfo {volume} {106}},\ \bibinfo {pages} {024903} (\bibinfo {year} {2022})}\BibitemShut {NoStop}%
\bibitem [{\citenamefont {Zhang}, \citenamefont {Stillinger},\ and\ \citenamefont {Torquato}(2016)}]{zhang2016transport}%
  \BibitemOpen
  \bibfield  {author} {\bibinfo {author} {\bibfnamefont {G.}~\bibnamefont {Zhang}}, \bibinfo {author} {\bibfnamefont {F.~H.}\ \bibnamefont {Stillinger}},\ and\ \bibinfo {author} {\bibfnamefont {S.}~\bibnamefont {Torquato}},\ }\bibfield  {title} {\enquote {\bibinfo {title} {Transport, geometrical, and topological properties of stealthy disordered hyperuniform two-phase systems},}\ }\href {https://doi.org/10.1063/1.4972862} {\bibfield  {journal} {\bibinfo  {journal} {The Journal of chemical physics}\ }\textbf {\bibinfo {volume} {145}} (\bibinfo {year} {2016})}\BibitemShut {NoStop}%
\bibitem [{\citenamefont {Shi}, \citenamefont {Jiao},\ and\ \citenamefont {Torquato}(2025)}]{shi2025three}%
  \BibitemOpen
  \bibfield  {author} {\bibinfo {author} {\bibfnamefont {W.}~\bibnamefont {Shi}}, \bibinfo {author} {\bibfnamefont {Y.}~\bibnamefont {Jiao}},\ and\ \bibinfo {author} {\bibfnamefont {S.}~\bibnamefont {Torquato}},\ }\bibfield  {title} {\enquote {\bibinfo {title} {Three-dimensional construction of hyperuniform, nonhyperuniform, and antihyperuniform disordered heterogeneous materials and their transport properties via spectral density functions},}\ }\href {https://doi.org/10.1103/PhysRevE.111.035310} {\bibfield  {journal} {\bibinfo  {journal} {Physical Review E}\ }\textbf {\bibinfo {volume} {111}},\ \bibinfo {pages} {035310} (\bibinfo {year} {2025})}\BibitemShut {NoStop}%
\bibitem [{\citenamefont {Liang}, \citenamefont {Wang},\ and\ \citenamefont {Song}(2024)}]{Liang_Wang_Song_2024}%
  \BibitemOpen
  \bibfield  {author} {\bibinfo {author} {\bibfnamefont {N.}~\bibnamefont {Liang}}, \bibinfo {author} {\bibfnamefont {Y.}~\bibnamefont {Wang}},\ and\ \bibinfo {author} {\bibfnamefont {B.}~\bibnamefont {Song}},\ }\bibfield  {title} {\enquote {\bibinfo {title} {Disordered hyperuniformity and thermal transport in monolayer amorphous carbon},}\ }\href {https://doi.org/10.1007/s11433-024-2523-4} {\bibfield  {journal} {\bibinfo  {journal} {Science China Physics, Mechanics \& Astronomy}\ }\textbf {\bibinfo {volume} {68}} (\bibinfo {year} {2024})}\BibitemShut {NoStop}%
\bibitem [{\citenamefont {Xu}\ \emph {et~al.}(2017)\citenamefont {Xu}, \citenamefont {Chen}, \citenamefont {Chen}, \citenamefont {Xu},\ and\ \citenamefont {Jiao}}]{xu2017microstructure}%
  \BibitemOpen
  \bibfield  {author} {\bibinfo {author} {\bibfnamefont {Y.}~\bibnamefont {Xu}}, \bibinfo {author} {\bibfnamefont {S.}~\bibnamefont {Chen}}, \bibinfo {author} {\bibfnamefont {P.-E.}\ \bibnamefont {Chen}}, \bibinfo {author} {\bibfnamefont {W.}~\bibnamefont {Xu}},\ and\ \bibinfo {author} {\bibfnamefont {Y.}~\bibnamefont {Jiao}},\ }\bibfield  {title} {\enquote {\bibinfo {title} {Microstructure and mechanical properties of hyperuniform heterogeneous materials},}\ }\href {https://doi.org/10.1103/PhysRevE.96.043301} {\bibfield  {journal} {\bibinfo  {journal} {Physical Review E}\ }\textbf {\bibinfo {volume} {96}},\ \bibinfo {pages} {043301} (\bibinfo {year} {2017})}\BibitemShut {NoStop}%
\bibitem [{\citenamefont {Torquato}(2022)}]{torquato2022extraordinary}%
  \BibitemOpen
  \bibfield  {author} {\bibinfo {author} {\bibfnamefont {S.}~\bibnamefont {Torquato}},\ }\bibfield  {title} {\enquote {\bibinfo {title} {Extraordinary disordered hyperuniform multifunctional composites},}\ }\href {https://doi.org/10.1177/00219983221116432} {\bibfield  {journal} {\bibinfo  {journal} {Journal of Composite Materials}\ }\textbf {\bibinfo {volume} {56}},\ \bibinfo {pages} {3635--3649} (\bibinfo {year} {2022})}\BibitemShut {NoStop}%
\bibitem [{\citenamefont {Khairunnisa}\ \emph {et~al.}(2025)\citenamefont {Khairunnisa}, \citenamefont {Guselnikova}, \citenamefont {Kang}, \citenamefont {Postnikov}, \citenamefont {Valiev}, \citenamefont {Hill}, \citenamefont {Nugraha}, \citenamefont {Yuliarto}, \citenamefont {Yamauchi},\ and\ \citenamefont {Henzie}}]{khairunnisa2025hyperuniform}%
  \BibitemOpen
  \bibfield  {author} {\bibinfo {author} {\bibfnamefont {S.~Z.}\ \bibnamefont {Khairunnisa}}, \bibinfo {author} {\bibfnamefont {O.}~\bibnamefont {Guselnikova}}, \bibinfo {author} {\bibfnamefont {Y.}~\bibnamefont {Kang}}, \bibinfo {author} {\bibfnamefont {P.~S.}\ \bibnamefont {Postnikov}}, \bibinfo {author} {\bibfnamefont {R.~R.}\ \bibnamefont {Valiev}}, \bibinfo {author} {\bibfnamefont {J.~P.}\ \bibnamefont {Hill}}, \bibinfo {author} {\bibfnamefont {N.}~\bibnamefont {Nugraha}}, \bibinfo {author} {\bibfnamefont {B.}~\bibnamefont {Yuliarto}}, \bibinfo {author} {\bibfnamefont {Y.}~\bibnamefont {Yamauchi}},\ and\ \bibinfo {author} {\bibfnamefont {J.}~\bibnamefont {Henzie}},\ }\bibfield  {title} {\enquote {\bibinfo {title} {Hyperuniform mesoporous gold films coated with halogen-bonding metal--organic frameworks for selective raman sensing of chlorinated hydrocarbons},}\ }\href {https://doi.org/10.1021/acsnano.5c09431} {\bibfield  {journal} {\bibinfo  {journal} {ACS nano}\ } (\bibinfo {year} {2025})}\BibitemShut
  {NoStop}%
\bibitem [{\citenamefont {Holm}\ \emph {et~al.}(2020)\citenamefont {Holm}, \citenamefont {Goodman}, \citenamefont {Stenlid}, \citenamefont {Aitbekova}, \citenamefont {Zelaya}, \citenamefont {Diroll}, \citenamefont {Johnston-Peck}, \citenamefont {Kao}, \citenamefont {Frank}, \citenamefont {Pettersson} \emph {et~al.}}]{holm2020nanoscale}%
  \BibitemOpen
  \bibfield  {author} {\bibinfo {author} {\bibfnamefont {A.}~\bibnamefont {Holm}}, \bibinfo {author} {\bibfnamefont {E.~D.}\ \bibnamefont {Goodman}}, \bibinfo {author} {\bibfnamefont {J.~H.}\ \bibnamefont {Stenlid}}, \bibinfo {author} {\bibfnamefont {A.}~\bibnamefont {Aitbekova}}, \bibinfo {author} {\bibfnamefont {R.}~\bibnamefont {Zelaya}}, \bibinfo {author} {\bibfnamefont {B.~T.}\ \bibnamefont {Diroll}}, \bibinfo {author} {\bibfnamefont {A.~C.}\ \bibnamefont {Johnston-Peck}}, \bibinfo {author} {\bibfnamefont {K.-C.}\ \bibnamefont {Kao}}, \bibinfo {author} {\bibfnamefont {C.~W.}\ \bibnamefont {Frank}}, \bibinfo {author} {\bibfnamefont {L.~G.}\ \bibnamefont {Pettersson}}, \emph {et~al.},\ }\bibfield  {title} {\enquote {\bibinfo {title} {Nanoscale spatial distribution of supported nanoparticles controls activity and stability in powder catalysts for co oxidation and photocatalytic h2 evolution},}\ }\href {https://doi.org/10.1021/jacs.0c03842} {\bibfield  {journal} {\bibinfo  {journal} {Journal of the American
  Chemical Society}\ }\textbf {\bibinfo {volume} {142}},\ \bibinfo {pages} {14481--14494} (\bibinfo {year} {2020})}\BibitemShut {NoStop}%
\bibitem [{\citenamefont {Tavakoli}\ \emph {et~al.}(2022)\citenamefont {Tavakoli}, \citenamefont {Spalding}, \citenamefont {Lambertz}, \citenamefont {Koppejan}, \citenamefont {Gkantzounis}, \citenamefont {Wan}, \citenamefont {Rohrich}, \citenamefont {Kontoleta}, \citenamefont {Koenderink}, \citenamefont {Sapienza} \emph {et~al.}}]{tavakoli2022over}%
  \BibitemOpen
  \bibfield  {author} {\bibinfo {author} {\bibfnamefont {N.}~\bibnamefont {Tavakoli}}, \bibinfo {author} {\bibfnamefont {R.}~\bibnamefont {Spalding}}, \bibinfo {author} {\bibfnamefont {A.}~\bibnamefont {Lambertz}}, \bibinfo {author} {\bibfnamefont {P.}~\bibnamefont {Koppejan}}, \bibinfo {author} {\bibfnamefont {G.}~\bibnamefont {Gkantzounis}}, \bibinfo {author} {\bibfnamefont {C.}~\bibnamefont {Wan}}, \bibinfo {author} {\bibfnamefont {R.}~\bibnamefont {Rohrich}}, \bibinfo {author} {\bibfnamefont {E.}~\bibnamefont {Kontoleta}}, \bibinfo {author} {\bibfnamefont {A.~F.}\ \bibnamefont {Koenderink}}, \bibinfo {author} {\bibfnamefont {R.}~\bibnamefont {Sapienza}}, \emph {et~al.},\ }\bibfield  {title} {\enquote {\bibinfo {title} {Over 65\% sunlight absorption in a 1 $\mu$m si slab with hyperuniform texture},}\ }\href {https://doi.org/10.1021/acsphotonics.1c01668} {\bibfield  {journal} {\bibinfo  {journal} {ACS photonics}\ }\textbf {\bibinfo {volume} {9}},\ \bibinfo {pages} {1206--1217} (\bibinfo {year}
  {2022})}\BibitemShut {NoStop}%
\bibitem [{\citenamefont {Mackay}\ \emph {et~al.}(2024)\citenamefont {Mackay}, \citenamefont {Marbach}, \citenamefont {Sprinkle},\ and\ \citenamefont {Thorneywork}}]{mackay2024countoscope}%
  \BibitemOpen
  \bibfield  {author} {\bibinfo {author} {\bibfnamefont {E.~K.}\ \bibnamefont {Mackay}}, \bibinfo {author} {\bibfnamefont {S.}~\bibnamefont {Marbach}}, \bibinfo {author} {\bibfnamefont {B.}~\bibnamefont {Sprinkle}},\ and\ \bibinfo {author} {\bibfnamefont {A.~L.}\ \bibnamefont {Thorneywork}},\ }\bibfield  {title} {\enquote {\bibinfo {title} {The countoscope: measuring self and collective dynamics without trajectories},}\ }\href {https://doi.org/10.1103/PhysRevX.14.041016} {\bibfield  {journal} {\bibinfo  {journal} {Physical Review X}\ }\textbf {\bibinfo {volume} {14}},\ \bibinfo {pages} {041016} (\bibinfo {year} {2024})}\BibitemShut {NoStop}%
\bibitem [{\citenamefont {Garz{\'o}}, \citenamefont {Brito},\ and\ \citenamefont {Soto}(2018)}]{garzo2018enskog}%
  \BibitemOpen
  \bibfield  {author} {\bibinfo {author} {\bibfnamefont {V.}~\bibnamefont {Garz{\'o}}}, \bibinfo {author} {\bibfnamefont {R.}~\bibnamefont {Brito}},\ and\ \bibinfo {author} {\bibfnamefont {R.}~\bibnamefont {Soto}},\ }\bibfield  {title} {\enquote {\bibinfo {title} {Enskog kinetic theory for a model of a confined quasi-two-dimensional granular fluid},}\ }\href {https://doi.org/10.1103/PhysRevE.98.052904} {\bibfield  {journal} {\bibinfo  {journal} {Physical Review E}\ }\textbf {\bibinfo {volume} {98}},\ \bibinfo {pages} {052904} (\bibinfo {year} {2018})}\BibitemShut {NoStop}%
\bibitem [{\citenamefont {Landau}\ and\ \citenamefont {Lifshitz}(2013)}]{landau2013statistical}%
  \BibitemOpen
  \bibfield  {author} {\bibinfo {author} {\bibfnamefont {L.~D.}\ \bibnamefont {Landau}}\ and\ \bibinfo {author} {\bibfnamefont {E.~M.}\ \bibnamefont {Lifshitz}},\ }\href {https://www.sciencedirect.com/book/9780080570464/statistical-physics} {\emph {\bibinfo {title} {Statistical Physics: Volume 5}}},\ Vol.~\bibinfo {volume} {5}\ (\bibinfo  {publisher} {Elsevier},\ \bibinfo {year} {2013})\BibitemShut {NoStop}%
\bibitem [{\citenamefont {Fruchart}, \citenamefont {Scheibner},\ and\ \citenamefont {Vitelli}(2023)}]{fruchart2023odd}%
  \BibitemOpen
  \bibfield  {author} {\bibinfo {author} {\bibfnamefont {M.}~\bibnamefont {Fruchart}}, \bibinfo {author} {\bibfnamefont {C.}~\bibnamefont {Scheibner}},\ and\ \bibinfo {author} {\bibfnamefont {V.}~\bibnamefont {Vitelli}},\ }\bibfield  {title} {\enquote {\bibinfo {title} {Odd viscosity and odd elasticity},}\ }\href {https://doi.org/10.1146/annurev-conmatphys-040821-125506} {\bibfield  {journal} {\bibinfo  {journal} {Annual Review of Condensed Matter Physics}\ }\textbf {\bibinfo {volume} {14}},\ \bibinfo {pages} {471--510} (\bibinfo {year} {2023})}\BibitemShut {NoStop}%
\bibitem [{\citenamefont {Brito}, \citenamefont {Risso},\ and\ \citenamefont {Soto}(2013)}]{brito2013hydrodynamic}%
  \BibitemOpen
  \bibfield  {author} {\bibinfo {author} {\bibfnamefont {R.}~\bibnamefont {Brito}}, \bibinfo {author} {\bibfnamefont {D.}~\bibnamefont {Risso}},\ and\ \bibinfo {author} {\bibfnamefont {R.}~\bibnamefont {Soto}},\ }\bibfield  {title} {\enquote {\bibinfo {title} {Hydrodynamic modes in a confined granular fluid},}\ }\href {https://doi.org/10.1103/PhysRevE.87.022209} {\bibfield  {journal} {\bibinfo  {journal} {Physical Review E—Statistical, Nonlinear, and Soft Matter Physics}\ }\textbf {\bibinfo {volume} {87}},\ \bibinfo {pages} {022209} (\bibinfo {year} {2013})}\BibitemShut {NoStop}%
\bibitem [{\citenamefont {Sekimoto}(2010)}]{sekimoto2010stochastic}%
  \BibitemOpen
  \bibfield  {author} {\bibinfo {author} {\bibfnamefont {K.}~\bibnamefont {Sekimoto}},\ }\href {https://doi.org/10.1007/978-3-642-05411-2} {\enquote {\bibinfo {title} {Stochastic energetics},}\ } (\bibinfo {year} {2010})\BibitemShut {NoStop}%
\bibitem [{\citenamefont {Cates}\ and\ \citenamefont {Nardini}(2025)}]{cates2024active}%
  \BibitemOpen
  \bibfield  {author} {\bibinfo {author} {\bibfnamefont {M.~E.}\ \bibnamefont {Cates}}\ and\ \bibinfo {author} {\bibfnamefont {C.}~\bibnamefont {Nardini}},\ }\bibfield  {title} {\enquote {\bibinfo {title} {Active phase separation: new phenomenology from non-equilibrium physics},}\ }\href {https://doi.org/10.1088/1361-6633/add278} {\bibfield  {journal} {\bibinfo  {journal} {Reports on Progress in Physics}\ }\textbf {\bibinfo {volume} {88}},\ \bibinfo {pages} {056601} (\bibinfo {year} {2025})}\BibitemShut {NoStop}%
\bibitem [{\citenamefont {Marconi}\ \emph {et~al.}(2008)\citenamefont {Marconi}, \citenamefont {Puglisi}, \citenamefont {Rondoni},\ and\ \citenamefont {Vulpiani}}]{marconi2008fluctuation}%
  \BibitemOpen
  \bibfield  {author} {\bibinfo {author} {\bibfnamefont {U.~M.~B.}\ \bibnamefont {Marconi}}, \bibinfo {author} {\bibfnamefont {A.}~\bibnamefont {Puglisi}}, \bibinfo {author} {\bibfnamefont {L.}~\bibnamefont {Rondoni}},\ and\ \bibinfo {author} {\bibfnamefont {A.}~\bibnamefont {Vulpiani}},\ }\bibfield  {title} {\enquote {\bibinfo {title} {Fluctuation--dissipation: response theory in statistical physics},}\ }\href {https://doi.org/10.1016/j.physrep.2008.02.002} {\bibfield  {journal} {\bibinfo  {journal} {Physics reports}\ }\textbf {\bibinfo {volume} {461}},\ \bibinfo {pages} {111--195} (\bibinfo {year} {2008})}\BibitemShut {NoStop}%
\bibitem [{\citenamefont {Bonachela}\ and\ \citenamefont {Munoz}(2009)}]{bonachela2009self}%
  \BibitemOpen
  \bibfield  {author} {\bibinfo {author} {\bibfnamefont {J.~A.}\ \bibnamefont {Bonachela}}\ and\ \bibinfo {author} {\bibfnamefont {M.~A.}\ \bibnamefont {Munoz}},\ }\bibfield  {title} {\enquote {\bibinfo {title} {Self-organization without conservation: true or just apparent scale-invariance?}}\ }\href {https://doi.org/10.1088/1742-5468/2009/09/P09009} {\bibfield  {journal} {\bibinfo  {journal} {Journal of Statistical Mechanics: Theory and Experiment}\ }\textbf {\bibinfo {volume} {2009}},\ \bibinfo {pages} {P09009} (\bibinfo {year} {2009})}\BibitemShut {NoStop}%
\bibitem [{\citenamefont {Grinstein}, \citenamefont {Lee},\ and\ \citenamefont {Sachdev}(1990)}]{grinstein1990conservation}%
  \BibitemOpen
  \bibfield  {author} {\bibinfo {author} {\bibfnamefont {G.}~\bibnamefont {Grinstein}}, \bibinfo {author} {\bibfnamefont {D.-H.}\ \bibnamefont {Lee}},\ and\ \bibinfo {author} {\bibfnamefont {S.}~\bibnamefont {Sachdev}},\ }\bibfield  {title} {\enquote {\bibinfo {title} {Conservation laws, anisotropy, and ‘‘self-organized criticality’’in noisy nonequilibrium systems},}\ }\href {https://doi.org/10.1103/PhysRevLett.64.1927} {\bibfield  {journal} {\bibinfo  {journal} {Physical review letters}\ }\textbf {\bibinfo {volume} {64}},\ \bibinfo {pages} {1927} (\bibinfo {year} {1990})}\BibitemShut {NoStop}%
\bibitem [{\citenamefont {Garrido}\ \emph {et~al.}(1990)\citenamefont {Garrido}, \citenamefont {Lebowitz}, \citenamefont {Maes},\ and\ \citenamefont {Spohn}}]{garrido1990long}%
  \BibitemOpen
  \bibfield  {author} {\bibinfo {author} {\bibfnamefont {P.~L.}\ \bibnamefont {Garrido}}, \bibinfo {author} {\bibfnamefont {J.~L.}\ \bibnamefont {Lebowitz}}, \bibinfo {author} {\bibfnamefont {C.}~\bibnamefont {Maes}},\ and\ \bibinfo {author} {\bibfnamefont {H.}~\bibnamefont {Spohn}},\ }\bibfield  {title} {\enquote {\bibinfo {title} {Long-range correlations for conservative dynamics},}\ }\href {https://doi.org/10.1103/PhysRevA.42.1954} {\bibfield  {journal} {\bibinfo  {journal} {Physical Review A}\ }\textbf {\bibinfo {volume} {42}},\ \bibinfo {pages} {1954} (\bibinfo {year} {1990})}\BibitemShut {NoStop}%
\bibitem [{\citenamefont {Van~Noije}\ \emph {et~al.}(1999)\citenamefont {Van~Noije}, \citenamefont {Ernst}, \citenamefont {Trizac},\ and\ \citenamefont {Pagonabarraga}}]{van1999randomly}%
  \BibitemOpen
  \bibfield  {author} {\bibinfo {author} {\bibfnamefont {T.}~\bibnamefont {Van~Noije}}, \bibinfo {author} {\bibfnamefont {M.}~\bibnamefont {Ernst}}, \bibinfo {author} {\bibfnamefont {E.}~\bibnamefont {Trizac}},\ and\ \bibinfo {author} {\bibfnamefont {I.}~\bibnamefont {Pagonabarraga}},\ }\bibfield  {title} {\enquote {\bibinfo {title} {Randomly driven granular fluids: Large-scale structure},}\ }\href {https://doi.org/10.1103/PhysRevE.59.4326} {\bibfield  {journal} {\bibinfo  {journal} {Physical Review E}\ }\textbf {\bibinfo {volume} {59}},\ \bibinfo {pages} {4326} (\bibinfo {year} {1999})}\BibitemShut {NoStop}%
\bibitem [{\citenamefont {Plati}\ and\ \citenamefont {Puglisi}(2021)}]{plati2021long}%
  \BibitemOpen
  \bibfield  {author} {\bibinfo {author} {\bibfnamefont {A.}~\bibnamefont {Plati}}\ and\ \bibinfo {author} {\bibfnamefont {A.}~\bibnamefont {Puglisi}},\ }\bibfield  {title} {\enquote {\bibinfo {title} {Long range correlations and slow time scales in a boundary driven granular model},}\ }\href {https://doi.org/10.1038/s41598-021-93091-1} {\bibfield  {journal} {\bibinfo  {journal} {Scientific Reports}\ }\textbf {\bibinfo {volume} {11}},\ \bibinfo {pages} {14206} (\bibinfo {year} {2021})}\BibitemShut {NoStop}%
\bibitem [{\citenamefont {Simha}\ and\ \citenamefont {Ramaswamy}(2002)}]{simha2002hydrodynamic}%
  \BibitemOpen
  \bibfield  {author} {\bibinfo {author} {\bibfnamefont {R.~A.}\ \bibnamefont {Simha}}\ and\ \bibinfo {author} {\bibfnamefont {S.}~\bibnamefont {Ramaswamy}},\ }\bibfield  {title} {\enquote {\bibinfo {title} {Hydrodynamic fluctuations and instabilities in ordered suspensions of self-propelled particles},}\ }\href {https://doi.org/10.1103/PhysRevLett.89.058101} {\bibfield  {journal} {\bibinfo  {journal} {Physical review letters}\ }\textbf {\bibinfo {volume} {89}},\ \bibinfo {pages} {058101} (\bibinfo {year} {2002})}\BibitemShut {NoStop}%
\bibitem [{\citenamefont {Kundu}, \citenamefont {Hirschberg},\ and\ \citenamefont {Mukamel}(2016)}]{kundu2016long}%
  \BibitemOpen
  \bibfield  {author} {\bibinfo {author} {\bibfnamefont {A.}~\bibnamefont {Kundu}}, \bibinfo {author} {\bibfnamefont {O.}~\bibnamefont {Hirschberg}},\ and\ \bibinfo {author} {\bibfnamefont {D.}~\bibnamefont {Mukamel}},\ }\bibfield  {title} {\enquote {\bibinfo {title} {Long range correlations in stochastic transport with energy and momentum conservation},}\ }\href {https://doi.org/10.1088/1742-5468/2016/03/033108} {\bibfield  {journal} {\bibinfo  {journal} {Journal of Statistical Mechanics: Theory and Experiment}\ }\textbf {\bibinfo {volume} {2016}},\ \bibinfo {pages} {033108} (\bibinfo {year} {2016})}\BibitemShut {NoStop}%
\bibitem [{\citenamefont {Spohn}(1983)}]{spohn1983long}%
  \BibitemOpen
  \bibfield  {author} {\bibinfo {author} {\bibfnamefont {H.}~\bibnamefont {Spohn}},\ }\bibfield  {title} {\enquote {\bibinfo {title} {Long range correlations for stochastic lattice gases in a non-equilibrium steady state},}\ }\href {https://doi.org/10.1088/0305-4470/16/18/029} {\bibfield  {journal} {\bibinfo  {journal} {Journal of Physics A: Mathematical and General}\ }\textbf {\bibinfo {volume} {16}},\ \bibinfo {pages} {4275} (\bibinfo {year} {1983})}\BibitemShut {NoStop}%
\bibitem [{\citenamefont {Dorfman}, \citenamefont {Kirkpatrick},\ and\ \citenamefont {Sengers}(1994)}]{dorfman1994generic}%
  \BibitemOpen
  \bibfield  {author} {\bibinfo {author} {\bibfnamefont {J.}~\bibnamefont {Dorfman}}, \bibinfo {author} {\bibfnamefont {T.}~\bibnamefont {Kirkpatrick}},\ and\ \bibinfo {author} {\bibfnamefont {J.}~\bibnamefont {Sengers}},\ }\bibfield  {title} {\enquote {\bibinfo {title} {Generic long-range correlations in molecular fluids},}\ }\href {https://doi.org/10.1146/annurev.pc.45.100194.001241} {\bibfield  {journal} {\bibinfo  {journal} {Annual Review of Physical Chemistry}\ }\textbf {\bibinfo {volume} {45}},\ \bibinfo {pages} {213--239} (\bibinfo {year} {1994})}\BibitemShut {NoStop}%
\bibitem [{\citenamefont {Grinstein}(1991)}]{grinstein1991generic}%
  \BibitemOpen
  \bibfield  {author} {\bibinfo {author} {\bibfnamefont {G.}~\bibnamefont {Grinstein}},\ }\bibfield  {title} {\enquote {\bibinfo {title} {Generic scale invariance in classical nonequilibrium systems},}\ }\href {https://doi.org/10.1063/1.348003} {\bibfield  {journal} {\bibinfo  {journal} {Journal of applied physics}\ }\textbf {\bibinfo {volume} {69}},\ \bibinfo {pages} {5441--5446} (\bibinfo {year} {1991})}\BibitemShut {NoStop}%
\bibitem [{\citenamefont {Bak}(1992)}]{bak1992self}%
  \BibitemOpen
  \bibfield  {author} {\bibinfo {author} {\bibfnamefont {P.}~\bibnamefont {Bak}},\ }\bibfield  {title} {\enquote {\bibinfo {title} {Self-organized criticality in non-conservative models},}\ }\href {https://doi.org/10.1016/0378-4371(92)90503-I} {\bibfield  {journal} {\bibinfo  {journal} {Physica A: Statistical Mechanics and its Applications}\ }\textbf {\bibinfo {volume} {191}},\ \bibinfo {pages} {41--46} (\bibinfo {year} {1992})}\BibitemShut {NoStop}%
\bibitem [{\citenamefont {Giusfredi}, \citenamefont {Iubini},\ and\ \citenamefont {Politi}(2024)}]{giusfredi2024localization}%
  \BibitemOpen
  \bibfield  {author} {\bibinfo {author} {\bibfnamefont {M.}~\bibnamefont {Giusfredi}}, \bibinfo {author} {\bibfnamefont {S.}~\bibnamefont {Iubini}},\ and\ \bibinfo {author} {\bibfnamefont {P.}~\bibnamefont {Politi}},\ }\bibfield  {title} {\enquote {\bibinfo {title} {Localization in boundary-driven lattice models},}\ }\href {https://doi.org/10.1007/s10955-024-03324-6} {\bibfield  {journal} {\bibinfo  {journal} {Journal of Statistical Physics}\ }\textbf {\bibinfo {volume} {191}},\ \bibinfo {pages} {119} (\bibinfo {year} {2024})}\BibitemShut {NoStop}%
\bibitem [{\citenamefont {Tan}\ \emph {et~al.}(2022)\citenamefont {Tan}, \citenamefont {Mietke}, \citenamefont {Li}, \citenamefont {Chen}, \citenamefont {Higinbotham}, \citenamefont {Foster}, \citenamefont {Gokhale}, \citenamefont {Dunkel},\ and\ \citenamefont {Fakhri}}]{tan2022odd}%
  \BibitemOpen
  \bibfield  {author} {\bibinfo {author} {\bibfnamefont {T.~H.}\ \bibnamefont {Tan}}, \bibinfo {author} {\bibfnamefont {A.}~\bibnamefont {Mietke}}, \bibinfo {author} {\bibfnamefont {J.}~\bibnamefont {Li}}, \bibinfo {author} {\bibfnamefont {Y.}~\bibnamefont {Chen}}, \bibinfo {author} {\bibfnamefont {H.}~\bibnamefont {Higinbotham}}, \bibinfo {author} {\bibfnamefont {P.~J.}\ \bibnamefont {Foster}}, \bibinfo {author} {\bibfnamefont {S.}~\bibnamefont {Gokhale}}, \bibinfo {author} {\bibfnamefont {J.}~\bibnamefont {Dunkel}},\ and\ \bibinfo {author} {\bibfnamefont {N.}~\bibnamefont {Fakhri}},\ }\bibfield  {title} {\enquote {\bibinfo {title} {Odd dynamics of living chiral crystals},}\ }\href {https://doi.org/10.1038/s41586-022-04889-6} {\bibfield  {journal} {\bibinfo  {journal} {Nature}\ }\textbf {\bibinfo {volume} {607}},\ \bibinfo {pages} {287--293} (\bibinfo {year} {2022})}\BibitemShut {NoStop}%
\bibitem [{\citenamefont {Massana-Cid}\ \emph {et~al.}(2021)\citenamefont {Massana-Cid}, \citenamefont {Levis}, \citenamefont {Hern{\'a}ndez}, \citenamefont {Pagonabarraga},\ and\ \citenamefont {Tierno}}]{massana2021arrested}%
  \BibitemOpen
  \bibfield  {author} {\bibinfo {author} {\bibfnamefont {H.}~\bibnamefont {Massana-Cid}}, \bibinfo {author} {\bibfnamefont {D.}~\bibnamefont {Levis}}, \bibinfo {author} {\bibfnamefont {R.~J.~H.}\ \bibnamefont {Hern{\'a}ndez}}, \bibinfo {author} {\bibfnamefont {I.}~\bibnamefont {Pagonabarraga}},\ and\ \bibinfo {author} {\bibfnamefont {P.}~\bibnamefont {Tierno}},\ }\bibfield  {title} {\enquote {\bibinfo {title} {Arrested phase separation in chiral fluids of colloidal spinners},}\ }\href {https://doi.org/10.1103/PhysRevResearch.3.L042021} {\bibfield  {journal} {\bibinfo  {journal} {Physical Review Research}\ }\textbf {\bibinfo {volume} {3}},\ \bibinfo {pages} {L042021} (\bibinfo {year} {2021})}\BibitemShut {NoStop}%
\bibitem [{\citenamefont {Caprini}\ and\ \citenamefont {Marini Bettolo~Marconi}(2025)}]{caprini2025Bubble}%
  \BibitemOpen
  \bibfield  {author} {\bibinfo {author} {\bibfnamefont {L.}~\bibnamefont {Caprini}}\ and\ \bibinfo {author} {\bibfnamefont {U.}~\bibnamefont {Marini Bettolo~Marconi}},\ }\bibfield  {title} {\enquote {\bibinfo {title} {Bubble phase induced by odd interactions in chiral systems},}\ }\href {https://doi.org/10.1063/5.0262594} {\bibfield  {journal} {\bibinfo  {journal} {The Journal of Chemical Physics}\ }\textbf {\bibinfo {volume} {162}},\ \bibinfo {pages} {161101} (\bibinfo {year} {2025})}\BibitemShut {NoStop}%
\bibitem [{\citenamefont {Caporusso}, \citenamefont {Gonnella},\ and\ \citenamefont {Levis}(2024)}]{caporusso2024phase}%
  \BibitemOpen
  \bibfield  {author} {\bibinfo {author} {\bibfnamefont {C.~B.}\ \bibnamefont {Caporusso}}, \bibinfo {author} {\bibfnamefont {G.}~\bibnamefont {Gonnella}},\ and\ \bibinfo {author} {\bibfnamefont {D.}~\bibnamefont {Levis}},\ }\bibfield  {title} {\enquote {\bibinfo {title} {Phase coexistence and edge currents in the chiral lennard-jones fluid},}\ }\href {https://doi.org/10.1103/PhysRevLett.132.168201} {\bibfield  {journal} {\bibinfo  {journal} {Physical Review Letters}\ }\textbf {\bibinfo {volume} {132}},\ \bibinfo {pages} {168201} (\bibinfo {year} {2024})}\BibitemShut {NoStop}%
\bibitem [{\citenamefont {Groot}\ and\ \citenamefont {Warren}(1997)}]{groot1997dissipative}%
  \BibitemOpen
  \bibfield  {author} {\bibinfo {author} {\bibfnamefont {R.~D.}\ \bibnamefont {Groot}}\ and\ \bibinfo {author} {\bibfnamefont {P.~B.}\ \bibnamefont {Warren}},\ }\bibfield  {title} {\enquote {\bibinfo {title} {Dissipative particle dynamics: Bridging the gap between atomistic and mesoscopic simulation},}\ }\href {https://doi.org/10.1063/1.474784} {\bibfield  {journal} {\bibinfo  {journal} {The Journal of chemical physics}\ }\textbf {\bibinfo {volume} {107}},\ \bibinfo {pages} {4423--4435} (\bibinfo {year} {1997})}\BibitemShut {NoStop}%
\bibitem [{\citenamefont {de~Graaf~Sousa}\ \emph {et~al.}(2025)\citenamefont {de~Graaf~Sousa}, \citenamefont {Andersen}, \citenamefont {Ardaševa},\ and\ \citenamefont {Doostmohammadi}}]{sousa2025selfpropulsiveactivenematics}%
  \BibitemOpen
  \bibfield  {author} {\bibinfo {author} {\bibfnamefont {N.}~\bibnamefont {de~Graaf~Sousa}}, \bibinfo {author} {\bibfnamefont {S.~G.}\ \bibnamefont {Andersen}}, \bibinfo {author} {\bibfnamefont {A.}~\bibnamefont {Ardaševa}},\ and\ \bibinfo {author} {\bibfnamefont {A.}~\bibnamefont {Doostmohammadi}},\ }\href {https://arxiv.org/abs/2509.02386} {\enquote {\bibinfo {title} {Self-propulsive active nematics},}\ } (\bibinfo {year} {2025})\BibitemShut {NoStop}%
\bibitem [{\citenamefont {Ngo}\ \emph {et~al.}(2014)\citenamefont {Ngo}, \citenamefont {Peshkov}, \citenamefont {Aranson}, \citenamefont {Bertin}, \citenamefont {Ginelli},\ and\ \citenamefont {Chat\'e}}]{PhysRevLett.113.038302}%
  \BibitemOpen
  \bibfield  {author} {\bibinfo {author} {\bibfnamefont {S.}~\bibnamefont {Ngo}}, \bibinfo {author} {\bibfnamefont {A.}~\bibnamefont {Peshkov}}, \bibinfo {author} {\bibfnamefont {I.~S.}\ \bibnamefont {Aranson}}, \bibinfo {author} {\bibfnamefont {E.}~\bibnamefont {Bertin}}, \bibinfo {author} {\bibfnamefont {F.}~\bibnamefont {Ginelli}},\ and\ \bibinfo {author} {\bibfnamefont {H.}~\bibnamefont {Chat\'e}},\ }\bibfield  {title} {\enquote {\bibinfo {title} {Large-scale chaos and fluctuations in active nematics},}\ }\href {https://link.aps.org/doi/10.1103/PhysRevLett.113.038302} {\bibfield  {journal} {\bibinfo  {journal} {Phys. Rev. Lett.}\ }\textbf {\bibinfo {volume} {113}},\ \bibinfo {pages} {038302} (\bibinfo {year} {2014})}\BibitemShut {NoStop}%
\bibitem [{\citenamefont {Ramaswamy}, \citenamefont {Simha},\ and\ \citenamefont {Toner}(2003)}]{ramaswamy2003active}%
  \BibitemOpen
  \bibfield  {author} {\bibinfo {author} {\bibfnamefont {S.}~\bibnamefont {Ramaswamy}}, \bibinfo {author} {\bibfnamefont {R.~A.}\ \bibnamefont {Simha}},\ and\ \bibinfo {author} {\bibfnamefont {J.}~\bibnamefont {Toner}},\ }\bibfield  {title} {\enquote {\bibinfo {title} {Active nematics on a substrate: Giantnumber fluctuations and long-time tails},}\ }\href {https://doi.org/10.1209/epl/i2003-00346-7} {\bibfield  {journal} {\bibinfo  {journal} {Europhysics Letters}\ }\textbf {\bibinfo {volume} {62}},\ \bibinfo {pages} {196} (\bibinfo {year} {2003})}\BibitemShut {NoStop}%
\bibitem [{\citenamefont {Zhang}\ and\ \citenamefont {Fodor}(2023)}]{zhang2023pulsating}%
  \BibitemOpen
  \bibfield  {author} {\bibinfo {author} {\bibfnamefont {Y.}~\bibnamefont {Zhang}}\ and\ \bibinfo {author} {\bibfnamefont {{\'E}.}~\bibnamefont {Fodor}},\ }\bibfield  {title} {\enquote {\bibinfo {title} {Pulsating active matter},}\ }\href {https://doi.org/10.1103/PhysRevLett.131.238302} {\bibfield  {journal} {\bibinfo  {journal} {Physical Review Letters}\ }\textbf {\bibinfo {volume} {131}},\ \bibinfo {pages} {238302} (\bibinfo {year} {2023})}\BibitemShut {NoStop}%
\bibitem [{\citenamefont {Banerjee}\ \emph {et~al.}(2024)\citenamefont {Banerjee}, \citenamefont {Desaleux}, \citenamefont {Ranft},\ and\ \citenamefont {Fodor}}]{banerjee2024hydrodynamics}%
  \BibitemOpen
  \bibfield  {author} {\bibinfo {author} {\bibfnamefont {T.}~\bibnamefont {Banerjee}}, \bibinfo {author} {\bibfnamefont {T.}~\bibnamefont {Desaleux}}, \bibinfo {author} {\bibfnamefont {J.}~\bibnamefont {Ranft}},\ and\ \bibinfo {author} {\bibfnamefont {{\'E}.}~\bibnamefont {Fodor}},\ }\bibfield  {title} {\enquote {\bibinfo {title} {Hydrodynamics of pulsating active liquids},}\ }\href {https://doi.org/10.48550/arXiv.2407.19955} {\bibfield  {journal} {\bibinfo  {journal} {arXiv:2407.19955}\ } (\bibinfo {year} {2024})}\BibitemShut {NoStop}%
\bibitem [{\citenamefont {Tjhung}\ and\ \citenamefont {Berthier}(2017)}]{tjhung2017discontinuous}%
  \BibitemOpen
  \bibfield  {author} {\bibinfo {author} {\bibfnamefont {E.}~\bibnamefont {Tjhung}}\ and\ \bibinfo {author} {\bibfnamefont {L.}~\bibnamefont {Berthier}},\ }\bibfield  {title} {\enquote {\bibinfo {title} {Discontinuous fluidization transition in time-correlated assemblies of actively deforming particles},}\ }\href {https://doi.org/10.1103/PhysRevE.96.050601} {\bibfield  {journal} {\bibinfo  {journal} {Physical Review E}\ }\textbf {\bibinfo {volume} {96}},\ \bibinfo {pages} {050601} (\bibinfo {year} {2017})}\BibitemShut {NoStop}%
\bibitem [{\citenamefont {Kuroda}\ \emph {et~al.}(2023)\citenamefont {Kuroda}, \citenamefont {Matsuyama}, \citenamefont {Kawasaki},\ and\ \citenamefont {Miyazaki}}]{kuroda2023anomalous}%
  \BibitemOpen
  \bibfield  {author} {\bibinfo {author} {\bibfnamefont {Y.}~\bibnamefont {Kuroda}}, \bibinfo {author} {\bibfnamefont {H.}~\bibnamefont {Matsuyama}}, \bibinfo {author} {\bibfnamefont {T.}~\bibnamefont {Kawasaki}},\ and\ \bibinfo {author} {\bibfnamefont {K.}~\bibnamefont {Miyazaki}},\ }\bibfield  {title} {\enquote {\bibinfo {title} {Anomalous fluctuations in homogeneous fluid phase of active brownian particles},}\ }\href {https://doi.org/10.1103/PhysRevResearch.5.013077} {\bibfield  {journal} {\bibinfo  {journal} {Phys. Rev. Res.}\ }\textbf {\bibinfo {volume} {5}},\ \bibinfo {pages} {013077} (\bibinfo {year} {2023})}\BibitemShut {NoStop}%
\bibitem [{Note1()}]{Note1}%
  \BibitemOpen
  \bibinfo {note} {One could argue that these systems might be described by a generalized Gibbs ensemble or by a generalized microcanonical ensemble. For instance, if $\protect \mathcal {\protect \bm {Q}}_n$ is conserved, the microcanonical entropy would take the form~\cite {landau2013statistical}: $S(E, N, V, \protect \mathcal {\protect \bm {Q}}_n)=\log \left (\Omega (E, N, V, \protect \mathcal {\protect \bm {Q}}_n)\right ) = \log \left (\DOTSB \sum@ \slimits@ _{\protect \rm state} \delta (E - E^{\protect \rm state}) \delta (\protect \mathcal {\protect \bm {Q}}_n - \protect \mathcal {\protect \bm {Q}}_n^{\protect \rm state})\right )$. However, the additional constraints are likely subextensive and thus vanish in the thermodynamic limit~\cite {niiyama2009effect}. For example, in standard molecular dynamics simulations of systems with short-range interactions, the center-of-mass velocity, linear momentum, and angular momentum are naturally conserved. However, even when these quantities are not explicitly
  included in the microcanonical entropy, accurate predictions of statistical observables can still be obtained~\cite {calvo2002sampling}. Similarly, the conservation of these quantities depend on the boundary conditions, which are known to often be irrelevant in the thermodynamics limit. The question is more delicate for systems with long-range interactions~\cite {laliena1999effect}.}\BibitemShut {Stop}%
\bibitem [{\citenamefont {Babbar}\ \emph {et~al.}(2025)\citenamefont {Babbar}, \citenamefont {Sadki}, \citenamefont {Prakash},\ and\ \citenamefont {Sondhi}}]{babbar2025classical}%
  \BibitemOpen
  \bibfield  {author} {\bibinfo {author} {\bibfnamefont {A.}~\bibnamefont {Babbar}}, \bibinfo {author} {\bibfnamefont {Y.}~\bibnamefont {Sadki}}, \bibinfo {author} {\bibfnamefont {A.}~\bibnamefont {Prakash}},\ and\ \bibinfo {author} {\bibfnamefont {S.}~\bibnamefont {Sondhi}},\ }\bibfield  {title} {\enquote {\bibinfo {title} {Classical fractons: Local chaos, global broken ergodicity, and an arrow of time},}\ }\href {https://doi.org/10.1103/g2l1-s2vy} {\bibfield  {journal} {\bibinfo  {journal} {Physical Review B}\ }\textbf {\bibinfo {volume} {111}},\ \bibinfo {pages} {245134} (\bibinfo {year} {2025})}\BibitemShut {NoStop}%
\bibitem [{\citenamefont {Huang}\ \emph {et~al.}(2023)\citenamefont {Huang}, \citenamefont {Farrell}, \citenamefont {Friedman}, \citenamefont {Zane}, \citenamefont {Glorioso},\ and\ \citenamefont {Lucas}}]{huang2023generalized}%
  \BibitemOpen
  \bibfield  {author} {\bibinfo {author} {\bibfnamefont {X.}~\bibnamefont {Huang}}, \bibinfo {author} {\bibfnamefont {J.~H.}\ \bibnamefont {Farrell}}, \bibinfo {author} {\bibfnamefont {A.~J.}\ \bibnamefont {Friedman}}, \bibinfo {author} {\bibfnamefont {I.}~\bibnamefont {Zane}}, \bibinfo {author} {\bibfnamefont {P.}~\bibnamefont {Glorioso}},\ and\ \bibinfo {author} {\bibfnamefont {A.}~\bibnamefont {Lucas}},\ }\bibfield  {title} {\enquote {\bibinfo {title} {Generalized time-reversal symmetry and effective theories for nonequilibrium matter},}\ }\href {https://doi.org/10.48550/arXiv.2310.12233} {\bibfield  {journal} {\bibinfo  {journal} {arXiv:2310.12233}\ } (\bibinfo {year} {2023})}\BibitemShut {NoStop}%
\bibitem [{\citenamefont {Guo}, \citenamefont {Glorioso},\ and\ \citenamefont {Lucas}(2022)}]{guo2022fracton}%
  \BibitemOpen
  \bibfield  {author} {\bibinfo {author} {\bibfnamefont {J.}~\bibnamefont {Guo}}, \bibinfo {author} {\bibfnamefont {P.}~\bibnamefont {Glorioso}},\ and\ \bibinfo {author} {\bibfnamefont {A.}~\bibnamefont {Lucas}},\ }\bibfield  {title} {\enquote {\bibinfo {title} {Fracton hydrodynamics without time-reversal symmetry},}\ }\href {https://doi.org/10.1103/PhysRevLett.129.150603} {\bibfield  {journal} {\bibinfo  {journal} {Physical Review Letters}\ }\textbf {\bibinfo {volume} {129}},\ \bibinfo {pages} {150603} (\bibinfo {year} {2022})}\BibitemShut {NoStop}%
\bibitem [{\citenamefont {Han}, \citenamefont {Lake},\ and\ \citenamefont {Ro}(2024)}]{han2024scaling}%
  \BibitemOpen
  \bibfield  {author} {\bibinfo {author} {\bibfnamefont {J.~H.}\ \bibnamefont {Han}}, \bibinfo {author} {\bibfnamefont {E.}~\bibnamefont {Lake}},\ and\ \bibinfo {author} {\bibfnamefont {S.}~\bibnamefont {Ro}},\ }\bibfield  {title} {\enquote {\bibinfo {title} {Scaling and localization in multipole-conserving diffusion},}\ }\href {https://doi.org/10.1103/PhysRevLett.132.137102} {\bibfield  {journal} {\bibinfo  {journal} {Physical Review Letters}\ }\textbf {\bibinfo {volume} {132}},\ \bibinfo {pages} {137102} (\bibinfo {year} {2024})}\BibitemShut {NoStop}%
\bibitem [{\citenamefont {Gillespie}(2007)}]{Gillespie2007Stochastic}%
  \BibitemOpen
  \bibfield  {author} {\bibinfo {author} {\bibfnamefont {D.~T.}\ \bibnamefont {Gillespie}},\ }\bibfield  {title} {\enquote {\bibinfo {title} {Stochastic simulation of chemical kinetics},}\ }\href {https://doi.org/https://doi.org/10.1146/annurev.physchem.58.032806.104637} {\bibfield  {journal} {\bibinfo  {journal} {Annual Review of Physical Chemistry}\ }\textbf {\bibinfo {volume} {58}},\ \bibinfo {pages} {35--55} (\bibinfo {year} {2007})}\BibitemShut {NoStop}%
\bibitem [{\citenamefont {Pretko}, \citenamefont {Chen},\ and\ \citenamefont {You}(2020)}]{pretko2020fracton}%
  \BibitemOpen
  \bibfield  {author} {\bibinfo {author} {\bibfnamefont {M.}~\bibnamefont {Pretko}}, \bibinfo {author} {\bibfnamefont {X.}~\bibnamefont {Chen}},\ and\ \bibinfo {author} {\bibfnamefont {Y.}~\bibnamefont {You}},\ }\bibfield  {title} {\enquote {\bibinfo {title} {Fracton phases of matter},}\ }\href {https://doi.org/10.1142/S0217751X20300033} {\bibfield  {journal} {\bibinfo  {journal} {International Journal of Modern Physics A}\ }\textbf {\bibinfo {volume} {35}},\ \bibinfo {pages} {2030003} (\bibinfo {year} {2020})}\BibitemShut {NoStop}%
\bibitem [{\citenamefont {Gromov}, \citenamefont {Lucas},\ and\ \citenamefont {Nandkishore}(2020)}]{gromov2020fracton}%
  \BibitemOpen
  \bibfield  {author} {\bibinfo {author} {\bibfnamefont {A.}~\bibnamefont {Gromov}}, \bibinfo {author} {\bibfnamefont {A.}~\bibnamefont {Lucas}},\ and\ \bibinfo {author} {\bibfnamefont {R.~M.}\ \bibnamefont {Nandkishore}},\ }\bibfield  {title} {\enquote {\bibinfo {title} {Fracton hydrodynamics},}\ }\href {https://doi.org/10.1103/PhysRevResearch.2.033124} {\bibfield  {journal} {\bibinfo  {journal} {Physical Review Research}\ }\textbf {\bibinfo {volume} {2}},\ \bibinfo {pages} {033124} (\bibinfo {year} {2020})}\BibitemShut {NoStop}%
\bibitem [{\citenamefont {Glorioso}\ \emph {et~al.}(2022)\citenamefont {Glorioso}, \citenamefont {Guo}, \citenamefont {Rodriguez-Nieva},\ and\ \citenamefont {Lucas}}]{glorioso2022breakdown}%
  \BibitemOpen
  \bibfield  {author} {\bibinfo {author} {\bibfnamefont {P.}~\bibnamefont {Glorioso}}, \bibinfo {author} {\bibfnamefont {J.}~\bibnamefont {Guo}}, \bibinfo {author} {\bibfnamefont {J.~F.}\ \bibnamefont {Rodriguez-Nieva}},\ and\ \bibinfo {author} {\bibfnamefont {A.}~\bibnamefont {Lucas}},\ }\bibfield  {title} {\enquote {\bibinfo {title} {Breakdown of hydrodynamics below four dimensions in a fracton fluid},}\ }\href {https://doi.org/10.1038/s41567-022-01631-x} {\bibfield  {journal} {\bibinfo  {journal} {Nature Physics}\ }\textbf {\bibinfo {volume} {18}},\ \bibinfo {pages} {912--917} (\bibinfo {year} {2022})}\BibitemShut {NoStop}%
\bibitem [{\citenamefont {Le~Bellac}(2000)}]{le2000thermal}%
  \BibitemOpen
  \bibfield  {author} {\bibinfo {author} {\bibfnamefont {M.}~\bibnamefont {Le~Bellac}},\ }\href {https://doi.org/10.1017/CBO9780511721700} {\emph {\bibinfo {title} {Thermal field theory}}}\ (\bibinfo  {publisher} {Cambridge university press},\ \bibinfo {year} {2000})\BibitemShut {NoStop}%
\bibitem [{\citenamefont {Jos{\'e}}\ and\ \citenamefont {Saletan}(1998)}]{jose1998classical}%
  \BibitemOpen
  \bibfield  {author} {\bibinfo {author} {\bibfnamefont {J.~V.}\ \bibnamefont {Jos{\'e}}}\ and\ \bibinfo {author} {\bibfnamefont {E.~J.}\ \bibnamefont {Saletan}},\ }\href {https://doi.org/10.1017/CBO9780511803772} {\emph {\bibinfo {title} {Classical dynamics: a contemporary approach}}}\ (\bibinfo  {publisher} {Cambridge university press},\ \bibinfo {year} {1998})\BibitemShut {NoStop}%
\bibitem [{\citenamefont {Prakash}, \citenamefont {Goriely},\ and\ \citenamefont {Sondhi}(2024)}]{prakash2024classical}%
  \BibitemOpen
  \bibfield  {author} {\bibinfo {author} {\bibfnamefont {A.}~\bibnamefont {Prakash}}, \bibinfo {author} {\bibfnamefont {A.}~\bibnamefont {Goriely}},\ and\ \bibinfo {author} {\bibfnamefont {S.}~\bibnamefont {Sondhi}},\ }\bibfield  {title} {\enquote {\bibinfo {title} {Classical nonrelativistic fractons},}\ }\href {https://doi.org/10.1103/PhysRevB.109.054313} {\bibfield  {journal} {\bibinfo  {journal} {Physical Review B}\ }\textbf {\bibinfo {volume} {109}},\ \bibinfo {pages} {054313} (\bibinfo {year} {2024})}\BibitemShut {NoStop}%
\bibitem [{\citenamefont {Osborne}\ and\ \citenamefont {Lucas}(2022)}]{osborne2022infinite}%
  \BibitemOpen
  \bibfield  {author} {\bibinfo {author} {\bibfnamefont {A.}~\bibnamefont {Osborne}}\ and\ \bibinfo {author} {\bibfnamefont {A.}~\bibnamefont {Lucas}},\ }\bibfield  {title} {\enquote {\bibinfo {title} {Infinite families of fracton fluids with momentum conservation},}\ }\href {https://doi.org/10.1103/PhysRevB.105.024311} {\bibfield  {journal} {\bibinfo  {journal} {Physical Review B}\ }\textbf {\bibinfo {volume} {105}},\ \bibinfo {pages} {024311} (\bibinfo {year} {2022})}\BibitemShut {NoStop}%
\bibitem [{\citenamefont {Gabrielli}, \citenamefont {Joyce},\ and\ \citenamefont {Torquato}(2008)}]{gabrielli2008tilings}%
  \BibitemOpen
  \bibfield  {author} {\bibinfo {author} {\bibfnamefont {A.}~\bibnamefont {Gabrielli}}, \bibinfo {author} {\bibfnamefont {M.}~\bibnamefont {Joyce}},\ and\ \bibinfo {author} {\bibfnamefont {S.}~\bibnamefont {Torquato}},\ }\bibfield  {title} {\enquote {\bibinfo {title} {Tilings of space and superhomogeneous point processes},}\ }\href {https://doi.org/10.1103/PhysRevE.77.031125} {\bibfield  {journal} {\bibinfo  {journal} {Physical Review E—Statistical, Nonlinear, and Soft Matter Physics}\ }\textbf {\bibinfo {volume} {77}},\ \bibinfo {pages} {031125} (\bibinfo {year} {2008})}\BibitemShut {NoStop}%
\bibitem [{\citenamefont {Dorfman}, \citenamefont {van Beijeren},\ and\ \citenamefont {Kirkpatrick}(2021)}]{dorfman2021contemporary}%
  \BibitemOpen
  \bibfield  {author} {\bibinfo {author} {\bibfnamefont {J.~R.}\ \bibnamefont {Dorfman}}, \bibinfo {author} {\bibfnamefont {H.}~\bibnamefont {van Beijeren}},\ and\ \bibinfo {author} {\bibfnamefont {T.~R.}\ \bibnamefont {Kirkpatrick}},\ }\href {https://doi.org/10.1017/9781139025942} {\emph {\bibinfo {title} {Contemporary kinetic theory of matter}}}\ (\bibinfo  {publisher} {Cambridge University Press},\ \bibinfo {year} {2021})\BibitemShut {NoStop}%
\bibitem [{\citenamefont {Mukherjee}\ and\ \citenamefont {Pradhan}(2023)}]{mukherjee2023dynamic}%
  \BibitemOpen
  \bibfield  {author} {\bibinfo {author} {\bibfnamefont {A.}~\bibnamefont {Mukherjee}}\ and\ \bibinfo {author} {\bibfnamefont {P.}~\bibnamefont {Pradhan}},\ }\bibfield  {title} {\enquote {\bibinfo {title} {Dynamic correlations in the conserved manna sandpile},}\ }\href {https://doi.org/10.1103/PhysRevE.107.024109} {\bibfield  {journal} {\bibinfo  {journal} {Physical Review E}\ }\textbf {\bibinfo {volume} {107}},\ \bibinfo {pages} {024109} (\bibinfo {year} {2023})}\BibitemShut {NoStop}%
\bibitem [{\citenamefont {Henkel}, \citenamefont {Hinrichsen},\ and\ \citenamefont {L{\"u}beck}(2008)}]{henkel2008non}%
  \BibitemOpen
  \bibfield  {author} {\bibinfo {author} {\bibfnamefont {M.}~\bibnamefont {Henkel}}, \bibinfo {author} {\bibfnamefont {H.}~\bibnamefont {Hinrichsen}},\ and\ \bibinfo {author} {\bibfnamefont {S.}~\bibnamefont {L{\"u}beck}},\ }\href {https://link.springer.com/book/10.1007/978-1-4020-8765-3} {\emph {\bibinfo {title} {Non-Equilibrium Phase Transitions: Volume I: Absorbing Phase Transitions}}}\ (\bibinfo  {publisher} {Springer},\ \bibinfo {year} {2008})\BibitemShut {NoStop}%
\bibitem [{\citenamefont {Wiese}(2016)}]{wiese2016coherent}%
  \BibitemOpen
  \bibfield  {author} {\bibinfo {author} {\bibfnamefont {K.~J.}\ \bibnamefont {Wiese}},\ }\bibfield  {title} {\enquote {\bibinfo {title} {Coherent-state path integral versus coarse-grained effective stochastic equation of motion: From reaction diffusion to stochastic sandpiles},}\ }\href {https://doi.org/10.1103/PhysRevE.93.042117} {\bibfield  {journal} {\bibinfo  {journal} {Physical Review E}\ }\textbf {\bibinfo {volume} {93}},\ \bibinfo {pages} {042117} (\bibinfo {year} {2016})}\BibitemShut {NoStop}%
\bibitem [{\citenamefont {Wiese}(2024)}]{wiese2024hyperuniformity}%
  \BibitemOpen
  \bibfield  {author} {\bibinfo {author} {\bibfnamefont {K.~J.}\ \bibnamefont {Wiese}},\ }\bibfield  {title} {\enquote {\bibinfo {title} {Hyperuniformity in the manna model, conserved directed percolation and depinning},}\ }\href {https://doi.org/10.1103/PhysRevLett.133.067103} {\bibfield  {journal} {\bibinfo  {journal} {Physical Review Letters}\ }\textbf {\bibinfo {volume} {133}},\ \bibinfo {pages} {067103} (\bibinfo {year} {2024})}\BibitemShut {NoStop}%
\bibitem [{\citenamefont {Ma}\ \emph {et~al.}(2025)\citenamefont {Ma}, \citenamefont {Pausch}, \citenamefont {Pruessner},\ and\ \citenamefont {Cates}}]{ma2025hyperuniformity}%
  \BibitemOpen
  \bibfield  {author} {\bibinfo {author} {\bibfnamefont {X.}~\bibnamefont {Ma}}, \bibinfo {author} {\bibfnamefont {J.}~\bibnamefont {Pausch}}, \bibinfo {author} {\bibfnamefont {G.}~\bibnamefont {Pruessner}},\ and\ \bibinfo {author} {\bibfnamefont {M.~E.}\ \bibnamefont {Cates}},\ }\bibfield  {title} {\enquote {\bibinfo {title} {Hyperuniformity at the absorbing state transition: Perturbative rg for random organization},}\ }\href {https://doi.org/10.48550/arXiv.2507.07793} {\bibfield  {journal} {\bibinfo  {journal} {arXiv:2507.07793}\ } (\bibinfo {year} {2025})}\BibitemShut {NoStop}%
\bibitem [{\citenamefont {Caballero}(2024)}]{caballero2024cupss}%
  \BibitemOpen
  \bibfield  {author} {\bibinfo {author} {\bibfnamefont {F.}~\bibnamefont {Caballero}},\ }\bibfield  {title} {\enquote {\bibinfo {title} {cupss: a package for pseudo-spectral integration of stochastic pdes},}\ }\href {https://doi.org/10.48550/arXiv.2405.02410} {\bibfield  {journal} {\bibinfo  {journal} {arXiv:2405.02410}\ } (\bibinfo {year} {2024})}\BibitemShut {NoStop}%
\bibitem [{\citenamefont {Ikeda}\ and\ \citenamefont {Kuroda}(2024)}]{ikeda2024continuous}%
  \BibitemOpen
  \bibfield  {author} {\bibinfo {author} {\bibfnamefont {H.}~\bibnamefont {Ikeda}}\ and\ \bibinfo {author} {\bibfnamefont {Y.}~\bibnamefont {Kuroda}},\ }\bibfield  {title} {\enquote {\bibinfo {title} {Continuous symmetry breaking of low-dimensional systems driven by inhomogeneous oscillatory driving forces},}\ }\href {https://doi.org/10.1103/PhysRevE.110.024140} {\bibfield  {journal} {\bibinfo  {journal} {Physical Review E}\ }\textbf {\bibinfo {volume} {110}},\ \bibinfo {pages} {024140} (\bibinfo {year} {2024})}\BibitemShut {NoStop}%
\bibitem [{\citenamefont {Diessel}, \citenamefont {Kim},\ and\ \citenamefont {Altman}(2025)}]{diessel2025stabilization}%
  \BibitemOpen
  \bibfield  {author} {\bibinfo {author} {\bibfnamefont {O.~K.}\ \bibnamefont {Diessel}}, \bibinfo {author} {\bibfnamefont {J.}~\bibnamefont {Kim}},\ and\ \bibinfo {author} {\bibfnamefont {E.}~\bibnamefont {Altman}},\ }\bibfield  {title} {\enquote {\bibinfo {title} {Stabilization of long-range order in low-dimensional nonequilibrium $ o (n) $ models},}\ }\href {https://arxiv.org/abs/2507.01959} {\bibfield  {journal} {\bibinfo  {journal} {arXiv:2507.01959}\ } (\bibinfo {year} {2025})}\BibitemShut {NoStop}%
\bibitem [{\citenamefont {Cleaver}\ \emph {et~al.}(1996)\citenamefont {Cleaver}, \citenamefont {Care}, \citenamefont {Allen},\ and\ \citenamefont {Neal}}]{cleaver1996extension}%
  \BibitemOpen
  \bibfield  {author} {\bibinfo {author} {\bibfnamefont {D.~J.}\ \bibnamefont {Cleaver}}, \bibinfo {author} {\bibfnamefont {C.~M.}\ \bibnamefont {Care}}, \bibinfo {author} {\bibfnamefont {M.~P.}\ \bibnamefont {Allen}},\ and\ \bibinfo {author} {\bibfnamefont {M.~P.}\ \bibnamefont {Neal}},\ }\bibfield  {title} {\enquote {\bibinfo {title} {Extension and generalization of the gay-berne potential},}\ }\href {https://doi.org/10.1103/PhysRevE.54.559} {\bibfield  {journal} {\bibinfo  {journal} {Physical Review E}\ }\textbf {\bibinfo {volume} {54}},\ \bibinfo {pages} {559} (\bibinfo {year} {1996})}\BibitemShut {NoStop}%
\bibitem [{\citenamefont {Allen}\ and\ \citenamefont {Germano}(2006)}]{allen2006expressions}%
  \BibitemOpen
  \bibfield  {author} {\bibinfo {author} {\bibfnamefont {M.~P.}\ \bibnamefont {Allen}}\ and\ \bibinfo {author} {\bibfnamefont {G.}~\bibnamefont {Germano}},\ }\bibfield  {title} {\enquote {\bibinfo {title} {Expressions for forces and torques in molecular simulations using rigid bodies},}\ }\href {https://doi.org/10.1080/00268970601075238} {\bibfield  {journal} {\bibinfo  {journal} {Molecular Physics}\ }\textbf {\bibinfo {volume} {104}},\ \bibinfo {pages} {3225--3235} (\bibinfo {year} {2006})}\BibitemShut {NoStop}%
\bibitem [{\citenamefont {Anderson}, \citenamefont {Glaser},\ and\ \citenamefont {Glotzer}(2020)}]{anderson2020hoomd}%
  \BibitemOpen
  \bibfield  {author} {\bibinfo {author} {\bibfnamefont {J.~A.}\ \bibnamefont {Anderson}}, \bibinfo {author} {\bibfnamefont {J.}~\bibnamefont {Glaser}},\ and\ \bibinfo {author} {\bibfnamefont {S.~C.}\ \bibnamefont {Glotzer}},\ }\bibfield  {title} {\enquote {\bibinfo {title} {Hoomd-blue: A python package for high-performance molecular dynamics and hard particle monte carlo simulations},}\ }\href {https://doi.org/10.1016/j.commatsci.2019.109363} {\bibfield  {journal} {\bibinfo  {journal} {Computational Materials Science}\ }\textbf {\bibinfo {volume} {173}},\ \bibinfo {pages} {109363} (\bibinfo {year} {2020})}\BibitemShut {NoStop}%
\bibitem [{\citenamefont {Snook}(2006)}]{snook2006langevin}%
  \BibitemOpen
  \bibfield  {author} {\bibinfo {author} {\bibfnamefont {I.}~\bibnamefont {Snook}},\ }\href {https://www.sciencedirect.com/book/9780444521293/the-langevin-and-generalised-langevin-approach-to-the-dynamics-of-atomic-polymeric-and-colloidal-systems} {\emph {\bibinfo {title} {The Langevin and generalised Langevin approach to the dynamics of atomic, polymeric and colloidal systems}}}\ (\bibinfo  {publisher} {Elsevier},\ \bibinfo {year} {2006})\BibitemShut {NoStop}%
\bibitem [{\citenamefont {Lubachevsky}\ and\ \citenamefont {Stillinger}(1990)}]{Lubachevsky_Stillinger_1990}%
  \BibitemOpen
  \bibfield  {author} {\bibinfo {author} {\bibfnamefont {B.~D.}\ \bibnamefont {Lubachevsky}}\ and\ \bibinfo {author} {\bibfnamefont {F.~H.}\ \bibnamefont {Stillinger}},\ }\bibfield  {title} {\enquote {\bibinfo {title} {Geometric properties of random disk packings},}\ }\href {https://doi.org/10.1007/bf01025983} {\bibfield  {journal} {\bibinfo  {journal} {Journal of Statistical Physics}\ }\textbf {\bibinfo {volume} {60}},\ \bibinfo {pages} {561–583} (\bibinfo {year} {1990})}\BibitemShut {NoStop}%
\bibitem [{\citenamefont {Smallenburg}(2022)}]{smallenburg2022efficient}%
  \BibitemOpen
  \bibfield  {author} {\bibinfo {author} {\bibfnamefont {F.}~\bibnamefont {Smallenburg}},\ }\bibfield  {title} {\enquote {\bibinfo {title} {Efficient event-driven simulations of hard spheres},}\ }\href {https://doi.org/10.1140/epje/s10189-022-00180-8} {\bibfield  {journal} {\bibinfo  {journal} {The European Physical Journal E}\ }\textbf {\bibinfo {volume} {45}},\ \bibinfo {pages} {22} (\bibinfo {year} {2022})}\BibitemShut {NoStop}%
\bibitem [{\citenamefont {Dean}(1996)}]{dean1996langevin}%
  \BibitemOpen
  \bibfield  {author} {\bibinfo {author} {\bibfnamefont {D.~S.}\ \bibnamefont {Dean}},\ }\bibfield  {title} {\enquote {\bibinfo {title} {Langevin equation for the density of a system of interacting langevin processes},}\ }\href {https://doi.org/10.1088/0305-4470/29/24/001} {\bibfield  {journal} {\bibinfo  {journal} {Journal of Physics A: Mathematical and General}\ }\textbf {\bibinfo {volume} {29}},\ \bibinfo {pages} {L613} (\bibinfo {year} {1996})}\BibitemShut {NoStop}%
\bibitem [{\citenamefont {Brossollet}\ and\ \citenamefont {Biroli}(2025)}]{brossollet2025entropy}%
  \BibitemOpen
  \bibfield  {author} {\bibinfo {author} {\bibfnamefont {A.}~\bibnamefont {Brossollet}}\ and\ \bibinfo {author} {\bibfnamefont {G.}~\bibnamefont {Biroli}},\ }\bibfield  {title} {\enquote {\bibinfo {title} {Entropy production from density field theories for interacting particles systems},}\ }\href {https://doi.org/10.48550/arXiv.2507.15131} {\bibfield  {journal} {\bibinfo  {journal} {arXiv:2507.15131}\ } (\bibinfo {year} {2025})}\BibitemShut {NoStop}%
\bibitem [{\citenamefont {Klimontovich}(2012)}]{klimontovich2012statistical}%
  \BibitemOpen
  \bibfield  {author} {\bibinfo {author} {\bibfnamefont {Y.~L.}\ \bibnamefont {Klimontovich}},\ }\href {https://link.springer.com/book/10.1007/978-94-011-0175-2} {\emph {\bibinfo {title} {Statistical theory of open systems: Volume 1: A unified approach to kinetic description of processes in active systems}}},\ Vol.~\bibinfo {volume} {67}\ (\bibinfo  {publisher} {Springer Science \& Business Media},\ \bibinfo {year} {2012})\BibitemShut {NoStop}%
\bibitem [{\citenamefont {Barré}\ \emph {et~al.}(2014)\citenamefont {Barré}, \citenamefont {Chétrite}, \citenamefont {Muratori},\ and\ \citenamefont {Peruani}}]{Barré_Chétrite_Muratori_Peruani_2014}%
  \BibitemOpen
  \bibfield  {author} {\bibinfo {author} {\bibfnamefont {J.}~\bibnamefont {Barré}}, \bibinfo {author} {\bibfnamefont {R.}~\bibnamefont {Chétrite}}, \bibinfo {author} {\bibfnamefont {M.}~\bibnamefont {Muratori}},\ and\ \bibinfo {author} {\bibfnamefont {F.}~\bibnamefont {Peruani}},\ }\bibfield  {title} {\enquote {\bibinfo {title} {Motility-induced phase separation of active particles in the presence of velocity alignment},}\ }\href {https://doi.org/10.1007/s10955-014-1008-9} {\bibfield  {journal} {\bibinfo  {journal} {Journal of Statistical Physics}\ }\textbf {\bibinfo {volume} {158}},\ \bibinfo {pages} {589–600} (\bibinfo {year} {2014})}\BibitemShut {NoStop}%
\bibitem [{\citenamefont {Farrell}\ \emph {et~al.}(2012)\citenamefont {Farrell}, \citenamefont {Marchetti}, \citenamefont {Marenduzzo},\ and\ \citenamefont {Tailleur}}]{farrell2012pattern}%
  \BibitemOpen
  \bibfield  {author} {\bibinfo {author} {\bibfnamefont {F.~D.}\ \bibnamefont {Farrell}}, \bibinfo {author} {\bibfnamefont {M.~C.}\ \bibnamefont {Marchetti}}, \bibinfo {author} {\bibfnamefont {D.}~\bibnamefont {Marenduzzo}},\ and\ \bibinfo {author} {\bibfnamefont {J.}~\bibnamefont {Tailleur}},\ }\bibfield  {title} {\enquote {\bibinfo {title} {Pattern formation in self-propelled particles with density-dependent motility},}\ }\href {https://doi.org/10.1103/PhysRevLett.108.248101} {\bibfield  {journal} {\bibinfo  {journal} {Physical review letters}\ }\textbf {\bibinfo {volume} {108}},\ \bibinfo {pages} {248101} (\bibinfo {year} {2012})}\BibitemShut {NoStop}%
\bibitem [{\citenamefont {Solon}\ \emph {et~al.}(2015)\citenamefont {Solon}, \citenamefont {Stenhammar}, \citenamefont {Wittkowski}, \citenamefont {Kardar}, \citenamefont {Kafri}, \citenamefont {Cates},\ and\ \citenamefont {Tailleur}}]{solon2015pressure}%
  \BibitemOpen
  \bibfield  {author} {\bibinfo {author} {\bibfnamefont {A.~P.}\ \bibnamefont {Solon}}, \bibinfo {author} {\bibfnamefont {J.}~\bibnamefont {Stenhammar}}, \bibinfo {author} {\bibfnamefont {R.}~\bibnamefont {Wittkowski}}, \bibinfo {author} {\bibfnamefont {M.}~\bibnamefont {Kardar}}, \bibinfo {author} {\bibfnamefont {Y.}~\bibnamefont {Kafri}}, \bibinfo {author} {\bibfnamefont {M.~E.}\ \bibnamefont {Cates}},\ and\ \bibinfo {author} {\bibfnamefont {J.}~\bibnamefont {Tailleur}},\ }\bibfield  {title} {\enquote {\bibinfo {title} {Pressure and phase equilibria in interacting active brownian spheres},}\ }\href {https://doi.org/10.1103/PhysRevLett.114.198301} {\bibfield  {journal} {\bibinfo  {journal} {Physical review letters}\ }\textbf {\bibinfo {volume} {114}},\ \bibinfo {pages} {198301} (\bibinfo {year} {2015})}\BibitemShut {NoStop}%
\bibitem [{\citenamefont {Nakamura}\ and\ \citenamefont {Yoshimori}(2009)}]{nakamura2009derivation}%
  \BibitemOpen
  \bibfield  {author} {\bibinfo {author} {\bibfnamefont {T.}~\bibnamefont {Nakamura}}\ and\ \bibinfo {author} {\bibfnamefont {A.}~\bibnamefont {Yoshimori}},\ }\bibfield  {title} {\enquote {\bibinfo {title} {Derivation of the nonlinear fluctuating hydrodynamic equation from the underdamped langevin equation},}\ }\href {https://doi.org/10.1088/1751-8113/42/6/065001} {\bibfield  {journal} {\bibinfo  {journal} {Journal of Physics A: Mathematical and Theoretical}\ }\textbf {\bibinfo {volume} {42}},\ \bibinfo {pages} {065001} (\bibinfo {year} {2009})}\BibitemShut {NoStop}%
\bibitem [{\citenamefont {Cornalba}, \citenamefont {Shardlow},\ and\ \citenamefont {Zimmer}(2019)}]{Cornalba_Shardlow_Zimmer_2019}%
  \BibitemOpen
  \bibfield  {author} {\bibinfo {author} {\bibfnamefont {F.}~\bibnamefont {Cornalba}}, \bibinfo {author} {\bibfnamefont {T.}~\bibnamefont {Shardlow}},\ and\ \bibinfo {author} {\bibfnamefont {J.}~\bibnamefont {Zimmer}},\ }\bibfield  {title} {\enquote {\bibinfo {title} {A regularized dean--kawasaki model: Derivation and analysis},}\ }\href {https://doi.org/10.1137/18m1172697} {\bibfield  {journal} {\bibinfo  {journal} {SIAM Journal on Mathematical Analysis}\ }\textbf {\bibinfo {volume} {51}},\ \bibinfo {pages} {1137–1187} (\bibinfo {year} {2019})}\BibitemShut {NoStop}%
\bibitem [{\citenamefont {Lutsko}(2012)}]{Lutsko_2012}%
  \BibitemOpen
  \bibfield  {author} {\bibinfo {author} {\bibfnamefont {J.~F.}\ \bibnamefont {Lutsko}},\ }\bibfield  {title} {\enquote {\bibinfo {title} {A dynamical theory of nucleation for colloids and macromolecules},}\ }\href {http://dx.doi.org/1f0.1063/1.3677191} {\bibfield  {journal} {\bibinfo  {journal} {The Journal of Chemical Physics}\ }\textbf {\bibinfo {volume} {136}} (\bibinfo {year} {2012})}\BibitemShut {NoStop}%
\bibitem [{\citenamefont {Perez-Bastías}\ and\ \citenamefont {Soto}(2025)}]{perezbastías2025twofieldtheoryphasecoexistence}%
  \BibitemOpen
  \bibfield  {author} {\bibinfo {author} {\bibfnamefont {P.}~\bibnamefont {Perez-Bastías}}\ and\ \bibinfo {author} {\bibfnamefont {R.}~\bibnamefont {Soto}},\ }\href {https://arxiv.org/abs/2504.13327} {\enquote {\bibinfo {title} {Two-field theory for phase coexistence of active brownian particles},}\ } (\bibinfo {year} {2025})\BibitemShut {NoStop}%
\bibitem [{\citenamefont {Illien}(2024)}]{illien2024dean}%
  \BibitemOpen
  \bibfield  {author} {\bibinfo {author} {\bibfnamefont {P.}~\bibnamefont {Illien}},\ }\bibfield  {title} {\enquote {\bibinfo {title} {The dean-kawasaki equation and stochastic density functional theory},}\ }\href {https://doi.org/10.1088/1361-6633/adee2e} {\bibfield  {journal} {\bibinfo  {journal} {Reports on Progress in Physics}\ } (\bibinfo {year} {2024})}\BibitemShut {NoStop}%
\bibitem [{\citenamefont {te~Vrugt}, \citenamefont {L{\"o}wen},\ and\ \citenamefont {Wittkowski}(2020)}]{te2020classical}%
  \BibitemOpen
  \bibfield  {author} {\bibinfo {author} {\bibfnamefont {M.}~\bibnamefont {te~Vrugt}}, \bibinfo {author} {\bibfnamefont {H.}~\bibnamefont {L{\"o}wen}},\ and\ \bibinfo {author} {\bibfnamefont {R.}~\bibnamefont {Wittkowski}},\ }\bibfield  {title} {\enquote {\bibinfo {title} {Classical dynamical density functional theory: from fundamentals to applications},}\ }\href {https://doi.org/10.1080/00018732.2020.1854965} {\bibfield  {journal} {\bibinfo  {journal} {Advances in Physics}\ }\textbf {\bibinfo {volume} {69}},\ \bibinfo {pages} {121--247} (\bibinfo {year} {2020})}\BibitemShut {NoStop}%
\bibitem [{\citenamefont {Archer}\ and\ \citenamefont {Rauscher}(2004)}]{archer2004dynamical}%
  \BibitemOpen
  \bibfield  {author} {\bibinfo {author} {\bibfnamefont {A.~J.}\ \bibnamefont {Archer}}\ and\ \bibinfo {author} {\bibfnamefont {M.}~\bibnamefont {Rauscher}},\ }\bibfield  {title} {\enquote {\bibinfo {title} {Dynamical density functional theory for interacting brownian particles: stochastic or deterministic?}}\ }\href {https://doi.org/10.1088/0305-4470/37/40/001} {\bibfield  {journal} {\bibinfo  {journal} {Journal of Physics A: Mathematical and General}\ }\textbf {\bibinfo {volume} {37}},\ \bibinfo {pages} {9325} (\bibinfo {year} {2004})}\BibitemShut {NoStop}%
\bibitem [{\citenamefont {Espanol}(1995)}]{espanol1995hydrodynamics}%
  \BibitemOpen
  \bibfield  {author} {\bibinfo {author} {\bibfnamefont {P.}~\bibnamefont {Espanol}},\ }\bibfield  {title} {\enquote {\bibinfo {title} {Hydrodynamics from dissipative particle dynamics},}\ }\href {https://doi.org/10.1103/PhysRevE.52.1734} {\bibfield  {journal} {\bibinfo  {journal} {Physical Review E}\ }\textbf {\bibinfo {volume} {52}},\ \bibinfo {pages} {1734} (\bibinfo {year} {1995})}\BibitemShut {NoStop}%
\bibitem [{\citenamefont {Bouchet}, \citenamefont {Gawedzki},\ and\ \citenamefont {Nardini}(2016)}]{Bouchet_Gawȩdzki_Nardini_2016}%
  \BibitemOpen
  \bibfield  {author} {\bibinfo {author} {\bibfnamefont {F.}~\bibnamefont {Bouchet}}, \bibinfo {author} {\bibfnamefont {K.}~\bibnamefont {Gawedzki}},\ and\ \bibinfo {author} {\bibfnamefont {C.}~\bibnamefont {Nardini}},\ }\bibfield  {title} {\enquote {\bibinfo {title} {Perturbative calculation of quasi-potential in non-equilibrium diffusions: A mean-field example},}\ }\href {https://doi.org/10.1007/s10955-016-1503-2} {\bibfield  {journal} {\bibinfo  {journal} {Journal of Statistical Physics}\ }\textbf {\bibinfo {volume} {163}},\ \bibinfo {pages} {1157–1210} (\bibinfo {year} {2016})}\BibitemShut {NoStop}%
\bibitem [{\citenamefont {Dawsont}\ and\ \citenamefont {G{\"a}rtner}(1987)}]{dawsont1987large}%
  \BibitemOpen
  \bibfield  {author} {\bibinfo {author} {\bibfnamefont {D.~A.}\ \bibnamefont {Dawsont}}\ and\ \bibinfo {author} {\bibfnamefont {J.}~\bibnamefont {G{\"a}rtner}},\ }\bibfield  {title} {\enquote {\bibinfo {title} {Large deviations from the mckean-vlasov limit for weakly interacting diffusions},}\ }\href {https://doi.org/10.1080/17442508708833446} {\bibfield  {journal} {\bibinfo  {journal} {Stochastics: An International Journal of Probability and Stochastic Processes}\ }\textbf {\bibinfo {volume} {20}},\ \bibinfo {pages} {247--308} (\bibinfo {year} {1987})}\BibitemShut {NoStop}%
\bibitem [{\citenamefont {Dinelli}, \citenamefont {O’Byrne},\ and\ \citenamefont {Tailleur}(2024)}]{dinelli2024fluctuating}%
  \BibitemOpen
  \bibfield  {author} {\bibinfo {author} {\bibfnamefont {A.}~\bibnamefont {Dinelli}}, \bibinfo {author} {\bibfnamefont {J.}~\bibnamefont {O’Byrne}},\ and\ \bibinfo {author} {\bibfnamefont {J.}~\bibnamefont {Tailleur}},\ }\bibfield  {title} {\enquote {\bibinfo {title} {Fluctuating hydrodynamics of active particles interacting via taxis and quorum sensing: static and dynamics},}\ }\href {https://doi.org/10.1088/1751-8121/ad72bc} {\bibfield  {journal} {\bibinfo  {journal} {Journal of Physics A: Mathematical and Theoretical}\ }\textbf {\bibinfo {volume} {57}},\ \bibinfo {pages} {395002} (\bibinfo {year} {2024})}\BibitemShut {NoStop}%
\bibitem [{\citenamefont {Bon}, \citenamefont {Carof},\ and\ \citenamefont {Illien}(2025)}]{bon2025non}%
  \BibitemOpen
  \bibfield  {author} {\bibinfo {author} {\bibfnamefont {L.~L.}\ \bibnamefont {Bon}}, \bibinfo {author} {\bibfnamefont {A.}~\bibnamefont {Carof}},\ and\ \bibinfo {author} {\bibfnamefont {P.}~\bibnamefont {Illien}},\ }\bibfield  {title} {\enquote {\bibinfo {title} {Non-gaussian density fluctuations in the dean-kawasaki equation},}\ }\href {https://doi.org/10.48550/arXiv.2501.16206} {\bibfield  {journal} {\bibinfo  {journal} {arXiv:2501.16206}\ } (\bibinfo {year} {2025})}\BibitemShut {NoStop}%
\bibitem [{\citenamefont {Bixon}\ and\ \citenamefont {Zwanzig}(1969)}]{bixon1969boltzmann}%
  \BibitemOpen
  \bibfield  {author} {\bibinfo {author} {\bibfnamefont {M.}~\bibnamefont {Bixon}}\ and\ \bibinfo {author} {\bibfnamefont {R.}~\bibnamefont {Zwanzig}},\ }\bibfield  {title} {\enquote {\bibinfo {title} {Boltzmann-langevin equation and hydrodynamic fluctuations},}\ }\href {https://doi.org/10.1103/PhysRev.187.267} {\bibfield  {journal} {\bibinfo  {journal} {Physical Review}\ }\textbf {\bibinfo {volume} {187}},\ \bibinfo {pages} {267} (\bibinfo {year} {1969})}\BibitemShut {NoStop}%
\bibitem [{\citenamefont {Lefevre}\ and\ \citenamefont {Biroli}(2007)}]{lefevre2007dynamics}%
  \BibitemOpen
  \bibfield  {author} {\bibinfo {author} {\bibfnamefont {A.}~\bibnamefont {Lefevre}}\ and\ \bibinfo {author} {\bibfnamefont {G.}~\bibnamefont {Biroli}},\ }\bibfield  {title} {\enquote {\bibinfo {title} {Dynamics of interacting particle systems: stochastic process and field theory},}\ }\href {https://doi.org/10.1088/1742-5468/2007/07/P07024} {\bibfield  {journal} {\bibinfo  {journal} {Journal of Statistical Mechanics: Theory and Experiment}\ }\textbf {\bibinfo {volume} {2007}},\ \bibinfo {pages} {P07024} (\bibinfo {year} {2007})}\BibitemShut {NoStop}%
\bibitem [{\citenamefont {Riccardo}\ \emph {et~al.}(2019)\citenamefont {Riccardo}, \citenamefont {Riccardo}, \citenamefont {Ramirez-Pastor},\ and\ \citenamefont {Pasinetti}}]{riccardo2019multiple}%
  \BibitemOpen
  \bibfield  {author} {\bibinfo {author} {\bibfnamefont {J.~J.}\ \bibnamefont {Riccardo}}, \bibinfo {author} {\bibfnamefont {J.~L.}\ \bibnamefont {Riccardo}}, \bibinfo {author} {\bibfnamefont {A.~J.}\ \bibnamefont {Ramirez-Pastor}},\ and\ \bibinfo {author} {\bibfnamefont {P.~M.}\ \bibnamefont {Pasinetti}},\ }\bibfield  {title} {\enquote {\bibinfo {title} {Multiple exclusion statistics},}\ }\href {https://doi.org/10.1103/PhysRevLett.123.020602} {\bibfield  {journal} {\bibinfo  {journal} {Physical Review Letters}\ }\textbf {\bibinfo {volume} {123}},\ \bibinfo {pages} {020602} (\bibinfo {year} {2019})}\BibitemShut {NoStop}%
\bibitem [{\citenamefont {Wang}\ and\ \citenamefont {Hazzard}(2025)}]{wang2025particle}%
  \BibitemOpen
  \bibfield  {author} {\bibinfo {author} {\bibfnamefont {Z.}~\bibnamefont {Wang}}\ and\ \bibinfo {author} {\bibfnamefont {K.~R.}\ \bibnamefont {Hazzard}},\ }\bibfield  {title} {\enquote {\bibinfo {title} {Particle exchange statistics beyond fermions and bosons},}\ }\href {https://doi.org/10.1038/s41586-024-08262-7} {\bibfield  {journal} {\bibinfo  {journal} {Nature}\ }\textbf {\bibinfo {volume} {637}},\ \bibinfo {pages} {314--318} (\bibinfo {year} {2025})}\BibitemShut {NoStop}%
\bibitem [{\citenamefont {T{\"a}uber}(2014)}]{tauber2014critical}%
  \BibitemOpen
  \bibfield  {author} {\bibinfo {author} {\bibfnamefont {U.~C.}\ \bibnamefont {T{\"a}uber}},\ }\href {https://doi.org/10.1017/CBO9781139046213} {\emph {\bibinfo {title} {Critical dynamics: a field theory approach to equilibrium and non-equilibrium scaling behavior}}}\ (\bibinfo  {publisher} {Cambridge University Press},\ \bibinfo {year} {2014})\BibitemShut {NoStop}%
\bibitem [{\citenamefont {Niiyama}\ \emph {et~al.}(2009)\citenamefont {Niiyama}, \citenamefont {Shimizu}, \citenamefont {Kobayashi}, \citenamefont {Okushima},\ and\ \citenamefont {Ikeda}}]{niiyama2009effect}%
  \BibitemOpen
  \bibfield  {author} {\bibinfo {author} {\bibfnamefont {T.}~\bibnamefont {Niiyama}}, \bibinfo {author} {\bibfnamefont {Y.}~\bibnamefont {Shimizu}}, \bibinfo {author} {\bibfnamefont {T.~R.}\ \bibnamefont {Kobayashi}}, \bibinfo {author} {\bibfnamefont {T.}~\bibnamefont {Okushima}},\ and\ \bibinfo {author} {\bibfnamefont {K.~S.}\ \bibnamefont {Ikeda}},\ }\bibfield  {title} {\enquote {\bibinfo {title} {Effect of translational and angular momentum conservation on energy equipartition in microcanonical equilibrium in small clusters},}\ }\href {https://doi.org/10.1103/PhysRevE.79.051101} {\bibfield  {journal} {\bibinfo  {journal} {Physical Review E—Statistical, Nonlinear, and Soft Matter Physics}\ }\textbf {\bibinfo {volume} {79}},\ \bibinfo {pages} {051101} (\bibinfo {year} {2009})}\BibitemShut {NoStop}%
\bibitem [{\citenamefont {Calvo}, \citenamefont {Galindez},\ and\ \citenamefont {Gad{\'e}a}(2002)}]{calvo2002sampling}%
  \BibitemOpen
  \bibfield  {author} {\bibinfo {author} {\bibfnamefont {F.}~\bibnamefont {Calvo}}, \bibinfo {author} {\bibfnamefont {J.}~\bibnamefont {Galindez}},\ and\ \bibinfo {author} {\bibfnamefont {F.}~\bibnamefont {Gad{\'e}a}},\ }\bibfield  {title} {\enquote {\bibinfo {title} {Sampling the configuration space of finite atomic systems: How ergodic is molecular dynamics?}}\ }\href {https://doi.org/10.1021/jp013691+} {\bibfield  {journal} {\bibinfo  {journal} {The Journal of Physical Chemistry A}\ }\textbf {\bibinfo {volume} {106}},\ \bibinfo {pages} {4145--4152} (\bibinfo {year} {2002})}\BibitemShut {NoStop}%
\bibitem [{\citenamefont {Laliena}(1999)}]{laliena1999effect}%
  \BibitemOpen
  \bibfield  {author} {\bibinfo {author} {\bibfnamefont {V.}~\bibnamefont {Laliena}},\ }\bibfield  {title} {\enquote {\bibinfo {title} {Effect of angular momentum conservation in the phase transitions of collapsing systems},}\ }\href {https://doi.org/10.1103/PhysRevE.59.4786} {\bibfield  {journal} {\bibinfo  {journal} {Physical Review E}\ }\textbf {\bibinfo {volume} {59}},\ \bibinfo {pages} {4786} (\bibinfo {year} {1999})}\BibitemShut {NoStop}%
\end{thebibliography}%

\end{document}